\documentclass[a4paper,11pt]{article}
\usepackage{float}
\usepackage{amssymb}
\usepackage{graphicx}
\usepackage{hyperref}
\usepackage[cal=esstix]{mathalfa}
\usepackage{amsmath}
\usepackage [english]{babel}
\usepackage [autostyle, english = american]{csquotes}
\MakeOuterQuote{"}
\usepackage{amsfonts}
\usepackage{rotating}
\usepackage{pdflscape}
\usepackage[official]{eurosym}
\usepackage{longtable}
\usepackage[left=2.5cm,right=2.5cm,top=2cm,bottom=2cm]{geometry}
\usepackage{xspace}
\usepackage{array}
\usepackage{latexsym}
\usepackage{bm}
\usepackage{bbm}
\usepackage{xcolor}
\usepackage{color,soul}
\usepackage{microtype}
\usepackage{afterpage}

\usepackage[flushleft]{threeparttable}
\usepackage{rotating}
\usepackage{tabularx}
\usepackage[section]{placeins}

\usepackage[toc,page]{appendix}

\usepackage{adjustbox}

\usepackage{xspace}
\usepackage{authblk}

\usepackage[round]{natbib}
\usepackage{amssymb}

\usepackage{subfiles}
\usepackage{subcaption}

\newcolumntype{A}{>{\centering\arraybackslash}p{2cm}} 
\newcolumntype{D}{>{\centering\arraybackslash}p{1.25cm}}
\newcolumntype{B}{>{\raggedright\arraybackslash}p{4.3cm}}
\newcolumntype{C}{>{\centering\arraybackslash}p{1.7cm}} 

\begin{document}

\title{\Large \textbf{Nudging Nutrition: Lessons from the Danish ``Fat Tax''}}

\date{\vspace{-20pt} Version: \today{}. \\ }

\author{
Christian M. Dahl\footnote{Department of Economics, University of Southern Denmark.}
\hspace{9pt}
Nadja van 't Hoff\protect\footnotemark[1] 
\hspace{9pt}
Giovanni Mellace\protect\footnotemark[1]
\hspace{9pt}
Sinne Smed\footnote{Department of Food and Resource Economics, University of Copenhagen.}
}

\maketitle

\thispagestyle{empty}
\addtocounter{page}{-1}

\begin{abstract}
\noindent
In October 2011, Denmark introduced the world's first and, to date, only tax targeting saturated fat. However, this tax was subsequently abolished in January 2013. Leveraging exogenous variation from untaxed Northern-German consumers, we employ a difference-in-differences approach to estimate the causal effects of both the implementation and repeal of the tax on consumption and expenditure behavior across eight product categories targeted by the tax. Our findings reveal significant heterogeneity in the tax's impact across these products.
During the taxed period, there was a notable decline in consumption of bacon, liver sausage,  and cheese, particularly among low-income households. In contrast, expenditure on butter, cream, and margarine increased as prices rose. Interestingly, we do not observe any difference in expenditure increases between high and low-income households, suggesting that the latter were disproportionately affected by the tax.
After the repeal of the tax, we do not observe any significant decline in consumption. On the contrary, there was an overall increase in consumption for certain products, prompting concerns about unintended consequences resulting from the brief implementation of the tax. Finally, we find strong evidence on an overall increase purchases of butter  abroad for households living  less than 50 km from the German boarder but we do not find strong evidence of spatial heterogeneous effects of the tax. 

\bigskip 
{\small \noindent \textbf{Keywords:} Fat tax, consumer responses, difference-in-differences, health policy. } \bigskip 
\newline
{\small \noindent \textbf{JEL classification: D12, I18, H31, C21.} \quad }
\end{abstract}

\vfill

\footnotesize{We express our gratitude for the financial support provided by the Independent Research Fund Denmark, grant 6109-00075B, and Helsefonden, grant 18-B-0130.
Additionally, we acknowledge the valuable feedback received from Phillip Heiler, Volha Lazuka, Toru Kitagawa, Michael Lechner, Pedro Sant' Anna, Jeff Smith, Anthony Wray, and the participants of the 2021 DGPE workshop, the participants of the ``Applied Econometrics of Program Evaluation'' course, as well as the participants of the 2024 EALE conference. Special thanks are extended to Kasper Veje Jakobsen,  Lajza Prekazi,  and Joachim Dahl Steensbjerre for their exceptional research assistance.}

\newpage

\newpage

\section{Introduction}

Taxation has long been employed as a policy instrument to influence consumer behavior, correct market failures, and promote social welfare. In economics, taxes serve as incentives or disincentives, aiming to internalize externalities associated with the consumption or production of certain goods \citep{pigou1920economics}. In the context of public health, taxes on tobacco, alcohol, sugary beverages, and unhealthy foods are designed to reduce consumption of harmful products, thereby improving health outcomes and reducing associated healthcare costs \citep{chaloupka2000economics, allcott2019regressive}.

Economic theory suggests that imposing a tax on goods with negative externalities should lead to a decrease in their consumption, as higher prices deter consumers (Pigouvian taxes). Additionally, such taxes can generate government revenue that can be allocated to further mitigate the externality. However, the effectiveness of these taxes depends on the price elasticity of demand for the taxed goods, the availability of substitutes, and the ability of consumers and producers to adjust their behavior \citep{chetty2009salience}.

Empirical studies have documented both intended and unintended consequences of taxation policies. For instance, taxes on sugar-sweetened beverages (SSBs) have been shown to reduce consumption \citep{falbe2016impact, fichera2021consumers}, but also, in some contexts, to lead to substitution effects where consumers switch to other unhealthy products \citep{miracolo2021sin}. Cross-border shopping to evade taxes is another unintended consequence that can undermine the policy's effectiveness \citep{beatty2011state}. Moreover, the regressive nature of such taxes can disproportionately affect lower-income households, raising equity concerns \citep{allcott2019regressive, wright2017policy}.

The Danish "fat tax," implemented between October 2011 and January 2013, provides a compelling case study to examine both the intended and unintended effects of taxation as an incentive. Aimed at reducing the intake of saturated fats linked to cardiovascular diseases \citep{hooper2020reduction}, the tax was levied on foods with saturated fat content exceeding 2.3\%, including products like butter, margarine, meats, and dairy products, excluding drinking milk and milk-based yogurts (see, \cite{smed2012financial} for more institutional details). Despite its potential significance for public health, there is limited empirical evidence on the effectiveness of the Danish fat tax in altering consumer behavior, leaving policymakers with little guidance on its actual impact.

Previous studies on food taxes have often relied on simulation models using price elasticities to predict consumer responses \citep{chouinard2007fat, allais2010effects, nordstrom2009impact}, but these lack causal interpretations and may not capture real-world complexities \citep{powell2009food}. Moreover, many analyses assume full pass-through of the tax to consumer prices, which may not hold in practice due to strategic pricing by firms \citep{cawley2017pass}. Empirical evidence using real-world data is scarce, and existing studies on the Danish fat tax lack proper control groups, hindering causal inference \citep{jensen2013danish, bodker2015danish, smed2016effects,jensen2016effects}.

Our study addresses this gap by leveraging a unique dataset that includes both Danish households affected by the fat tax and a control group of Northern German households. Using a difference-in-differences (DID) approach \citep{callaway2021difference}, we estimate the causal effects of the fat tax on consumption and expenditure behaviors across different food groups. This methodology allows us to isolate the impact of the tax from other contemporaneous factors, providing robust evidence on both intended and unintended consequences.

Our findings contribute to the broader literature on taxation as an incentive by highlighting the nuanced effects of the Danish fat tax. Specifically, we find that while the tax was effective in reducing the consumption of only two products—cheese and liver sausage—it did not significantly reduce consumption of other high-fat products like butter, cream, and margarine. Instead, for these products, we observe significant increases in expenditure without corresponding reductions in consumption. This suggests that the tax led to higher prices that consumers, especially those with lower price sensitivity, were willing to pay, indicating an inelastic demand for these goods.

Furthermore, our analysis reveals unintended consequences such as cross-border shopping and the regressive nature of the tax. We find substantial evidence of increased cross-border purchases of butter by Danish households living close to the German border during the tax period. Despite this increase in cross-border shopping, the overall consumption of butter did not decrease significantly, implying that these avoidance behaviors mitigated the intended effect of the tax on reducing saturated fat intake. Additionally, we observe geographical variations in the tax's impact on expenditure, with households farther from the border experiencing higher price increases due to the tax being fully passed through to consumers in those areas.

Importantly, we also find that lower-income households were more affected by the tax, exhibiting significant reductions in consumption of certain products. This raises concerns about the equity implications of such policies, as the tax may disproportionately burden lower-income individuals.

By placing our results within the broader economic literature on taxation and consumer behavior, we shed light on the complexities of using taxes as incentives. Our study underscores the importance of considering demand elasticities, substitution effects, geographical factors, and heterogeneity among consumers when designing tax policies to achieve public health objectives without unintended adverse effects. Policymakers should account for the ease of cross-border shopping and the elasticity of demand in different regions to ensure the effectiveness and equity of such taxes.

The remainder of this paper is organized as follows: In Section \ref{sec:literature_background}, we discuss the institutional background of the Danish fat tax. Subsequently, Section \ref{sec:data_identification_estimation} offers an overview of the data, detailing our identification strategy and the underlying assumptions. The results are presented in Section \ref{sec:results}, where we focus on each product category individually and examine potential price mechanisms. Section \ref{sec:robustness_checks} addresses concerns related to our identification strategy, while Section\ref{sec:heterogeneity} explores potential heterogeneity in tax responses. In Section \ref{sec:cross-border}, we provide some descriptive evidence on cross-border shopping of taxed products, estimate geographical variations in the tax's impact, and potential spillover
effects on German households living close to the border. The paper concludes in Section \ref{sec:conclusion}.

\bigskip


\section{Background on the Danish fat tax}
\label{sec:literature_background}

In October 2011, Denmark introduced a tax targeting saturated fat, with the aim of improving dietary health and with the expectation of raising a tax revenue of 1 billion DKK (156.5 million USD\footnote{A conversion rate of 6.39 DKK per USD is used throughout the paper.}) \citep{bodker2015danish, smed2012financial}.
The tax was 16 DKK (2.51 USD) per kilo of saturated fat on foods exceeding a threshold limit of 2.3\% saturated fat. Including the 25\% value-added tax imposed on all products in Denmark, this increases to 20 DKK (3.13 USD) per kilo. 
The threshold was set such that drinking milk and milk-based yogurts were excluded from the tax. The tax predominantly affected products such as meat, dairy, oil, margarine, butter, and butter blends as these products contain considerable amounts of saturated fat. 
Producers had the option to use standardized coefficients for meat products instead of the actual fat content in the products. 
These coefficients were animal-specific (that is, there was one coefficient for pork products, one for beef, etc.), and based on the average content of saturated fat in the various types of meat consumed in Denmark \citep{smed2012financial}. 
Generally these coefficients were used instead of specific rates for different meat cuts since the latter option set higher requirements for documentation and thereby had higher administration costs. 
The administration of the fat tax was burdensome and estimated to have a retail and whole-sale cost of 200 million DKK (approximately 31 million USD) according to a report by the Danish Chamber of Commerce and the Confederation of Danish Industry \citep{bodker2015rise}.

The Danish Tax Committee anticipated an increase in consumer prices up to 7\% for meats, up to 33\% for dairy, up to 35\% for animal fat, up to 16\% for cooking oils, up to 24\% for margarine and up to 20\% for blended butter products \citep{bodker2015danish}. \cite{authority} analyzed the price changes of 27 taxed products and found that, for only six of these products, the consumer price increases exceeded what could be explained by the general food price index trends. For some of these products, they observed an overshifting of the tax, i.e., retailers increased the price on meat by more than what could be justified by the tax, increasing their profit margin by 1\% to 5\%. This finding is supported by \cite{jensen2016effects} who found a price increase of 13\% to 16\% for high-fat minced beef and cream products. These estimated increases were higher than the a priori expected tax-induced price increases. The same study found small negative effect on prices for lower-fat products \citep{jensen2016effects}. 

In an agreement in November 2012, the government decided that the fat tax should be repealed with effect from January 2013. The main reasons stated were increased consumer prices, large administration cost, and the endangerment of Danish jobs \citep{authority}.\footnote{For more information on the Danish tax on saturated fat see \cite{smed2012financial} and \cite{ecorys2014food}. For more information on the repeal of the Danish tax on saturated fat see \cite{vallgaarda2015danish}.}

\bigskip


\section{Identification strategy, data, estimation, and inference}
\label{sec:data_identification_estimation}

\subsection{Identification strategy}
\label{sec:identification}

Our study's key strength lies in observing a suitable control group—Northern German households from Schleswig-Holstein and Hamburg—which enables us to identify the causal effect of the Danish fat tax on consumption behavior. This control group helps remove time trends and unobserved demand shocks \citep{cawley2019economics}, and their geographical proximity and cultural similarity to Danish households increase the likelihood of experiencing similar external shocks.
The regions of Schleswig and Holstein (Now Schleswig-Holstein in Germany) as well as Northern Schleswig (now Southern Jutland in Denmark) have been under alternating  Danish and German governance since the Middle Ages. This intertwined governance led to a unique blend of Danish and German cultures in these areas, leaving Danish minorities in German territories and German minorities in Danish regions. 

Today, the Danish-German border region is therefore characterized by significant cultural and economic integration. The Danish minority in South Schleswig, Germany, and the German minority in North Schleswig, Denmark, enjoy protections and support from both governments. The Copenhagen-Bonn Declarations of 1955 formalized minority rights, allowing for the preservation of language, education, and cultural practices. These minorities operate their own schools, libraries, churches, and cultural institutions, fostering strong cultural ties across the border. Economic connections are also robust, with cross-border trade and labor mobility contributing to the regional economy. Shared industries, agricultural practices, and retail markets further intertwine the regions economically.

 These deep cultural and economic ties between Denmark and Northern Germany make the latter an appropriate control group for our study. The similarities in consumer behavior, dietary patterns, and market dynamics enhance the validity of our DID approach.


\subsection{Data}
\label{sec:data}

To estimate the impact of the Danish fat tax on consumption behavior, we leverage household panel data from GfK Consumer Scan, which encompasses households located in Denmark as well as households in Northern Germany, specifically Schleswig-Holstein and Hamburg.\footnote{For more information on the Danish data set, see \cite{smed2008empirical}.} 
It is worth noting that these Danish household data are also used in the analysis of \cite{jensen2013danish} and \cite{smed2016effects}, yet they lacked the Northern-German households as a control group and the length of the data periods considered differ.

GfK maintains a representative panel with respect to geographic location, education, and age. \footnote{See,  \href{https://panel.gfk.com/scan-dk/hjem}{https://panel.gfk.com/scan-dk/hjem}} When a household leaves the panel, it is replaced by a similar household with respect to the aforementioned characteristics. 
Participants in the panel are equipped with a hand scanning device, enabling them to document detailed information on their weekly grocery shopping.
The diary keeper reports the volume of foods purchased for the household, the expenditure, date and time of the purchase, among other details. 
While surveys and self-reported data are often susceptible to recall bias, this is less likely to be a concern when using scanning devices.
A scanning device can ease registration of purchase information, which subsequently might be experienced as less burdensome to the households than registering food consumption manually.
Moreover, our data provides an advantage over store scanner data by incorporating purchases made abroad. However, we acknowledge that our data do not account for food acquired outside of stores, such as in restaurants.
The purchase diaries are submitted to GfK weekly and thoroughly checked for correct completion and consistency. As highlighted by \cite{smed2008empirical}, a comparison of the budget-shares of foods from the GfK consumer panel with budget-shares revealed from data of Statistics Denmark shows no major biases for the distribution of expenses between basic food categories, even though the home-scan panel generally under-report in terms of absolute values. 

In our analysis, we consider a wide variety of product categories targeted to a different degree by the tax, encompassing bacon, butter and butter blends (referred to as butter), hard cheeses (referred to as cheese), cream, liver sausage, margarine, salami, and sour cream. \footnote{We found no significant differences between butter and butter blends, so we combined them into a single category.}
The product categories were first and foremost chosen based on their substantial contribution to the saturated fat intake in Denmark, secondly based on comparability  between Germany and Denmark as well as an adequate number of observations.

Our main analysis uses data covering the period from January 2009 until December 2014. The data are aggregated to a quarterly household panel, encompassing 11 quarters before the tax introduction and 13 quarters after the tax introduction.   
Furthermore, we create product-specific panels where households that never purchase from the considered product category are excluded. If the number of zero purchases for a given product changes systematically over time differentially between Denmark and Northern Germany, our results could be affected by sample selection bias.  However, Figure \ref{fig:zero_observations} below shows similar pre-trends in the average number of observations with zero outcome between Denmark and Northern Germany. In general, for all product, we do not find any systematic change in the difference in the number of zeros even in the post-treatment period. This may be due to our use of quarterly data, which significantly reduces the number of excluded households and the potential for sample selection bias.  

\begin{figure}[!htbp]
    \centering
    
   \begin{subfigure}{0.45\textwidth}
        \centering
        \includegraphics[width=\textwidth]{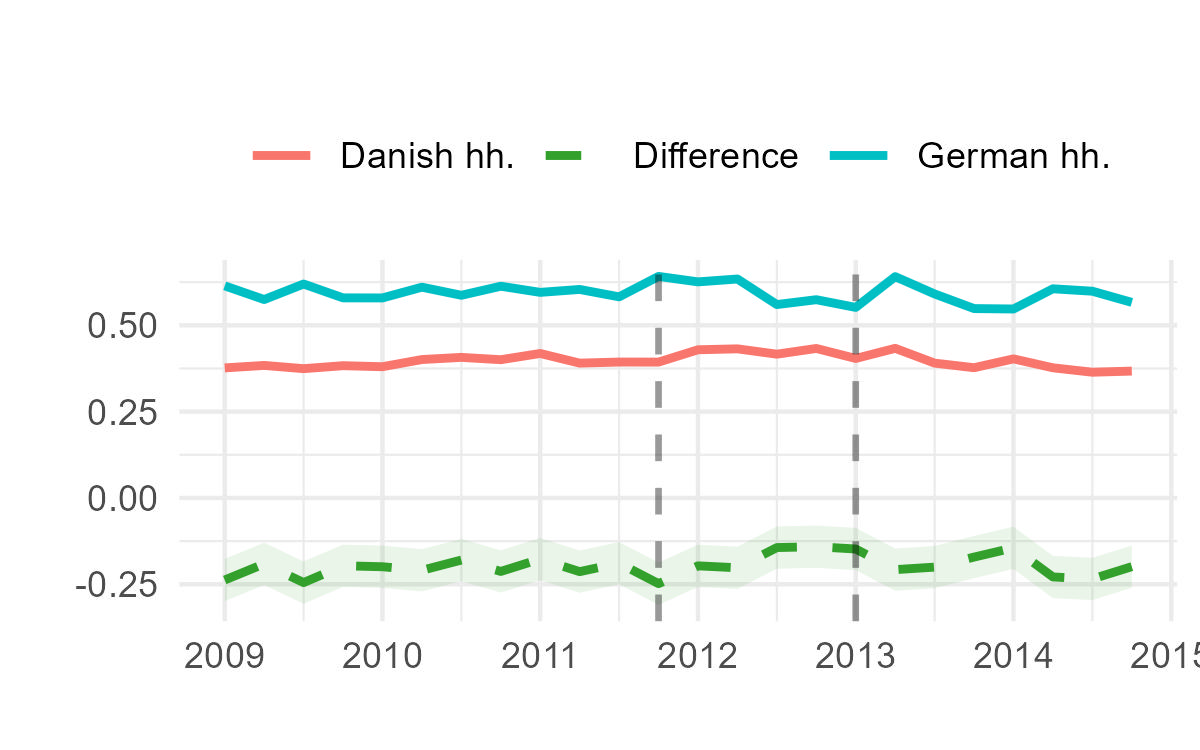}
        \caption{Bacon.}
        \label{fig:zero_observations_bacon}
   \end{subfigure}
    \hfill
   \begin{subfigure}{0.45\textwidth}
        \centering
        \includegraphics[width=\textwidth]{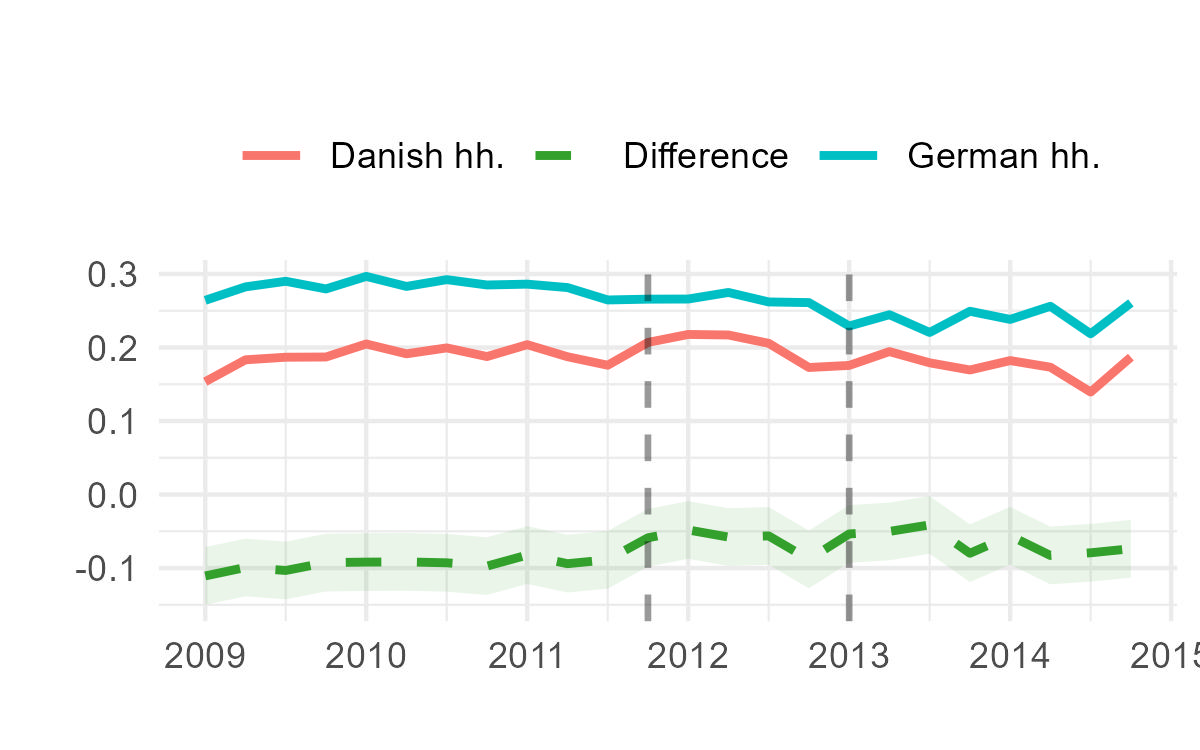}
        \caption{Butter.}
        \label{fig:zero_observations_butter}
   \end{subfigure}
    
     \begin{subfigure}{0.45\textwidth}
        \centering
        \includegraphics[width=\textwidth]{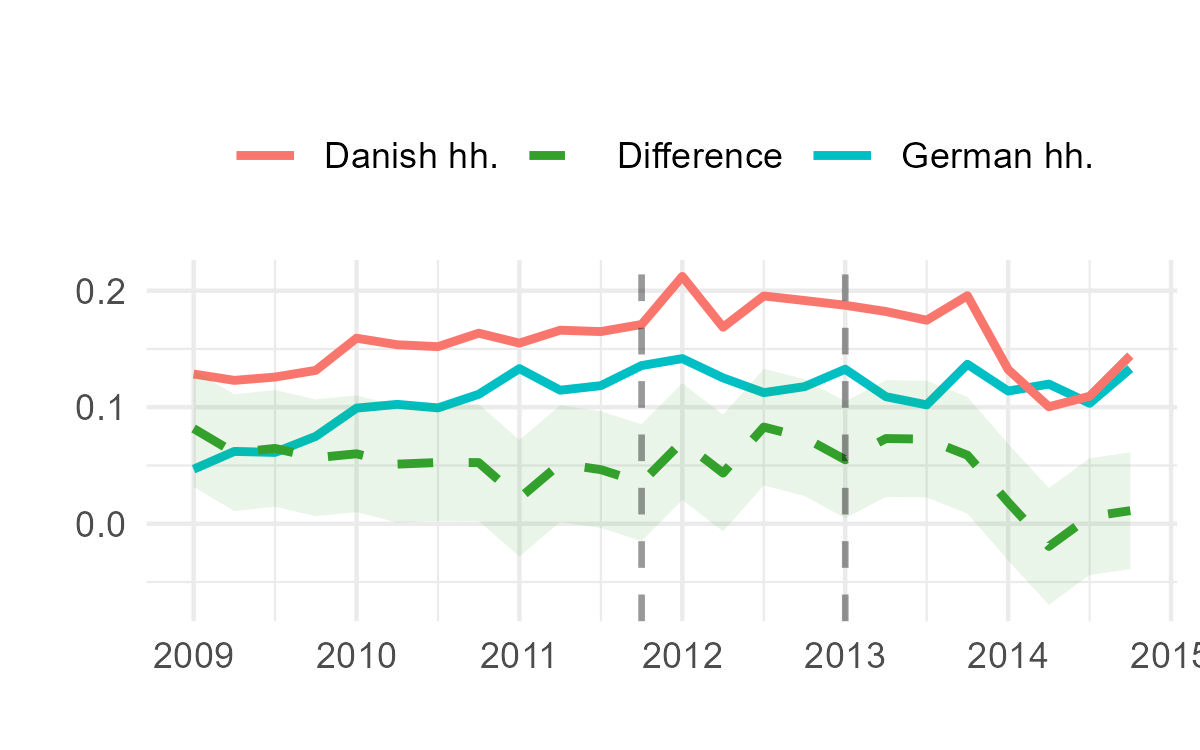}
        \caption{Cheese.}
        \label{fig:zero_observations_cheese}
   \end{subfigure}
    \hfill
     \begin{subfigure}{0.45\textwidth}
        \centering
        \includegraphics[width=\textwidth]{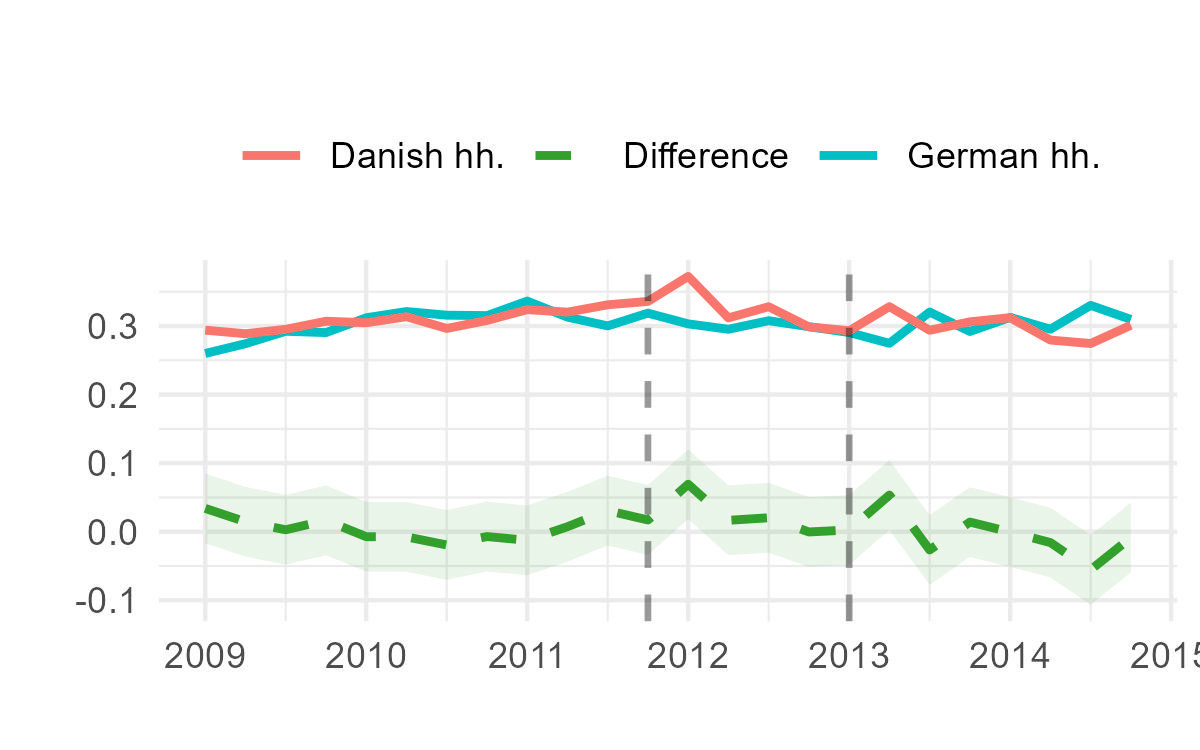}
        \caption{Cream.}
        \label{fig:zero_observations_cream}
   \end{subfigure}

     \begin{subfigure}{0.45\textwidth}
        \centering
        \includegraphics[width=\textwidth]{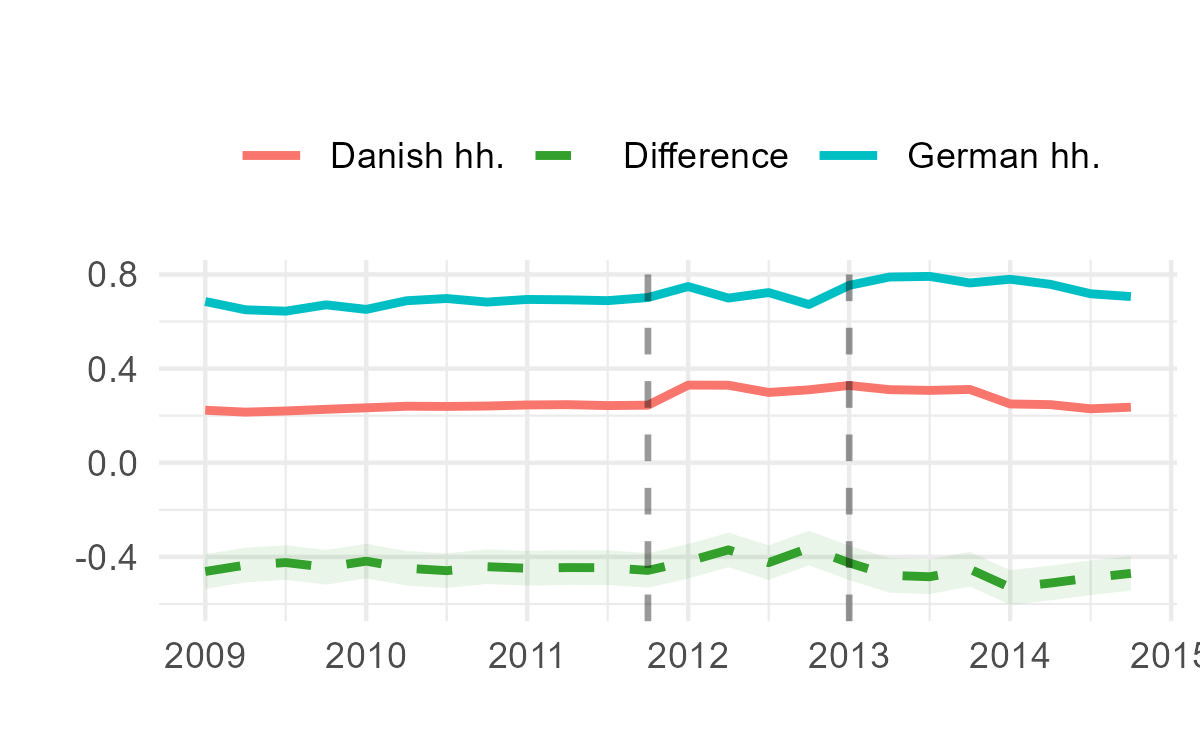}
        \caption{Liver sausage.}
        \label{fig:zero_observations_liver}
   \end{subfigure}
    \hfill
     \begin{subfigure}{0.45\textwidth}
        \centering
        \includegraphics[width=\textwidth]{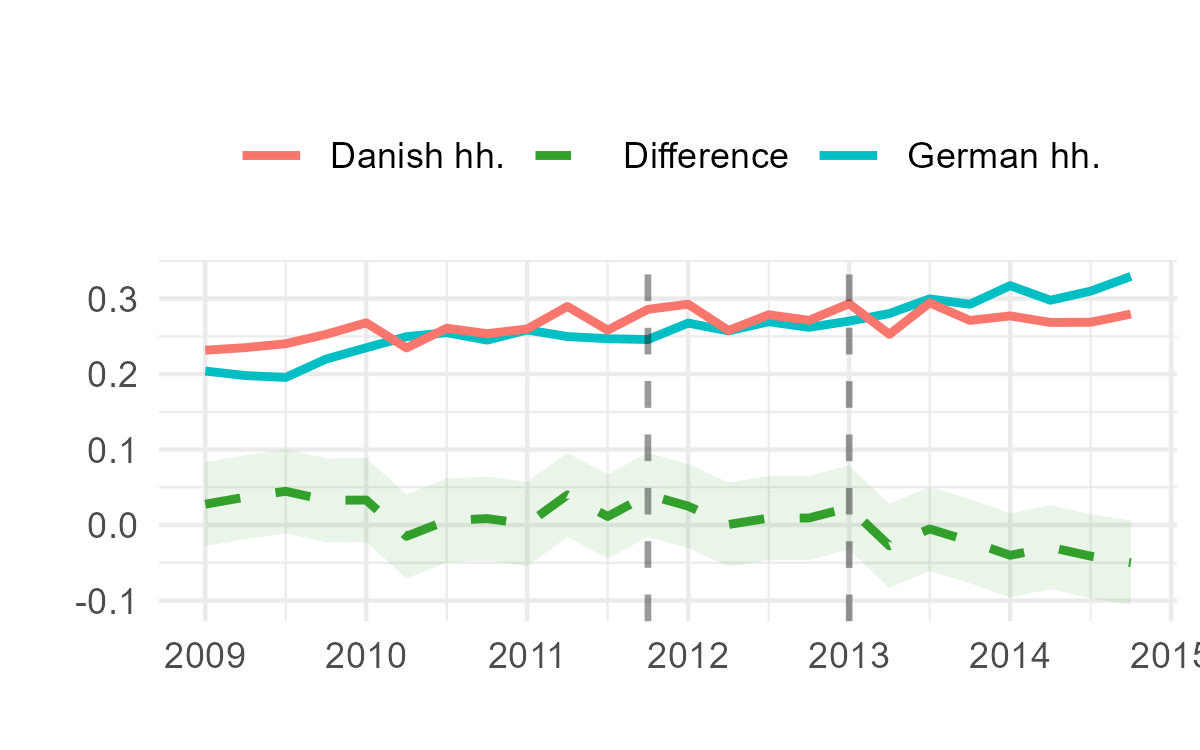}
        \caption{Margarine.}
        \label{fig:zero_observations_margarine}
   \end{subfigure}

         \begin{subfigure}{0.45\textwidth}
        \centering
        \includegraphics[width=\textwidth]{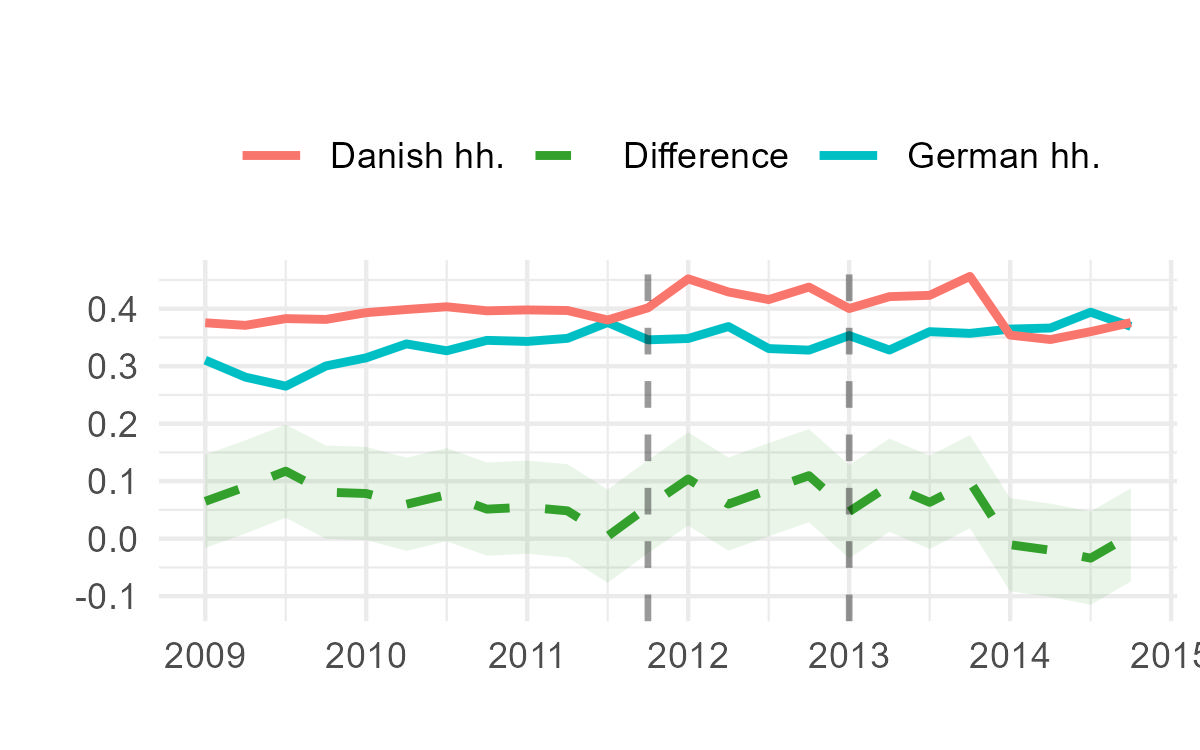}
        \caption{Salami.}
        \label{fig:zero_observations_salami}
   \end{subfigure}
    \hfill
     \begin{subfigure}{0.45\textwidth}
        \centering
        \includegraphics[width=\textwidth]{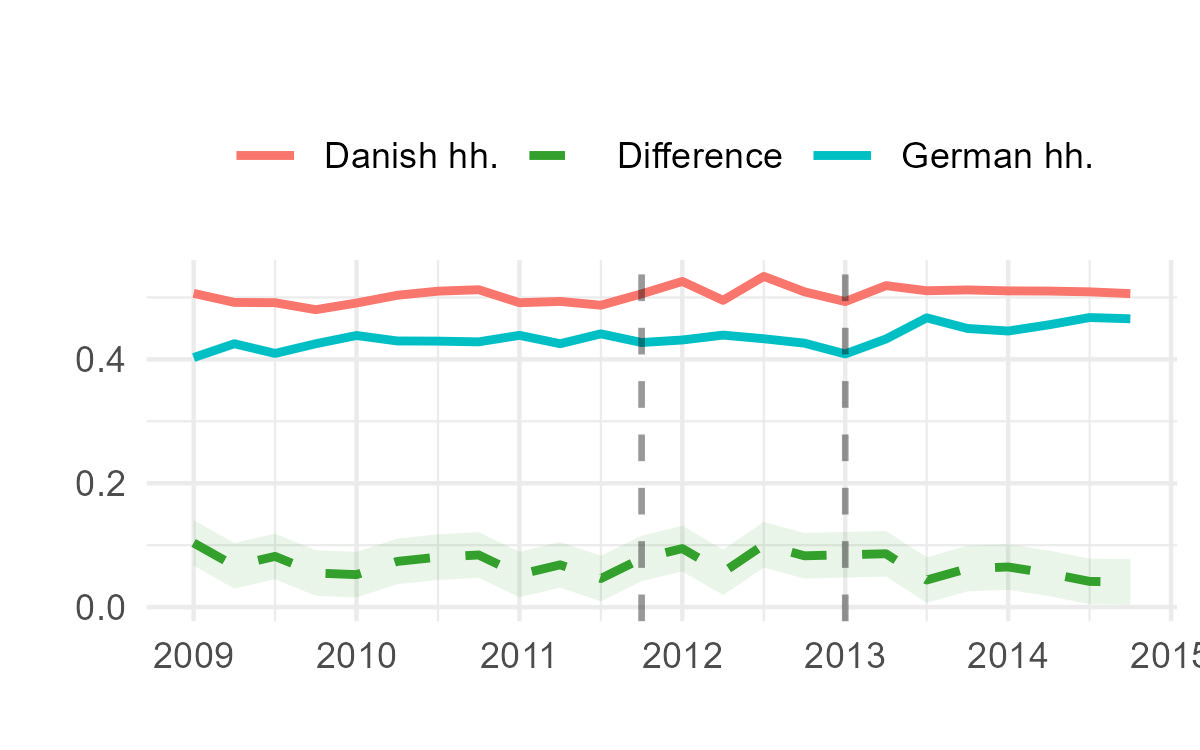}
        \caption{Sour cream.}
        \label{fig:zero_observations_Sour_cream}
    \end{subfigure}
 \caption{Trends in the average number of zero-outcome observations across various products between Denmark and Northern Germany. \label{fig:zero_observations} }  
\end{figure}

Figures \ref{fig:data_chart_dk} and \ref{fig:data_chart_de} in Appendix \ref{app:flowcharts} illustrate the data preparation steps and the number of available observations for each step.
Our final analysis incorporates between 2,056 and 2,526 unique Danish households, depending on the product category, and 553 to 1,616 German households.

The data further contain background information on the participating households, which is based on an annual questionnaire filled in by the diary keeper. This includes information on the geographic region of residence, the age of the head of the household, the number of children below age 15, the household size, income level, gender of the diary keeper, and education level.   We harmonize the variables that are reported differently in the Danish and German data sets, such as e.g. income and education.\footnote{The education level is based on the International Standard Classification of Education (ISCED).} To control for income, we harmonize the income variable, as it is reported in different intervals in the two datasets. To ensure a valid comparison of households with similar income levels, we generate separate quintiles for Germany and Denmark using the midpoints of each income range, incorporating all available households in each country. These quintiles categorize households into five income levels: very low, low, medium, high, and very high income.
For our analysis, we use the most recently observed pre-tax characteristics for a household. That is, if a household's income is observed in both the years 2009 and 2010, we use the income level observed in 2010.
Pre-treatment characteristics of the households in our study are summarized in Table \ref{tab:sum_stat} in Appendix \ref{app:summary_characteristics}.\footnote{\cite{smed2016effects} use the same data and compare the descriptive statistics to official numbers from Statistics Denmark. They observe that the panel contains relatively more households in the capital and households with a lower level of educational attainment.}
There are some notable differences between the Northern-German and Danish households. For instance, the head of the household tend to be older in the Germany data compared to the Danish data. Additionally, there is a noticeable gender distribution distinction among diary keepers, with 78\% being female in Denmark as opposed to only 59\% in Germany. Moreover, households in Germany tend to be larger compared to those in Denmark. 
Since the household characteristics listed in the table are included as control variables in our difference-in-differences strategy, we address the concern that differences in consumption trends between Germany and Denmark might be caused by imbalances in sample composition. 

To investigate consumers' response to the introduction of the fat tax, we define three distinct outcomes of interest. Firstly, we define three variables to investigate consumption and expenditure behavior: (1) the average quarterly household consumption, measured by the average weight or volume in gr or ml purchased per quarter per household, (2) the average number of packages purchased per household, and (3) the average quarterly total expenditure per household in Euro cents. 
Note that we use product purchases as a proxy for consumption, recognizing that a purchase does not necessarily indicate consumption.
To investigate the price changes caused by the tax, we construct a variable for the average quarterly price paid per 100 units (gr or ml depending on the product) in Euro cents, allowing us to investigate whether the tax was passed on to consumers by retailers and producers.\footnote{Danish expenditure is converted from DKK to Euro cents using the rate 7.4405.} 
Additionally, since quarters have different number of days we  re-weight our outcomes such that each quarter consist of the same amount of days. Finally, we partial out the effect of seasonal trends from all outcomes. To achieve this we simply run two separate linear regressions, one for each country, for each outcome on quarterly dummies and we use the residuals of these regressions as outcomes.\footnote{The pre-tax and tax average quantities of our outcomes of interest are presented in Table \ref{tab:dep_quants} in Appendix \ref{app:pretax_tax_dep}.} Outliers are removed using Tukey's fences with the constant $k=3$.

\bigskip


\subsection{Causal Effects Definition and Estimation}
\label{sec:estimation}

The quarterly panel we use ranges from 2009 to 2014, encompassing 20 quarters. Denote a particular quarter with $q$ where $q \in \{2009\textls{-}Q1,\allowbreak 2009\textls{-}Q2,\allowbreak 2009\textls{-}Q3, \allowbreak ..., \allowbreak 2014\textls{-}Q4\}$.
Denote with $Y_{iq}$ the outcome of interest for household $i$ in quarter $q$. As described in Section \ref{sec:data}, three different outcomes concerning consumption behavior (weight, packages, and expenditure) are studied. In addition we also estimate the effect of the tax on average prices paid per 100 units (gr or ml depending on the product).
The tax, our treatment, was introduced at the beginning of the fourth quarter of 2011, meaning that the third quarter of 2011 ($2011\text{-}Q3$) was the last pre-treatment quarter, and it was abolished at the beginning of the first quarter of 2013; thus, $2012\text{-}Q4$ is the last period where the tax was still in place. Let $D_{i}=1$ if unit $i$ is a Danish household, and $D_{i}=0$ otherwise. Define $Y_{pre}$ as the reference period, e.g., $2011\text{-}Q3$. It is important to note that the reference period varies depending on the causal effect of interest, and the precise reference period used will be explicitly stated on a case-by-case basis.
Ignoring for the moment pre-treatment covariates, when all the treated units are treated at the same time, as in our case, the $ATT(q)$ are identical to the ATTs in the two-period DID approach and are identified by:
\begin{equation}\label{eq:estimand}
    ATT(q) = E(Y_q-Y_{pre}|D=1)-E(Y_q-Y_{pre}|D=0).
\end{equation}
When controlling for pre-treatment covariates, denoted by $X$, the ATT in each quarter is identified by the following expression:
\begin{equation}\label{eq:dr_did}
    ATT(q) = E \left[ \left(
        \frac{D}{E(D)} - \frac{\frac{p(X)(1-D)}{1-p(X)}}{E\left[ \frac{p(X)(1-D)}{1-p(X)} \right]}\right)
    (Y_q-Y_{pre} - m_q(X)) \right],
\end{equation}
where the propensity score $p(X)$ denotes the probability of being subject to the tax, conditional on the pre-tax households characteristics $X$, and $m_q(X)=E(Y_q-Y_{pre}|X,D=0)$ denotes the population outcome regression for the Northern-German households in quarter $q$.

Equations (\ref{eq:estimand}) and (\ref{eq:dr_did}) provide the ATT for each individual quarter. 
A straightforward approach to establish the overall average causal effect for a specific period is to aggregate the time average treatment effects, $ATT(q)$, of that period. For example, we can aggregate the $ATT(q)$ in the tax period as follows:
\begin{align}\label{eq:overall_att}
    ATT = \frac{1}{5}\sum_{q\in \{ 2011\textls{-}Q4,...,2012\textls{-}Q4\}} ATT(q).
\end{align}
Simply put, this is the mean over the $ATT(q)$ for the period that the tax was in effect.

We adopt the estimator for event-study designs proposed by \cite{callaway2021difference}, which extends the doubly robust estimator of \cite{sant2020doubly} to the setting with multiple time periods.\footnote{Note that \cite{callaway2021difference} additionally consider the setting with variation in treatment timing (staggered adoption), where individuals belong to a certain group depending on the timing of their treatment. However, in our context, the number of groups equals one as the tax was simultaneously introduced and repealed for all individuals.} \footnote{We tailored the R package of \cite{did_package} to our setting and express our gratitude to the original authors for making their code publicly available.} The main advantage of this approach over other estimators, such as the common two-way fixed effects estimator (see, for instance, \cite{roth2023s} for a recent review of the DID literature), is that it allows to control for observed covariates in a flexible way.

Estimation proceeds in two steps. In the first step, the propensity score $p(x)$ and outcome model $m_q(X)$ are estimated using logistic and linear regressions, respectively. In the second step, the fitted values of the two models are plugged into the sample analogue of Equation (\ref{eq:dr_did}).
The doubly robust DID estimator attains the semiparametric efficiency bound when both nuisance models are correctly specified. A strength of using \cite{callaway2021difference} is that they provide simultaneous confidence bands which account for multiple-testing and are obtained by a multiplier-type bootstrap procedure.

Some assumptions that are standard in literature need to be satisfied for Equations (\ref{eq:estimand}) to (\ref{eq:overall_att}) to be valid.
The first assumption rules out anticipation effects.  This implies that Danish households did not adjust their consumption behavior in anticipation of the tax, e.g., by hoarding. As the tax was adopted approximately half a year before its implementation, we cannot exclude the presence of hoarding. We rigorously study the presence of  anticipation effects of the fat tax in Section \ref{sec:plausibility_assumptions}. 

The second assumption is the parallel trends assumption.
This assumption states that in the absence of the Danish fat tax, the outcome trends in Denmark would have followed the same trends as the ones observed in Northern-Germany during the tax period. 
If the composition of the treatment and control groups are different in terms of household characteristics, then this might cause the German and Danish trends to diverge in case of heterogeneous treatment effects, leading to a violation of the unconditional parallel trends assumption \citep{roth2023s}.
To address this concern, we relax the parallel trends assumption slightly and allow for the presence of covariate-specific trends by conditioning on pre-tax household characteristics.
When controlling for observed covariates  we also need to impose the standard overlap assumption, i.e., it must hold that for all $X$ that $p(X)<0$. This assumption ensures that for each Danish household, we can find German households with similar characteristics. 
We investigate the plausibility of the parallel trends and overlap assumptions in Section \ref{sec:plausibility_assumptions}.

It is crucial to note that tests for the parallel trends assumption offer insights into pre-tax trend similarities. However, the parallel trends assumption is essentially untestable as the post treatment trends that Denmark would have experienced had the tax not been introduced  cannot be observed. This means that this assumption could be perfectly plausible in the pre-tax period, while violated in the post-tax period, and vice-versa. 
Nonetheless, observing similar pre-tax trends provides evidence in favor of the parallel trends assumption.

In Section \ref{sec:robustness_checks}, we describe the robustness  checks we have performed.  First, we exclude Danish households living close to the German border, since the introduction of a tax can lead to increased border trade to avoid the tax.
Second, we investigate how much our results rely on controlling for household characteristics by estimating the effects unconditionally. 

\bigskip


\section{Results}
\label{sec:results}


\subsection{Plausibility of identifying assumptions}
\label{sec:plausibility_assumptions}

We first examine the plausibility of the no anticipation assumption which is crucial for DID. As mentioned in Section \ref{sec:identification}, this assumption posits that households did not change their behavior in anticipation of the tax implementation. 
To explore this, we estimate the effect of the tax in the two quarters preceding its introduction, namely the second and third quarters of 2011. The tax was announced on March 17, 2011, and introduced in October, 2011.
The results reported in Table \ref{tab:anticipation_quarters} in Appendix \ref{app:ae-twfe}  do not show any statistically significant anticipation effect. Nevertheless, we take a conservative stand and, to avoid any potential bias due to anticipation effects, we exclude the second and third quarter of 2011 from the analysis. We find qualitatively similar results including these two quarters.\footnote{The results are available from the authors upon request.} 

A second aforementioned crucial assumption for the credibility of our estimates is the parallel trends assumption. 
An advantage of having multiple pre-tax time periods is the opportunity to gain some insight into the plausibility of this assumption.
We assess the plausibility in two different ways. 
First, Table \ref{tab:main_results} contains the p-values of a Wald test for parallel trends, where the assumption of parallel trends in the pre-tax period is tested under the null hypothesis that they are parallel. Notably, the p-values derived from this approach do not raise concerns, as we fail to reject the null hypothesis of parallel trends, with the exception of  butter in the expenditure outcome.
Second, the results of pseudo treatment effects for the pre-tax periods are shown in Appendix \ref{app:dynamic_treatments}. Reassuringly, the uniform confidence intervals in all pre-tax periods contain zero, offering further evidence in support of the parallel trends assumption for all product categories and all outcomes. 
Altogether, inspection of the pre-tax trends lends strong support of the plausibility of parallel trends between Danish and Northern-German households in the tax period.

Finally, we address the overlap assumption, which is commonly tested by ensuring sufficient overlap in the treated and control units' propensity score distributions. Histograms displaying the fitted propensity scores for the Danish and Northern-German households can be found in Appendix \ref{app:fitted_prop_scores}.
The histograms show good overlap.

In conclusion, we do not find any substantial empirical evidence against the plausibility of the assumptions underlying our DID identification strategy.

\bigskip


\subsection{Two-way fixed effects estimates}\label{app:two_way_fixed_est}
The Two-Way Fixed Effects (TWFE) estimates reported in Table \ref{tab:twe} in Appendix \ref{app:ae-twfe}  indicate that the Danish fat tax had differential impacts across various high-fat food products. The tax effectively reduced consumption and expenditure for items like bacon, cheese, liver sausage, and salami, demonstrating a significant decrease in their intake. Conversely, for products such as butter and sour cream, expenditure increased despite minor reductions in consumption, suggesting inelastic demand where consumers were willing to pay higher prices without substantially altering their purchasing habits.

\subsection{The causal effect of the fat tax on consumer responses during the tax period}
As mentioned in the introduction, we employ the \citeauthor{callaway2021difference} estimator in our main specification instead of the traditional TWFE  model. This choice is motivated by the estimator's ability to provide more robust inference and to flexibly incorporate covariates, thereby enhancing the reliability and validity of our results, particularly in the presence of treatment effect heterogeneity.

The results depicted in Panel A of Table \ref{tab:main_results}
 are generally consistent with those obtained from the TWFE estimates but with some notable differences. Specifically, we observe a statistically significant decrease in the consumption of cheese and liver sausage during the tax period. There is an average reduction of approximately 13.36\% for cheese (p-value = 0.058) and 10.56\% for liver sausage (p-value = 0.019) compared to the pre-tax quarter. The decrease in cheese consumption is significant at the 10\% level, while the reduction in liver sausage consumption is significant at the 5\% level.

For salami, although there is a reduction of 14.38\%, the p-value is 0.270, indicating that this decrease is not statistically significant. Similarly, we do not observe significant changes in the consumption of bacon, butter, cream, margarine, or sour cream during the tax period.

In Panel B of Table \ref{tab:main_results}, we see substantial increases in expenditure for butter, cream, margarine, and sour cream, despite the absence of any discernible impact on their consumption. Specifically, expenditures rose by  2.35 euro cents (25.20\%) for butter (p-value = 0.000), 0.71 (14.49\%) for cream (p-value = 0.019), 0.85 euro cents (15.42\%) for margarine (p-value = 0.005), and 0.27 euro cents (16.03\%) for sour cream (p-value = 0.092). These increases are statistically significant, with butter and margarine significant at the 1\% level, cream at the 5\% level, and sour cream at the 10\% level.

Conversely, we observe a decrease in expenditure for liver sausage by 0.55 euro cents (9.19\%), which is marginally significant (p-value = 0.052). This aligns with the decreased consumption of liver sausage. Expenditure decreases for bacon and salami are not statistically significant, indicating that the tax did not significantly affect spending on these products.

These results suggest that the tax led to higher prices for butter, cream, margarine, and sour cream, which consumers were willing to pay without significantly reducing their consumption—indicating inelastic demand for these products. In contrast, the tax effectively reduced both the consumption and expenditure of cheese and liver sausage.

\begin{table}[bt]
\centering
\caption{The causal effect of the Danish fat tax on the average quarterly consumption and expenditure during the tax period (October 2011 to December 2012).}
\label{tab:main_results}
\begin{adjustbox}{max width=\textwidth}
\begin{tabular}{B *{8}{C}}
  \hline \hline
  & (1) & (2) & (3) & (4) & (5) & (6) & (7) & (8) \\
 & Bacon & Butter & Cheese & Cream & Liver sausage & Margarine & Salami & Sour cream \\ 
  \hline
  \multicolumn{9}{l}{\textbf{Panel A: Quarterly average weight consumed in gr/ml}} \\ \hline
  
  \textit{Danish fat tax (ATT)} & $-64.363$ & 73.292 & $-345.752^*$ & 27.663 & $-115.680^{**}$ & 114.433 & $-54.267$ & 27.768 \\ 
  \textit{(se)} & (31.976) & (81.409) & (127.169) & (63.224) & (33.229) & (95.311) & (34.480) & (37.123) \\ 
  \textit{p-value} & 0.166 & 0.536 & 0.058 & 0.755 & 0.019 & 0.414 & 0.270 & 0.602\\
  \textit{\% change} & $-11.39$\% & 4.81\% & $-13.36$\% & 2.01\% & $-10.56$\% & 6.24\% & $-14.38$\% & 4.85\% \\ 
  \textit{obs.} & 28,874 & 40,929 & 32,020 & 39,274 & 29,417 & 38,687 & 30,800 & 36,121  \\ 
  \textit{p-val. pre-trend} & 0.092 & 0.384 & 0.333 & 0.455 & 0.196 & 0.772 & 0.978 & 0.208 \\ 
    \hline
  \multicolumn{9}{l}{\textbf{Panel B: Quarterly average expenditure in Euro cents}} \\ \hline
  \textit{Danish fat tax (ATT)} & $-26.947$ & $234.879^{***}$ & $-243.321^{*}$ & $70.823^{**}$ & $-54.601^{*}$ & $84.643^{***}$ & $-53.808$ & $27.131^{*}$ \\ 
  \textit{(se)} & (22.878) & (36.546) & (104.101) & (21.156) & (19.624) & (20.909) & (34.438) & (10.745) \\ 
  \textit{p-value} & 0.430 & 0.000 & 0.086 & 0.019 & 0.052 & 0.005 & 0.275 & 0.092\\
  \textit{\% change} & -7.36\% & 25.20\% & -10.91\% & 14.49\% & -9.19\% & 15.42\% & -11.98\% & 16.03\% \\ 
  \textit{obs.} & 28,874 & 40,929 & 32,020 & 39,274 & 29,417 & 38,687 & 30,800 & 36,121 \\  
  \textit{p-val. pre-trend} & 0.175 & 0.000 & 0.501 & 0.748 & 0.304 & 0.603 & 0.994 & 0.075 \\ 
   \hline \hline
   \multicolumn{9}{p{1.3\textwidth}}{\normalsize Significance level: * $p<0.1$ ** $p<0.05$ *** $p<0.01$. Standard errors clustered at unit level. \textit{\% change} compares the estimated tax average to the pre-tax sample mean for Danish households. The estimates are obtained using the doubly robust DID approach, controlling for age of head of household, number of children under 15, gender, ISCED, income level, and household size.}\\ 
\end{tabular}
\end{adjustbox}
\end{table}

Our findings highlight the nuanced effects of the fat tax across different products. While the tax was effective in reducing consumption of certain high-fat items like cheese and liver sausage, it did not achieve the same outcome for other products such as butter and cream. Instead, consumers continued purchasing these goods despite higher prices, leading to increased expenditures. This underscores the importance of considering demand elasticities and potential substitution effects when designing taxation policies aimed at altering consumer behavior to achieve public health objectives.

\bigskip

\subsection{The causal effect of the fat tax on consumer responses after its abolishment}

The fat tax was abolished in January 2013. To gain insight into potential habit formation induced by the tax, we study the two years following its abolishment, comparing consumption and expenditure to the pre-tax period. We present the effects of the abolishment of the fat tax on consumption and expenditure in Table \ref{tab:results_abolishment}.

For butter, margarine, and sour cream, we find that consumption significantly increased after the tax was abolished, as shown in Panel A of Table \ref{tab:results_abolishment}. Specifically, butter consumption increased by 256.95 grams per quarter (a 16.05\% increase, p-value = 0.023), margarine consumption increased by 325.94 grams per quarter (an 18.05\% increase, p-value = 0.030), and sour cream consumption increased by 90.65 grams per quarter (a 15.41\% increase, p-value = 0.083).

These increases in consumption were accompanied by significant increases in expenditure for these products, as shown in Panel B of Table \ref{tab:results_abolishment}. Expenditure on butter increased by 162.86 euro cents (16.92\%, p-value = 0.003), and expenditure on sour cream increased by 37.95 euro cents (21.65\%, p-value = 0.010). Expenditure on cream also showed an increase of 57.76 euro cents (11.39\%, p-value = 0.055), which is marginally significant.

For other products such as cheese, liver sausage, and salami, we do not observe significant changes in consumption or expenditure after the tax was abolished. The ATT estimates for these products are not statistically significant, indicating that consumption levels returned to or remained at pre-tax levels.

Overall, despite the short-term reduction in consumption observed during the tax period for some products, these effects did not persist after the tax was abolished. Instead, the consumption of certain high-fat products increased, surpassing pre-tax levels in some cases. This suggests that the tax did not lead to lasting changes in consumer behavior.

These findings are consistent with previous research, such as \citet{allais2010effects}, which indicates that long-term effects of food taxes on consumption can take several years to establish. In our study, the abolishment of the tax led to a rebound in consumption, highlighting the challenges in achieving sustained dietary changes through taxation alone.

In summary, the fat tax failed to create any long-term reductions in the consumption of the targeted products. Its abolishment was followed by significant increases in consumption and expenditure for products like butter, margarine, and sour cream. This underscores the importance of considering not only the immediate impact of such taxes but also their long-term effectiveness.

\begin{table}[!hbt]
\centering
\caption{The causal effect of the Danish fat tax on the average quarterly consumption and expenditure during the two years after abolishment of the tax (January 2013 to December 2014), using the pre-tax period (January 2011 to March 2011) as reference period.}
\label{tab:results_abolishment}
\begin{adjustbox}{max width=\textwidth}
\begin{tabular}{B *{8}{C}}
  \hline \hline
  & (1) & (2) & (3) & (4) & (5) & (6) & (7) & (8) \\
 & Bacon & Butter & Cheese & Cream & Liver sausage & Margarine & Salami & Sour cream \\
  \hline
  \multicolumn{9}{l}{\textbf{Panel A: Quarterly average weight consumed in gr/ml}} \\ \hline
  \textit{Danish fat tax (ATT)} & -18.080 & $256.953^{**}$ & 29.770 & 117.048 & -5.385 & $325.944^{**}$ & 19.274 & $90.648^{*}$ \\ 
  \textit{(se)} & (30.325) & (77.633) & (128.601) & (69.297) & (33.609) & (101.442) & (33.354) & (34.894) \\ 
  \textit{p-value} & 0.692 & 0.023 & 0.874 & 0.242 & 0.914 & 0.030 & 0.699 & 0.083\\
  \textit{\% change} & -3.13\% & 16.05\% & 1.14\% & 8.31\% & -0.49\% & 18.05\% & 4.90\% & 15.41\% \\ 
  \textit{obs.} & 32,648 & 45,295 & 35,949 & 43,648 & 33,178 & 42,954 & 34,711 & 40,299 \\ 
  \textit{p-val. pre-trend} & 0.051 & 0.362 & 0.251 & 0.369 & 0.168 & 0.853 & 0.980 & 0.138 \\ \hline
  \multicolumn{9}{l}{\textbf{Panel B: Quarterly average expenditure in Euro cents}} \\ \hline 
  \textit{Danish fat tax (ATT)} & 3.263 & $162.860^{***}$ & 9.141 & $57.764^{*}$ & -3.958 & 39.009 & 7.371 & $37.947^{**}$ \\ 
  \textit{(se)} & (24.635) & (36.912) & (98.288) & (20.108) & (19.625) & (22.154) & (33.382) & (9.741) \\ 
  \textit{p-value} & 0.928 & 0.003 & 0.949 & 0.055 & 0.892 & 0.229 & 0.880 & 0.010\\
  \textit{\% change} & 0.86\% & 16.92\% & 0.40\% & 11.39\% & -0.67\% & 7.66\% & 1.59\% & 21.65\% \\ 
  \textit{obs.} & 32,648 & 45,295 & 35,949 & 43,648 & 33,178 & 42,954 & 34,711 & 40,299 \\ 
  \textit{p-val. pre-trend} & 0.124 & 0.000 & 0.230 & 0.668 & 0.212 & 0.734 & 0.995 & 0.050  \\ 
   \hline \hline
   \multicolumn{9}{p{1.3\textwidth}}{\normalsize Significance level: * $p<0.1$ ** $p<0.05$ *** $p<0.01$. Standard errors clustered at unit level. \textit{\% change} compares the estimated tax average to the pre-tax sample mean for Danish households. The estimates are obtained using the doubly robust DID approach, controlling for age of head of household, number of children under 15, gender, ISCED, income level, and household size.}\\ 
\end{tabular}
\end{adjustbox}
\end{table}

\bigskip


\subsection{Producer responses}
\label{sec:mechanism}

\subsubsection{Price mechanism}

Since the tax was levied on producers, it remains unclear to what extent consumers experienced price increases for the taxed products. In this section, we aim to estimate the causal effect of the tax on the prices paid for each product. One limitation of this approach is that we can only observe prices for products that have been purchased, not the overall price level. However, the results presented in Table \ref{tab:results_price_mechanism} provide us with insights into whether consumers encountered higher prices for the products they chose to purchase.

Table \ref{tab:results_price_mechanism} shows that during the tax period, prices for butter, cream, margarine, and sour cream significantly increased in response to the tax by 23.44\%, 10.58\%, 11.35\%, and 12.31\%, respectively. These price increases are lower than those anticipated by the Danish Tax Committee, which expected increases in consumer prices up to 35\% for animal fat, 33\% for dairy, and 24\% for margarine.

These price increases largely explain our previous finding that consumer expenditure on these products increased during the tax period. The rise in prices corresponded with notable expenditure increases for butter, cream, margarine, and sour cream, as consumers continued purchasing these products despite higher prices, indicating inelastic demand.

Interestingly, we do not find significant price increases for other products during the tax period. The prices for bacon, cheese, liver sausage, and salami did not change significantly, suggesting that the tax was not fully passed through to consumers for these products, or that market dynamics mitigated the price impact.

After the tax was abolished, as shown in Panel B of Table \ref{tab:results_price_mechanism}, we find that the price increases persisted for butter and sour cream, with prices remaining 3.22\% and 8.15\% higher than pre-tax levels, respectively. These increases are statistically significant at the 5\% and 1\% levels, respectively. In contrast, the price of margarine decreased by 5.01\% compared to pre-tax levels (p-value = 0.064), indicating that any price increase during the tax period was reversed after the tax was removed.

Moreover, we observe a substantial decrease in the price of liver sausage after the tax was abolished, with prices dropping by 23.49\% compared to pre-tax levels (p-value = 0.005). This suggests that factors other than the tax influenced the price of liver sausage during the post-tax period.

Overall, these findings indicate that the tax led to significant price increases for certain products, which were not entirely reversed after the tax was abolished. The persistence of higher prices for butter and sour cream suggests potential price rigidity or other market factors maintaining higher prices. The absence of significant price changes for other products during the tax period highlights the complexity of tax pass-through in different product markets.

\begin{table}[bt]
\centering
\caption{The causal effect of the Danish fat tax on the quarterly average price paid per 100 units.}
\label{tab:results_price_mechanism}
\begin{adjustbox}{max width=\textwidth}
\begin{tabular}{B *{8}{C}}
  \hline \hline
  & (1) & (2) & (3) & (4) & (5) & (6) & (7) & (8) \\
 & Bacon & Butter & Cheese & Cream & Liver sausage & Margarine & Salami & Sour cream \\ 
  \hline
  \multicolumn{9}{l}{\textbf{Panel A: Tax period versus pre-tax quarter}} \\ \hline
  \textit{Danish fat tax (ATT)} & -4.375 & $14.935^{***}$ & 3.600 & $3.968^{***}$ & -2.798 & $3.566^{***}$ & 0.875 & $3.863^{***}$ \\ 
  \textit{(se)} & (3.356) & (0.608) & (1.697) & (0.396) & (4.070) & (0.484) & (3.551) & (0.625) \\ 
  \textit{p-value} & 0.362 & 0.000 & 0.152 & 0.000 & 0.628 & 0.000 & 0.868 & 0.000\\
  \textit{\% change} & -6.41\% & 23.44\% & 3.85\% & 10.58\% & -4.50\% & 11.35\% & 0.66\% & 12.31\% \\ 
  \textit{obs.} & 15,723 & 32,189 & 26,964 & 27,735 & 19,521 & 28,682 & 19,094 & 19,590 \\ 
  \textit{p-val. pre-trend} & 0.628 & 0.000 & 0.586 & 0.009 & 0.892 & 0.441 & 0.805 & 0.000 \\ 
\hline
  \multicolumn{8}{l}{\textbf{Panel B:  Post-tax period versus pre-tax quarter}} \\ \hline
  \textit{Danish fat tax (ATT)} & -7.215 & $2.039^{**}$ & 0.126 & -0.093 & $-14.521^{***}$ & $-1.483^{*}$ & $-9.527^{*}$ & $2.585^{***}$ \\ 
  \textit{(se)} & (3.205) & (0.633) & (1.764) & (0.381) & (3.558) & (0.526) & (3.502) & (0.577) \\ 
  \textit{p-value} & 0.121 & 0.031 & 0.962 & 0.868 & 0.005 & 0.064 & 0.065 & 0.002\\
  \textit{\% change} & -10.37\% & 3.22\% & 0.13\% & -0.24\% & -23.49\% & -5.01\% & -7.33\% & 8.15\% \\ 
  \textit{obs.} & 18,034 & 36,058 & 30,553 & 30,979 & 21,942 & 31,466 & 21,746 & 21,739 \\ 
  \textit{p-val. pre-trend} & 0.745 & 0.000 & 0.674 & 0.009 & 0.964 & 0.557 & 0.728 & 0.000 \\
   \hline \hline
   \multicolumn{9}{p{1.3\textwidth}}{\normalsize Significance level: * $p<0.1$ ** $p<0.05$ *** $p<0.01$. Standard errors clustered at unit level. \textit{\% change} compares the estimated tax average to the pre-tax sample mean for Danish households. The estimates are obtained using the doubly robust DID approach, controlling for age of head of household, number of children under 15, gender, ISCED, income level, and household size.}\\ 
\end{tabular}
\end{adjustbox}
\end{table}

\subsubsection{Package sizes}
Since the Danish fat tax was levied on producers based on the saturated fat content of their products, it is important to examine not only consumer responses but also potential adjustments made by producers. One plausible producer response to the tax is to reduce package sizes.  Smaller package sizes could allow producers to keep the overall price per package lower, even if the price per unit increases due to the tax. This strategy might help prevent a decline in sales volumes by making products appear more affordable to price-sensitive consumers.

In Table \ref{tab:packages}, we present the causal effects of the Danish fat tax on the average package sizes of various high-fat food products. During the tax period, significant reductions in average package sizes are observed for several products. For bacon, the average package size decreased by 31.46 grams, representing a 16.35\% reduction compared to the pre-tax period, with the result being statistically significant at the 5\% level (p-value = 0.026). This suggests that producers may have reduced the package sizes of bacon in response to the tax, potentially to keep the total price per package at a level acceptable to consumers.

Similarly, for cheese, there is a significant decrease in average package size by 66.18 grams, an 11.69\% reduction (p-value = 0.000). This substantial decrease indicates a strategic adjustment by producers, as the reduction in package size could help maintain sales by making the products appear less expensive, despite the higher per-unit cost.

For liver sausage, the average package size decreased by 20.24 grams, corresponding to a 8.22\% reduction, which is statistically significant at the 5\% level (p-value = 0.035). This finding aligns with the pattern observed for bacon and cheese, suggesting a consistent producer response across different types of processed meats affected by the tax.

In the case of salami, there is a marginally significant reduction in average package size by 16.05 grams, a 12.26\% decrease (p-value = 0.095). While the result is significant at the 10\% level, it still points toward a possible producer strategy of offering smaller packages during the tax period.

For other products, such as butter and margarine, the reductions in average package sizes during the tax period are not statistically significant. Butter shows a decrease of 11.77 grams (5.75\%, p-value = 0.104), and margarine decreases by 20.68 grams (5.88\%, p-value = 0.177). The lack of statistical significance suggests that producers may not have adjusted package sizes for these products in response to the tax, or that the adjustments were not substantial enough to be detected.

After the tax was abolished, some of the reductions in package sizes persisted into the post-tax period. For bacon, the average package size remained significantly lower than in the pre-tax period, with a reduction of 29.79 grams (15.74\%, p-value = 0.032). This persistence suggests that the changes made during the tax period may have led to lasting shifts in producer strategies or consumer preferences.

In the case of butter, the average package size decreased significantly by 20.55 grams in the post-tax period, a 10.11\% reduction compared to the pre-tax period (p-value = 0.003). This significant decrease, which was not statistically significant during the tax period, indicates that producers adjusted package sizes after the tax was abolished. One possible explanation is that producers delayed adjustments during the tax period due to existing packaging inventories or contracts and implemented changes afterward to optimize production and pricing strategies.

For cheese, the significant reduction in average package size continued into the post-tax period, with a decrease of 58.64 grams (10.64\%, p-value = 0.002). The sustained smaller package sizes suggest that both producers and consumers may have adapted to these changes, leading to a new norm in packaging practices for cheese products.

For liver sausage, margarine, salami, cream, and sour cream, the changes in average package sizes during the post-tax period are not statistically significant. 

The observed reductions in package sizes, particularly for bacon, cheese, and butter, can be interpreted as a producer response to the fat tax. By offering smaller packages, producers could keep the total price per package lower, making the products more attractive to consumers who are sensitive to price increases. This strategy might help sustain sales volumes despite the higher per-unit costs imposed by the tax. From a consumer perspective, smaller package sizes could lead to reduced consumption if consumers do not compensate by purchasing additional units. However, if consumers adjust by buying multiple smaller packages, the intended effect of the tax on reducing consumption may be diminished.

The persistence of smaller package sizes after the tax was abolished suggests that the changes may have resulted in lasting alterations to the market. Producers might have found that smaller packages are more profitable or better aligned with consumer preferences, or they may have decided not to revert to previous packaging sizes due to the costs associated with changing production processes.

These findings highlight the importance of considering producer behavior when evaluating the impact of taxation policies. Producer strategies, such as adjusting package sizes, can influence how taxes affect consumer prices and purchasing patterns. If producers respond to taxes by modifying product characteristics, the effectiveness of the tax in achieving public health objectives may be affected. Policymakers should take into account potential producer responses when designing tax policies and consider accompanying measures that address these responses to enhance the overall effectiveness of the intervention.

In conclusion, the Danish fat tax appears to have prompted producers to reduce package sizes for certain high-fat products, particularly bacon, cheese, and butter. These adjustments likely aimed to mitigate the impact of the tax on consumers by keeping the total price per package more affordable. The persistence of these changes after the tax was abolished indicates that the tax may have had lasting effects on market practices. Understanding producer responses is crucial for assessing the full impact of taxation policies and for developing comprehensive strategies to promote healthier consumption patterns.

\begin{table}[bt]
\centering
\caption{The causal effect of the Danish fat tax on the average package sizes.}
\label{tab:packages}
\begin{adjustbox}{max width=\textwidth}
\begin{tabular}{B *{8}{C}}
  \hline \hline
  & (1) & (2) & (3) & (4) & (5) & (6) & (7) & (8) \\
 & Bacon & Butter & Cheese & Cream & Liver sausage & Margarine & Salami & Sour cream \\ 
  \hline
  \multicolumn{8}{l}{\textbf{Panel A: Tax period versus pre-tax quarter}} \\ \hline
  \textit{Danish fat tax (ATT)} & $-31.460^{**}$ & $-11.773$ & $-66.176^{***}$ & 1.176 & $-20.241^{**}$ & $-20.681$ & $-16.048^{*}$ & $-5.893$ \\ 
  \textit{(se)} & (9.825) & (4.947) & (12.638) & (6.676) & (6.487) & (10.221) & (7.035) & (7.121) \\ 
  \textit{p-value} & 0.026 & 0.104 & 0.000 & 0.906 & 0.035 & 0.177 & 0.095 & 0.562\\
  \textit{\% change} & -16.35\% & -5.75\% & -11.69\% & 0.41\% & -8.22\% & -5.88\% & -12.26\% & -2.55\% \\ 
  \textit{obs.} & 28,874 & 40,929 & 32,020 & 39,274 & 29,417 & 38,687 & 30,800 & 36,121  \\ 
  \textit{p-val. pre-trend} & 0.713 & 0.416 & 0.360 & 0.065 & 0.872 & 0.135 & 0.846 & 0.229  \\ 
   \hline
  \multicolumn{8}{l}{\textbf{Panel B: Post-tax period versus pre-tax quarter}} \\ \hline
  \textit{Danish fat tax (ATT)} & $-29.789^{**}$ & $-20.545^{***}$ & $-58.641^{***}$ & 11.990 & $-4.845$ & $-3.388$ & $-9.116$ & 0.291 \\ 
  \textit{(se)} & (9.627) & (4.717) & (13.244) & (6.367) & (6.123) & (9.605) & (7.010) & (7.175) \\ 
  \textit{p-value} & 0.032 & 0.003 & 0.002 & 0.202 & 0.606 & 0.814 & 0.389 & 0.978\\
  \textit{\% change} & -15.74\% & -10.11\% & -10.64\% & 4.14\% & -2.01\% & -0.98\% & -7.05\% & 0.13\% \\ 
  \textit{obs.} & 32,648 & 45,295 & 35,949 & 43,648 & 33,178 & 42,954 & 34,711 & 40,299 \\ 
  \textit{p-val. pre-trend} & 0.468 & 0.396 & 0.439 & 0.038 & 0.914 & 0.093 & 0.857 & 0.179 \\ 
   \hline \hline
   \multicolumn{9}{p{1.3\textwidth}}{\normalsize Significance level: * $p<0.1$ ** $p<0.05$ *** $p<0.01$. Standard errors clustered at unit level. \textit{\% change} compares the estimated tax average to the pre-tax sample mean for Danish households. The estimates are obtained using the doubly robust DID approach, controlling for age of head of household, number of children under 15, gender, ISCED, income level, and household size.}\\ 
\end{tabular}
\end{adjustbox}
\end{table}

\bigskip


\section{Robustness checks}
\label{sec:robustness_checks}

In this section, we conduct several robustness checks to assess the reliability of the main findings presented in the previous section, addressing the potential concerns with respect to our identification strategy stated in Section \ref{sec:identification}.
First, the issue of potential cross-border shopping merits attention, as it might obscure the effect of the tax. 
To investigate potential cross-border shopping effects, we exclude Southern-Danish households from South/West Jutland and Funen, focusing on Danish households that are less likely to engage in cross-border shopping and hence fully subject to the tax.
Interestingly, the results in Table \ref{tab:robust_southern_dk} in Appendix \ref{app:robust_southern_dk} exhibit similar patterns as our main results, indicating that changes in consumption behavior were similar for Danish households that were able to cross-border shop compared to Danish households that did not have this opportunity. 
Note that in our main results, the unintended effect of cross-border shopping is incorporated as we do include purchases made abroad.

Second, we consider concerns surrounding disparities in Danish and Northern-German household characteristics and assess whether differences in household characteristics between Danish and Northern-German households produced the effects of our main results in Section \ref{sec:results}. To gauge the influence of the covariates, we contrast the treatment effects obtained under the conditional parallel trends assumption with those derived under the unconditional parallel trends assumption. If doubly robust DID estimates based on the conditional parallel trends assumption resemble the estimates based on the unconditional parallel trends assumption, this would indicate that covariates contain limited information on consumer behavior trends and are unlikely to drive the results in Section \ref{sec:results}.
For brevity, we only included the results for the weight outcome, but similar observations apply to the other two main outcomes. Interested readers may obtain the complete set of results upon request.
Reassuringly, Figures \ref{fig:unconditional_pta_weight} in Appendix \ref{app:unconditional_pta_weight} and \ref{fig:unconditional_pta_expenditure} in Appendix \ref{app:unconditional_pta_expenditure} display extremely similar trends compared to the figures in Appendix \ref{app:dynamic_treatments}. This suggests that covariates have minimal influence on the trends. Upon comparing the results, it becomes evident that, unsurprisingly, incorporating covariates enhances precision.

\bigskip


\section{Heterogeneity analysis by income levels}
\label{sec:heterogeneity}

Criticisms of the tax being regressive suggest that low-income households may be disproportionately affected compared to wealthier ones. Groceries constitute a larger proportion of the income of poorer households, making them disproportionately affected by the tax if they do not adjust their consumption behavior \citep{allais2010effects, zhen2011habit}. Understanding how the tax affects different income groups is vital for targeted policy improvements, and extrapolating the tax's applicability to other countries. 

\begin{figure}[htbp] 
  \centering
  \begin{subfigure}{\textwidth}
    \centering
    \includegraphics[width=0.85\textwidth]{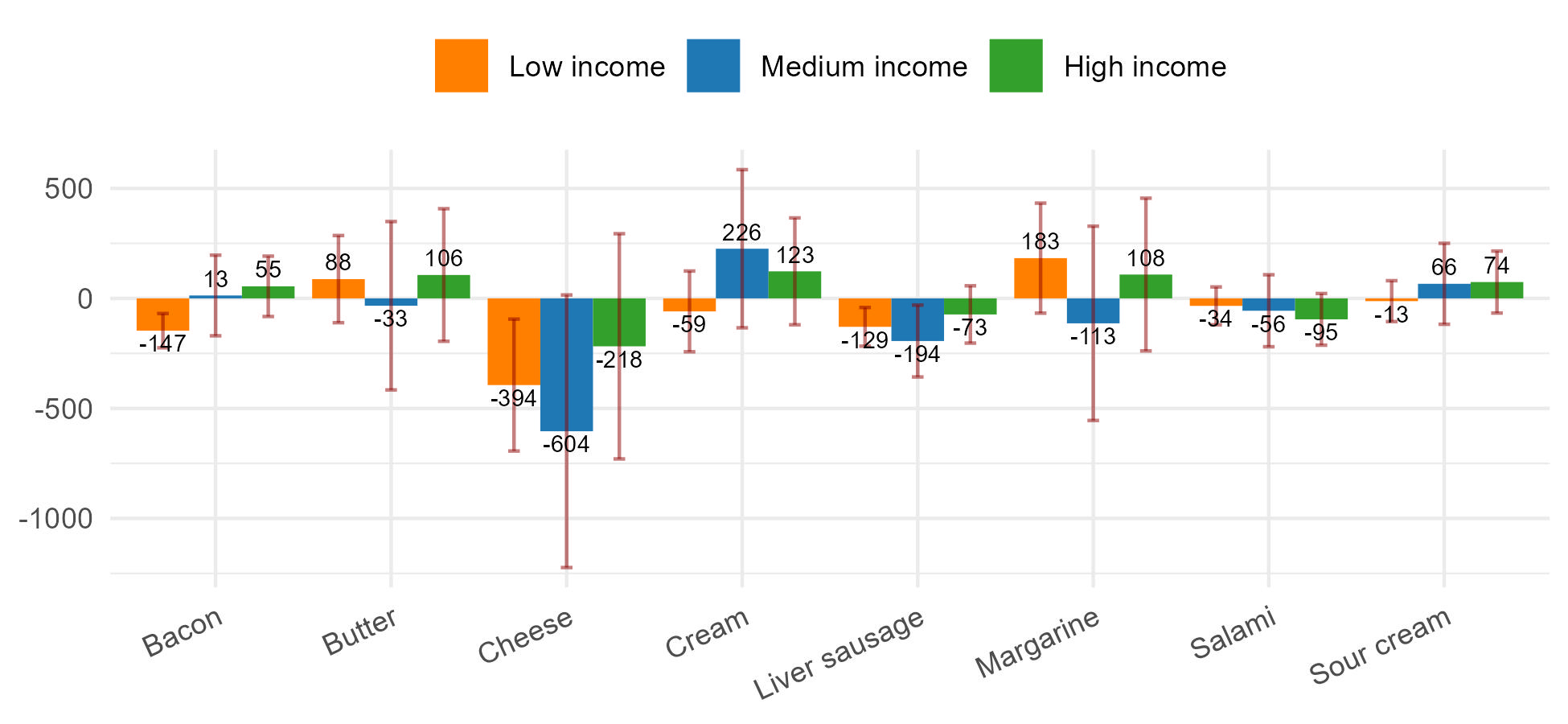} 
    \caption{Average quarterly weight consumed (gr/ml).}
    \label{fig:low_high_income_weight}
  \end{subfigure}
  \hfill
  \begin{subfigure}{\textwidth}
    \centering
    \includegraphics[width=0.85\textwidth]{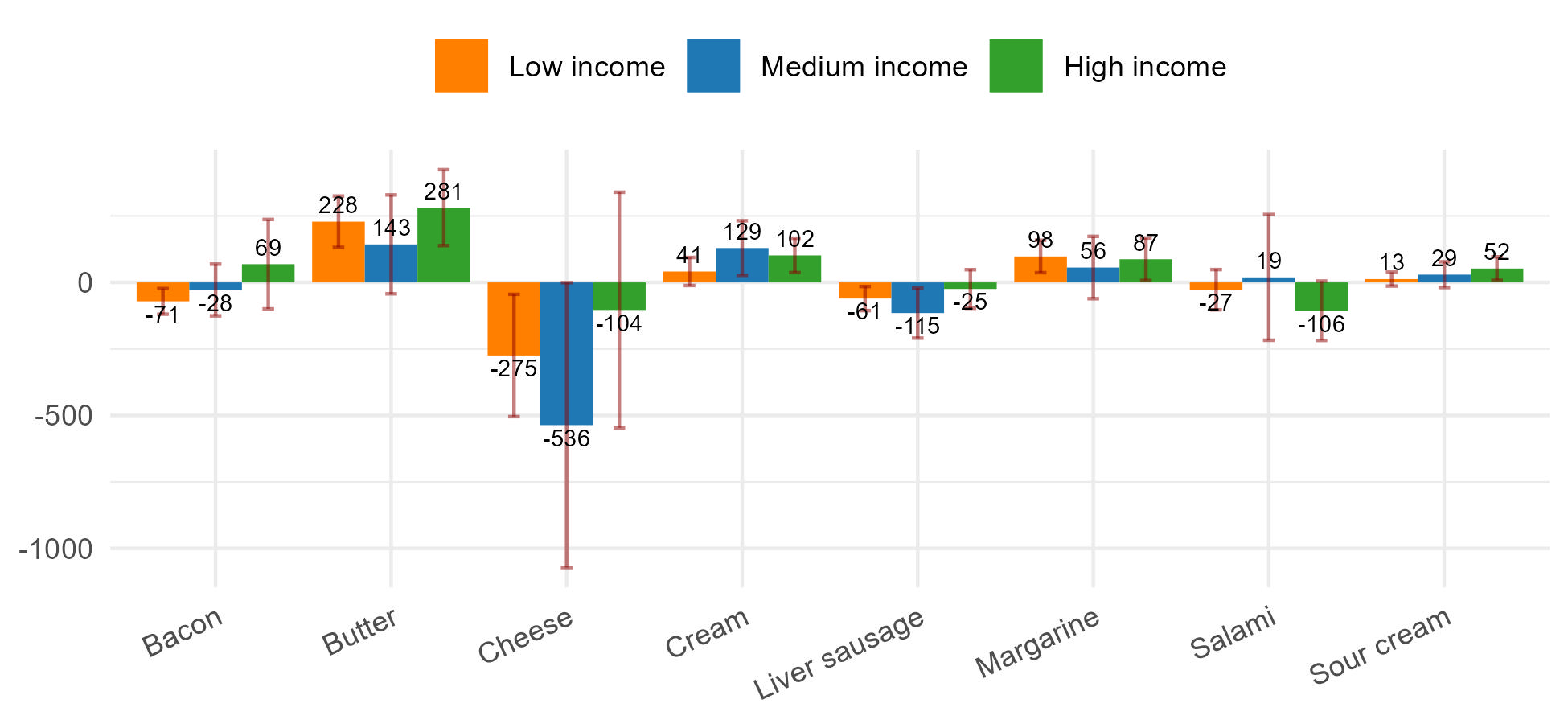}
    \caption{Average quarterly expenditure in Euro cents.}
    \label{fig:low_high_income_expenditure}
  \end{subfigure}  \caption{Doubly robust DID estimates of the causal effect of the Danish fat tax during the tax period (October 2011 to December 2012) for low, medium and high-income households. Low-income households are defined as households with very low income or low income. High-income households are defined as households with very high income or high income.  95\% confidence intervals in red. }
  \label{fig:low_high_income}
\end{figure}

\begin{figure}[htbp] 
  \centering
  \begin{subfigure}{\textwidth}
    \centering
    \includegraphics[width=0.85\textwidth]{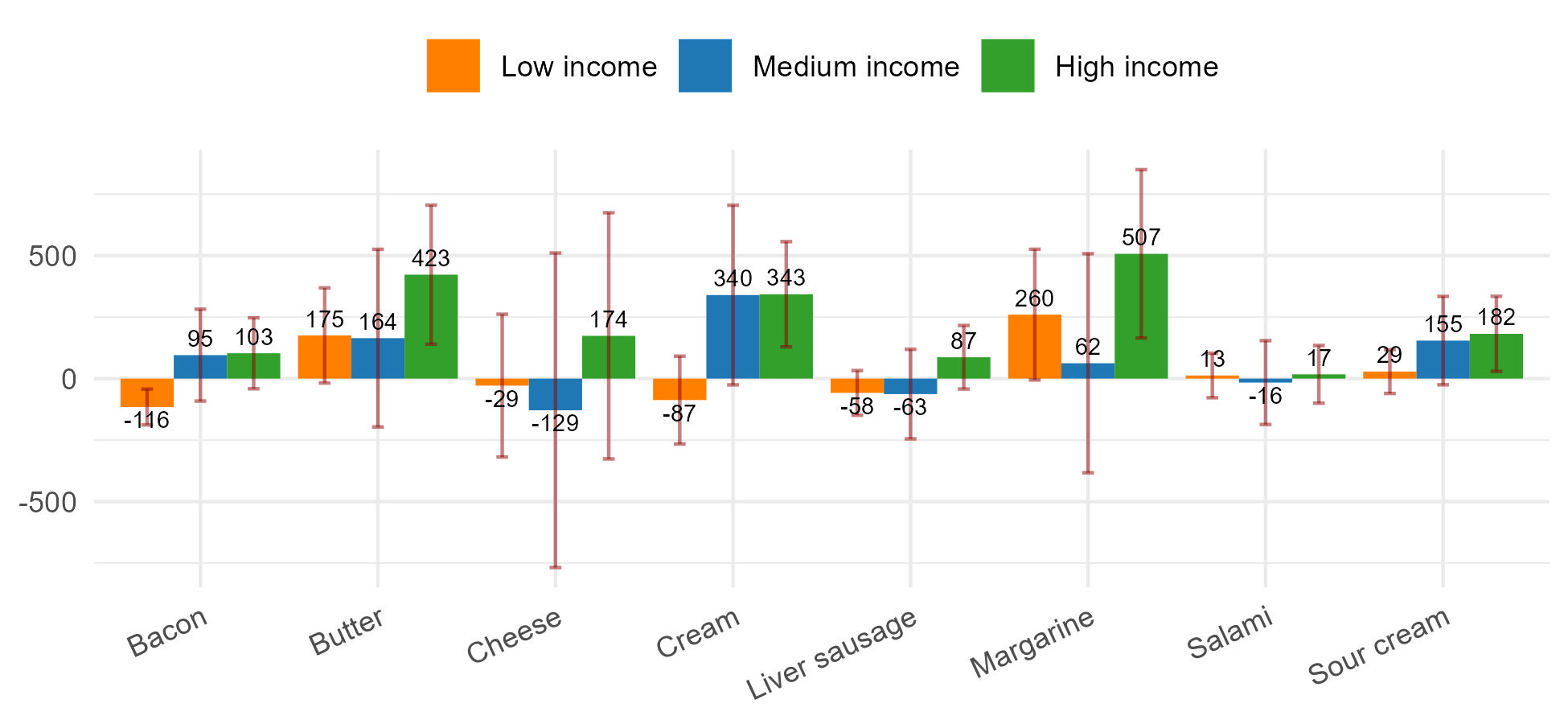} 
    \caption{Average quarterly weight consumed (gr/ml).}
    \label{fig:post_vs_pre_low_high_income_weight}
  \end{subfigure}
  \hfill
  \begin{subfigure}{\textwidth}
    \centering
    \includegraphics[width=0.85\textwidth]{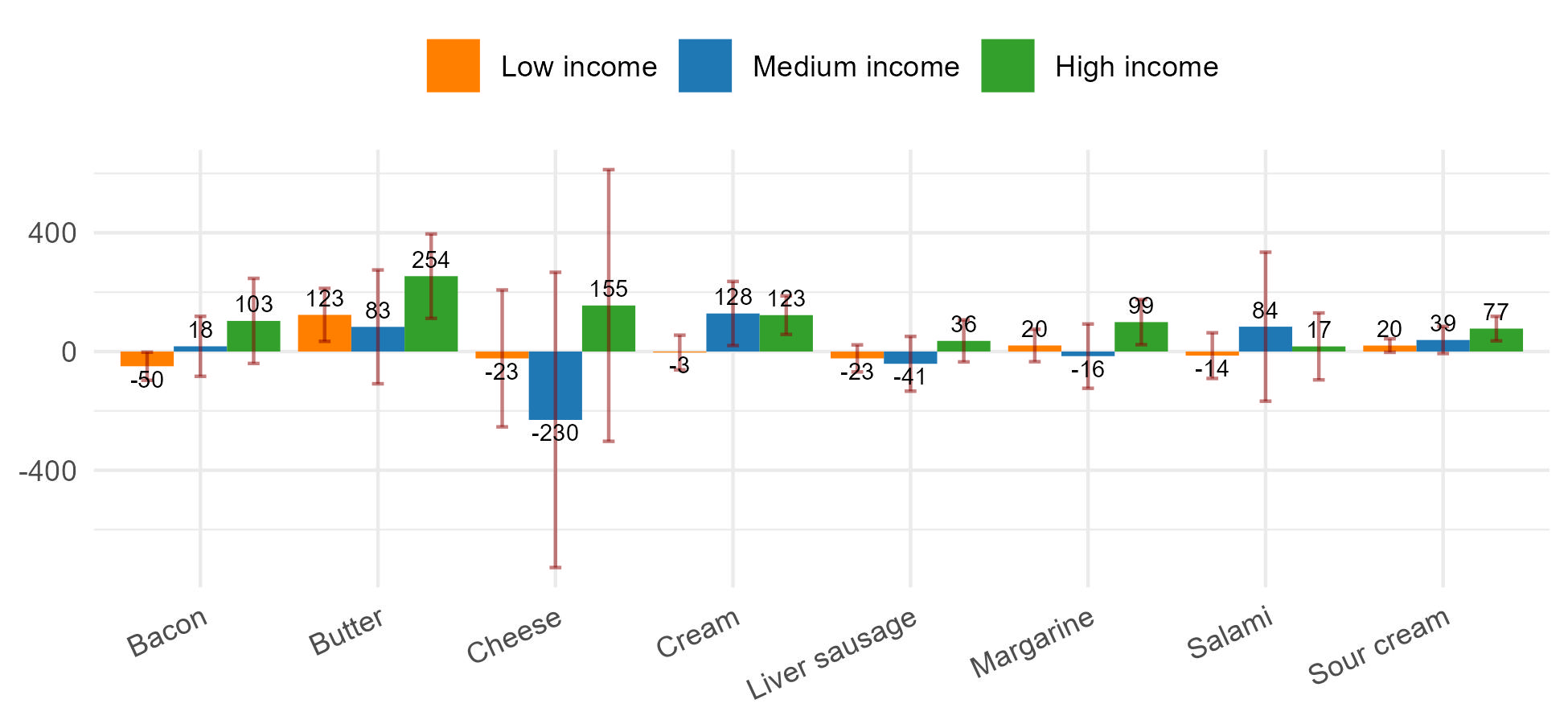}
    \caption{Average quarterly expenditure in Euro cents.}
    \label{fig:post_vs_pre_low_high_income_expenditure}
  \end{subfigure}
  \caption{Doubly robust DID estimates of the causal effect of the Danish fat tax during the post-tax period (January 2013 to December 2014) compared to the pre-tax quarter (January 2011 to March 2011) for low, medium and high-income households. Low-income households are defined as households with very low income or low income. High-income households are defined as households with very high income or high income.  95\% confidence intervals in red. }
  \label{fig:post_vs_pre_low_high_income}
\end{figure}

In order to investigate potential heterogeneity in the effects of the Danish fat tax, we 
re-estimate the model for the subsamples of households who have a low or very low income, middle income and high or very high income (refer to Section \ref{sec:data} to see how we define income categories).  
The results are presented in Figure \ref{fig:low_high_income}.
Interestingly, Panel (a) of Figure \ref{fig:low_high_income} shows a significant reduction in bacon consumption for low-income households, for whom we also observe a significant decrease in consumption of cheese and liver sausage. The effects are not statistically significant for these products, for neither medium-income nor high-income households. Moreover, we find that low-income and medium-income households also reduced their expenditure on cheese and liver sausage, as shown in Panel (b) of Figure \ref{fig:low_high_income}.

Similar to the main analysis, we observe an increase in expenditure for both butter and margarine. Notably, the increase is fairly similar between households with high and low incomes, but not statistically significant for medium-income households. This means that low-income households were disproportionately affected by the tax. 

The results in Figure \ref{fig:post_vs_pre_low_high_income} show that, after the abolishment of the tax, the increase in butter, cream, margarine, and sour cream consumption is largely driven by high-income households, with relatively weaker effects on low-income households.  

\bigskip

\section{The effect of the tax on cross-border shopping of butter and liver sausage, geo-spatial effect heterogeneity and spillover effects}
\label{sec:cross-border}
The introduction of the Danish fat tax in October 2011 created an economic incentive for Danish consumers, particularly those residing near the German border, to engage in cross-border shopping to circumvent the tax. This section presents descriptive evidence on how the tax influenced cross-border purchasing behavior for two high-fat products: butter and liver sausage. We re-estimate the tax's impact on these products based on geographic proximity to the German border, evaluating the extent to which cross-border shopping mitigated its intended effects on domestic consumption and expenditure. Additionally, we estimate potential spillover effects on German households living close to the border.

\begin{figure}[!htbp]
    \centering
    \begin{subfigure}{0.95\textwidth}
        \centering
        \includegraphics[width=0.85\textwidth]{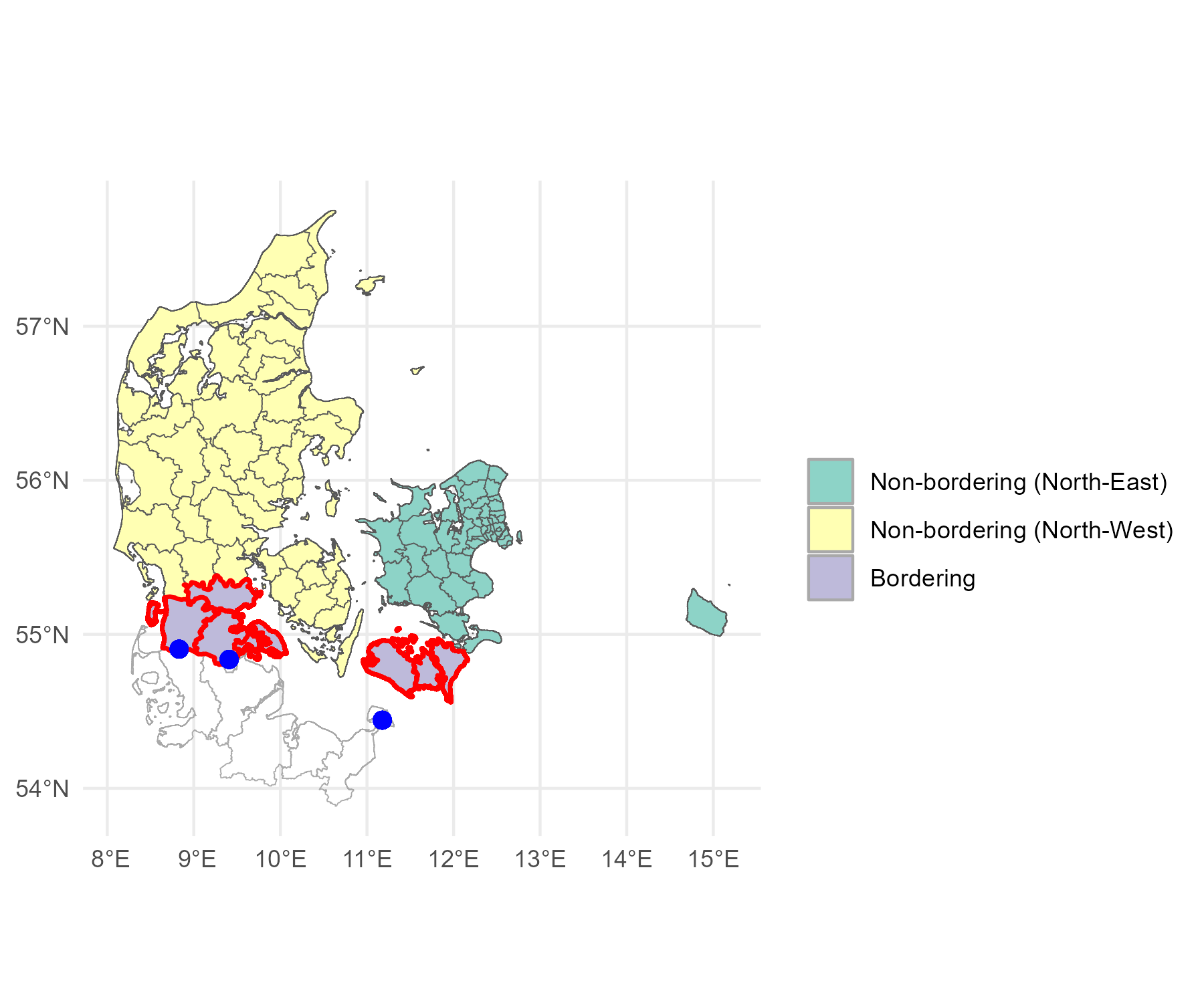}
        \caption{Denmark.}
        \label{fig:map_bordering_danes_and_shop_locations}
    \end{subfigure}
    \vspace{0.5cm}
    \begin{subfigure}{0.95\textwidth}
        \centering
        \includegraphics[width=0.85\textwidth]{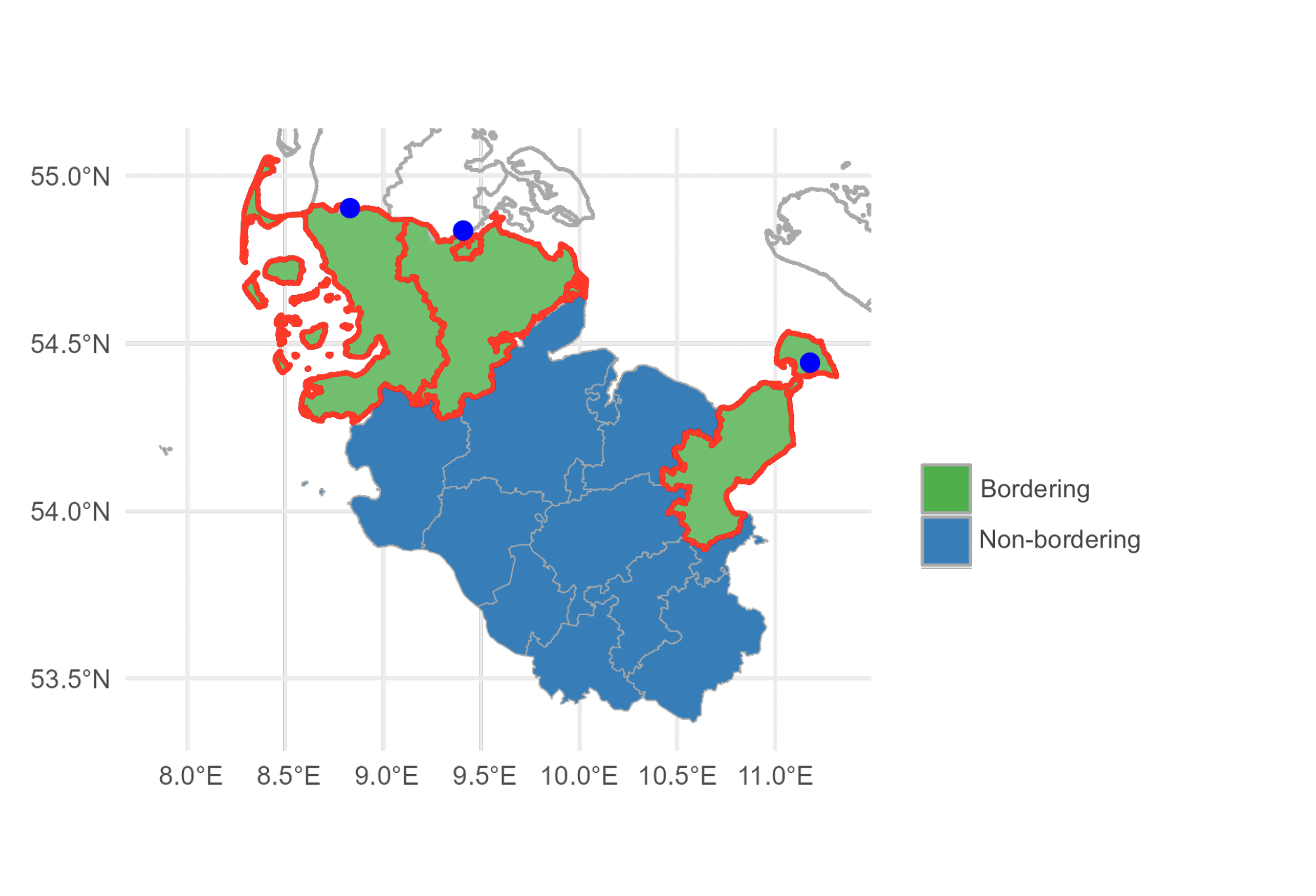}
        \caption{Northern Germany.}
        \label{fig:map_bordering_germans_and_shop_locations}
    \end{subfigure}
    \caption{Map with highlighted regions for bordering regions and blue dots are the three German shops used for determining minimum travel distance in km for Danish households. Regions are based on driving distance to the border.}
    \label{fig:maps_bordering_shop_location}
\end{figure}

To examine the geo-spatial differences in the tax's effect and assess these spillover effects, we incorporated information on driving distances to the border. For Danish households, we calculated the driving distances in kilometers from each postal code to the nearest of three German shops located close to the Danish border: (1) Dorfstraße 3, 25927 Aventoft; (2) Alte Zollstraße 44, 24955 Harrislee; and (3) Mummendorfer Weg 7, 23769 Fehmarn. These locations are indicated with blue dots in Figure \ref{fig:maps_bordering_shop_location}. Figure \ref{fig:map_bordering_danes_and_shop_locations} illustrates our division of Denmark into three geographical areas, categorized by their distance from the German border.

Similarly, for German households, we computed the driving distances from the administrative centers of their respective Kreise (administrative districts) to the nearest of the three specified shops. This is used to define border and non-border German regions (See Figure \ref{fig:map_bordering_germans_and_shop_locations}).
In our data, we can identify purchases made abroad, but we cannot be certain that these are specifically made in Germany. However, since the other neighboring country, Sweden, can only be accessed through passing the bridge or by taking a ferry which involves larger traveling costs, it is highly likely that the vast majority of these purchases are made in Germany. We should acknowledge the higher likelihood of misreporting purchases made abroad. Our descriptive analysis assumes that such misreporting behaviors affect butter and liver sausage similarly to all other products. Moreover, we assume that the introduction of the tax does not directly affect misreporting behaviors.

Table \ref{tab:fraction_per_distance_dk} presents the fraction of Danes in the sample based on the distance to the closest German shop. 
Only 2.93\% of our sample live within 50 km from the border. 
Table \ref{tab:fraction_per_distance_de} presents similar overview for Northern-German households.

\begin{table}[htbp]
  \begin{minipage}[t]{0.48\textwidth}
    \centering
    \caption{Fraction of Danes in sample based on distance to closest German shop.}
    \begin{tabular}{lrr}
      \hline
    Distance & Nr. households & Fraction \\ 
      \hline
      $<$50 km &  76 & 2.97\% \\ 
      50-100 km & 232 & 9.05\% \\ 
      100-150 km & 441 & 17.21\% \\ 
      150-200 km & 1124 & 43.85\% \\ 
      200-250 km & 417 & 16.27\% \\ 
      $>=$250 km & 273 & 10.65\% \\ 
       \hline
    \end{tabular}
    \label{tab:fraction_per_distance_dk}
  \end{minipage}
  \hfill
  \begin{minipage}[t]{0.48\textwidth}
    \centering
    \caption{Fraction of Germans in sample based on distance of administrative centers within each Kreis to closest German shop at the border.}
    \begin{tabular}{lrr}
      \hline
    Distance & Nr. households & Fraction \\ 
      \hline
    $<$50 km & 176 & 14.05\% \\ 
  50-100 km & 311 & 24.82\% \\ 
  100-150 km & 311 & 24.82\% \\ 
  $>=$150 km & 455 & 36.31\% \\
       \hline
    \end{tabular}
    \label{tab:fraction_per_distance_de}
  \end{minipage}
\end{table}

\subsection{Descriptive Evidence of Cross-Border Purchases}

\begin{figure}[htbp]
    \centering
    \begin{subfigure}{0.95\textwidth}
        \includegraphics[width=0.85\textwidth]{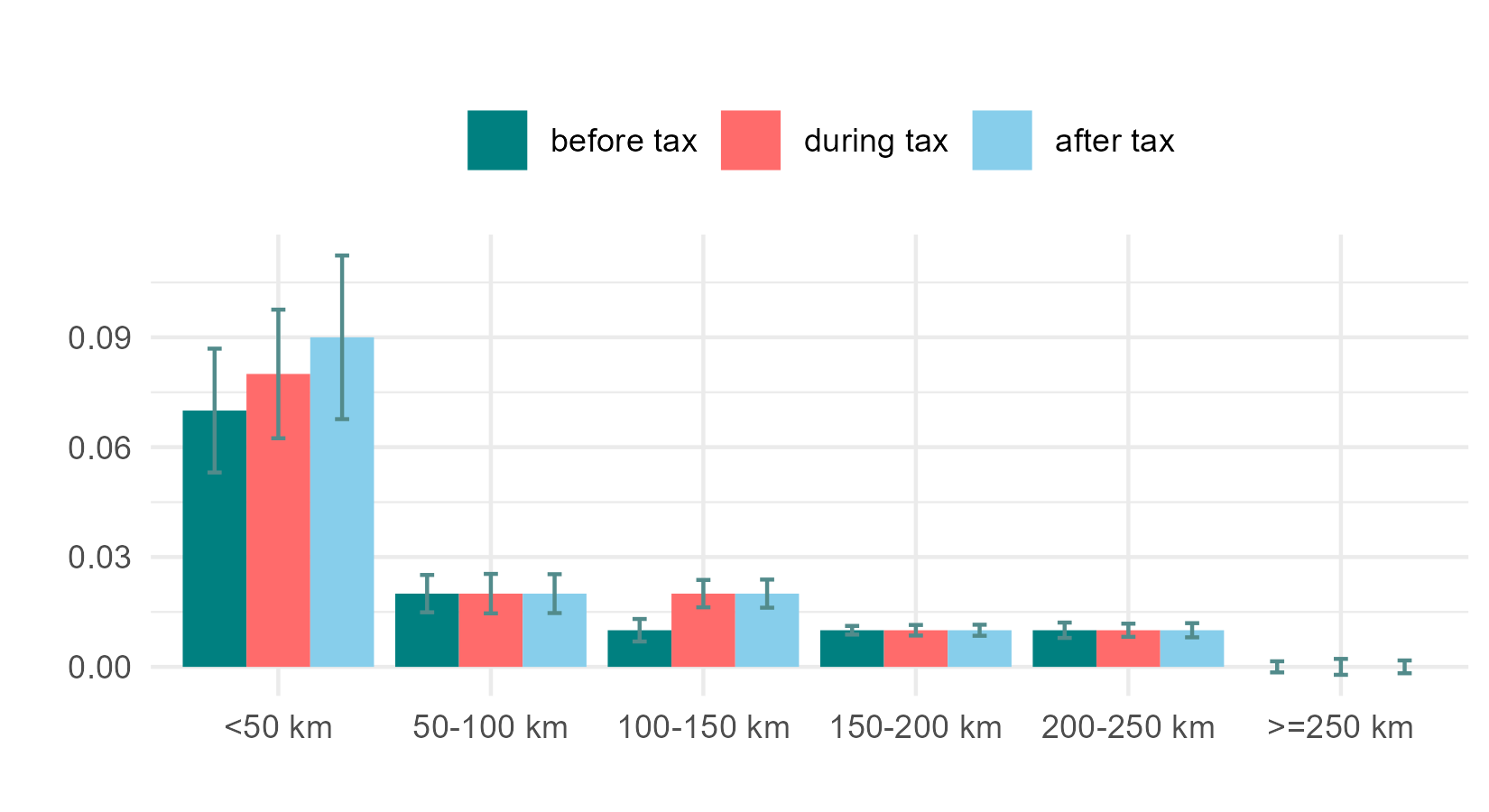}
        \caption{Average share of all products purchased abroad.}
        \label{fig:all}
    \end{subfigure}

    \begin{subfigure}{0.95\textwidth} 
        \includegraphics[width=0.85\textwidth]{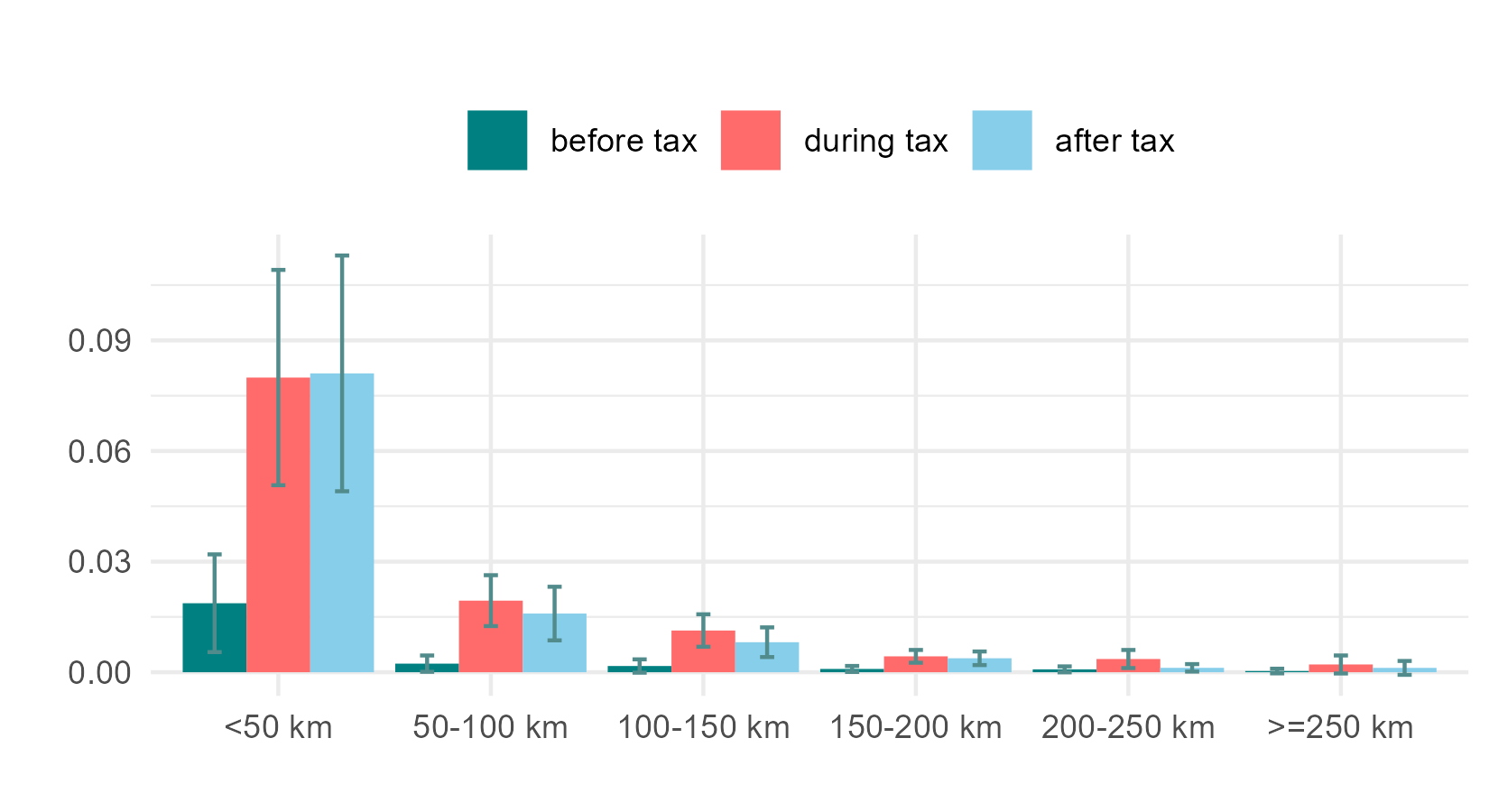}
        \caption{Average share of Danes' butter products purchased abroad.}
        \label{fig:butter}
    \end{subfigure}

    \begin{subfigure}{0.95\textwidth} 
        \includegraphics[width=0.85\textwidth]{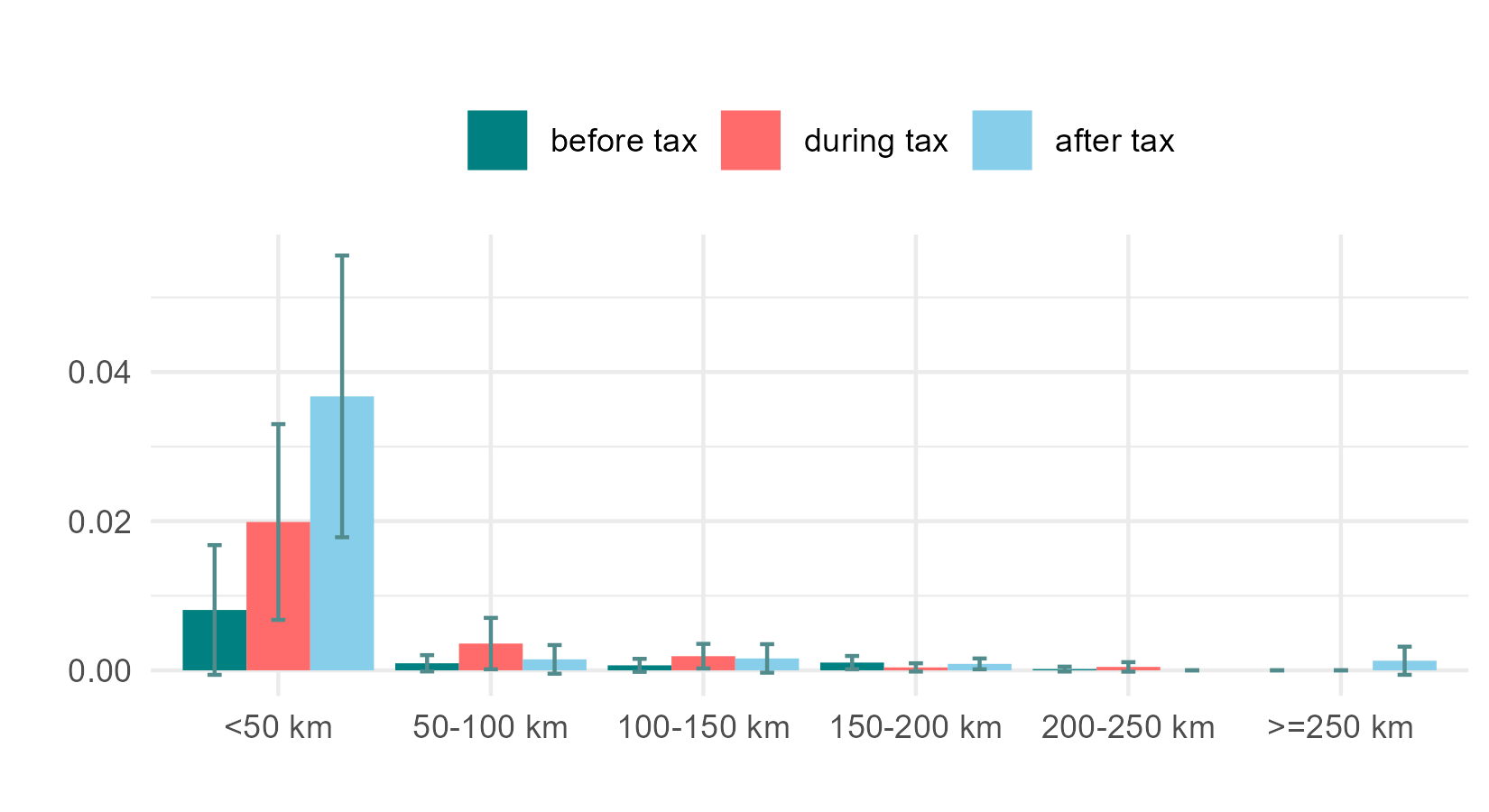}
        \caption{Average share of Danes' liver sausage products purchased abroad.}
        \label{fig:ls}
    \end{subfigure}

    \caption{X-axis: Minimum distance to German shop in km. Y-axis: Average consumption  five quarters before tax (green), five quarters during tax (red), five quarters after tax (blue).}
    \label{fig:share_per_distance_plot}
\end{figure}

To assess the impact of the fat tax on cross-border shopping, we begin by analyzing the proportion of butter and liver sausage products purchased abroad by Danish households and comparing it to the overall share of all products bought abroad. Figure \ref{fig:share_per_distance_plot} illustrates the temporal trends in the average share of these products bought abroad by households in bordering regions compared to those in non-bordering regions.

For butter, there is a noticeable increase in the share of products purchased abroad by households in bordering regions during the tax period (Figure \ref{fig:butter}). The share rises from approximately 2\% before the tax to about 8\% during the tax, indicating a significant shift towards cross-border shopping. This elevated level persists even after the tax is abolished, suggesting a lasting change in shopping habits.

A more modest increase is observed for liver sausage purchases (Figure \ref{fig:ls}). Households near the German border increased their share of liver sausage purchases abroad from approximately 1\% to nearly 2\% during the tax period. Interestingly, the level of cross-border purchasing seems to have risen to 4\% in the post-tax period, implying that this behavior was sustained even after the tax was lifted.

\begin{figure}[tbp]
    \centering
    \includegraphics[width=0.9\textwidth]{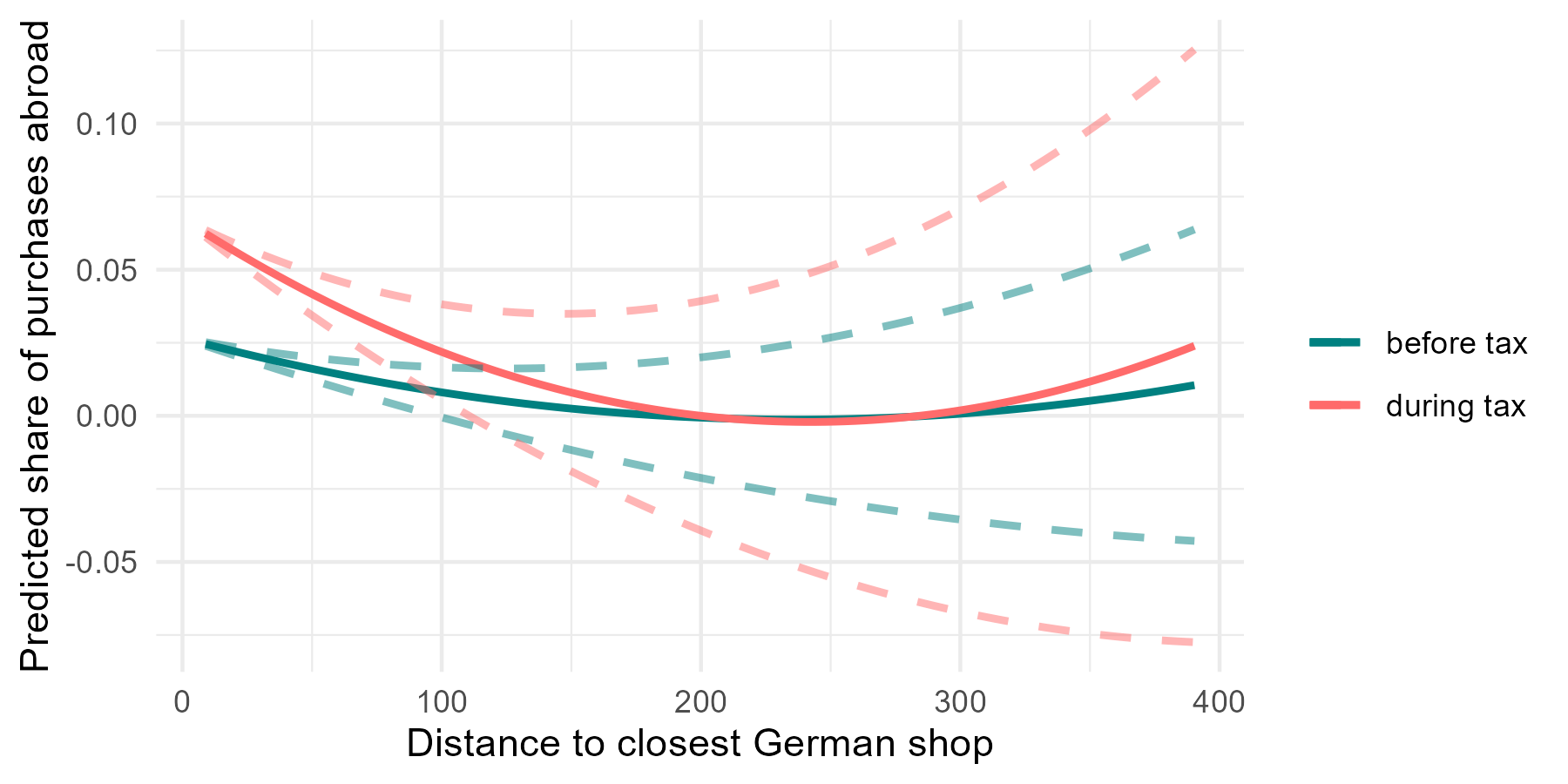}
    \caption{Plot of effect of distance on predicted share of butter purchased abroad before tax and during tax estimated with Equation (\ref{eq:random_eff_model}).}
    \label{fig:random_effects_model_plot}
\end{figure}

\begin{figure}[tbp]
    \centering
    \includegraphics[width=0.9\textwidth]{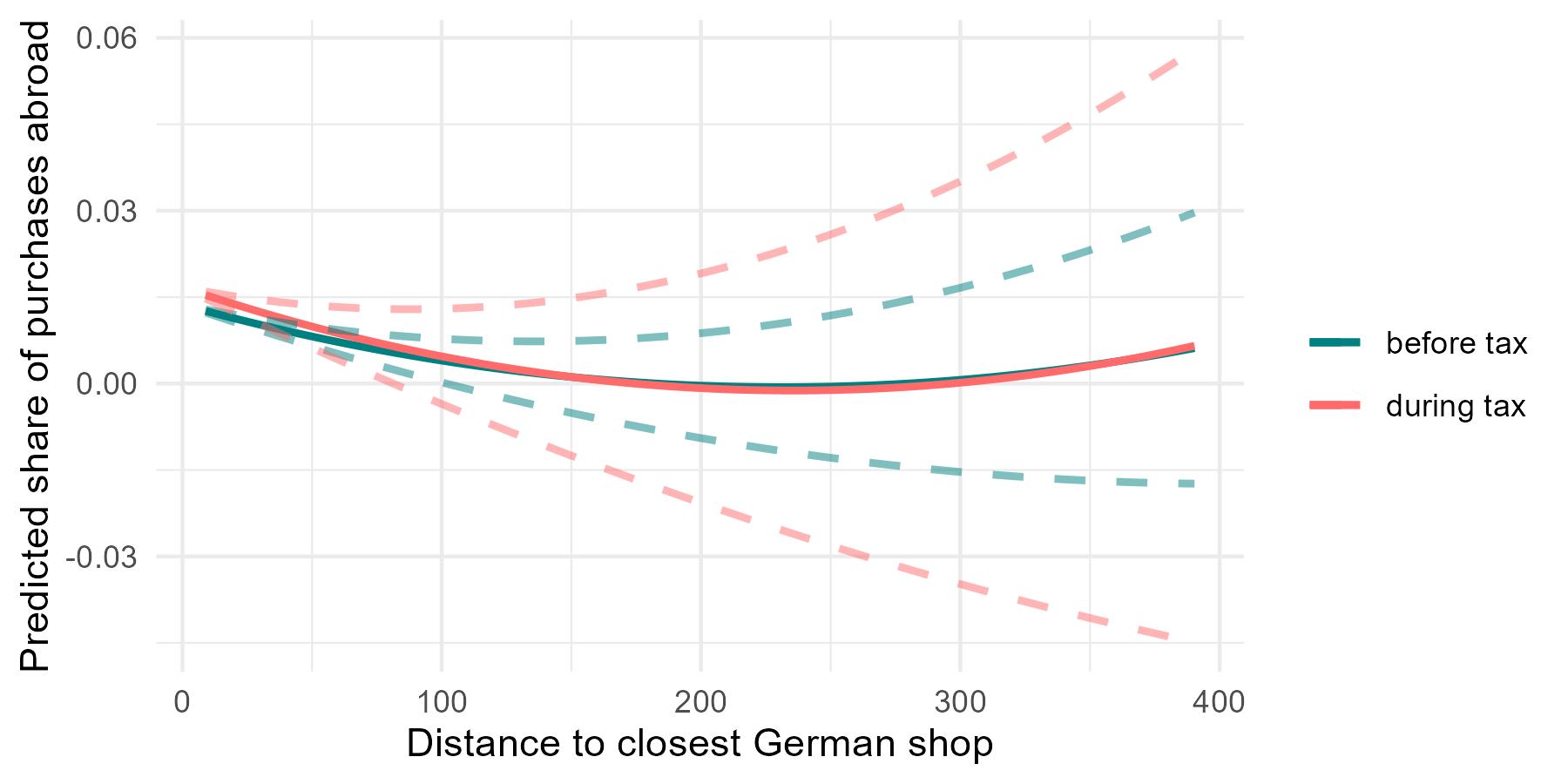}
    \caption{Plot of effect of distance on predicted share of liver sausage purchased abroad before tax and during tax estimated with Equation (\ref{eq:random_eff_model}).}
    \label{fig:random_effects_model_plot_ls}
\end{figure}

To quantify the relationship between geographic proximity to the German border and cross-border shopping, we employ a random effects model using only Danish households:\footnote{
Since our variable of interest, distance to the border, is constant over time, we rely on a random effects model, as a fixed effects model would not be suitable for time-invariant variables.}
\begin{align}\label{eq:random_eff_model}
    y_{iq} &= \alpha_i + \beta_1 dist_{i} + \beta_2 dist_{i}^2 + \beta_3 tax_{q} + \beta_4 \left(dist_{i} \times tax_{q} \right) + \beta_5 \left(dist_{i}^2 \times tax_{q} \right) + \beta_6 \textbf{X}_{i} + \varepsilon_{iq},
\end{align}
where $\textbf{X}_{i}$ contains household characteristics (age of head of household, number of kids under 15, gender of diary keeper, educational attainment (ISCED), income level, and household size), $tax$ is dummy for period where tax was in effect, $i$ index for household, $q$ index for quarter, $dist$ is the minimum distance to one of the three German shops, $y$ is the average share of butter purchased abroad per household per quarter.
The model reveals that households residing within approximately 70 kilometers of the German border significantly increased their share of butter purchases abroad during the tax period (Figure \ref{fig:random_effects_model_plot}). We do not find any significant change for liver sausage (Figure \ref{fig:random_effects_model_plot_ls}). 

The increase in cross-border purchases, particularly of butter, highlights a key limitation of the Danish fat tax: the ability of households living near the border to avoid the tax. This raises important questions about the design and implementation of such taxes in smaller countries or regions with accessible, untaxed neighboring markets. Policymakers aiming to use consumption taxes to influence behavior may need to consider geographical context and potential leakage effects, as observed here.

\subsection{Geo-spatial effect heterogeneity of the tax}

\begin{figure}[H]
    \centering
    \includegraphics[width=0.8\textwidth]{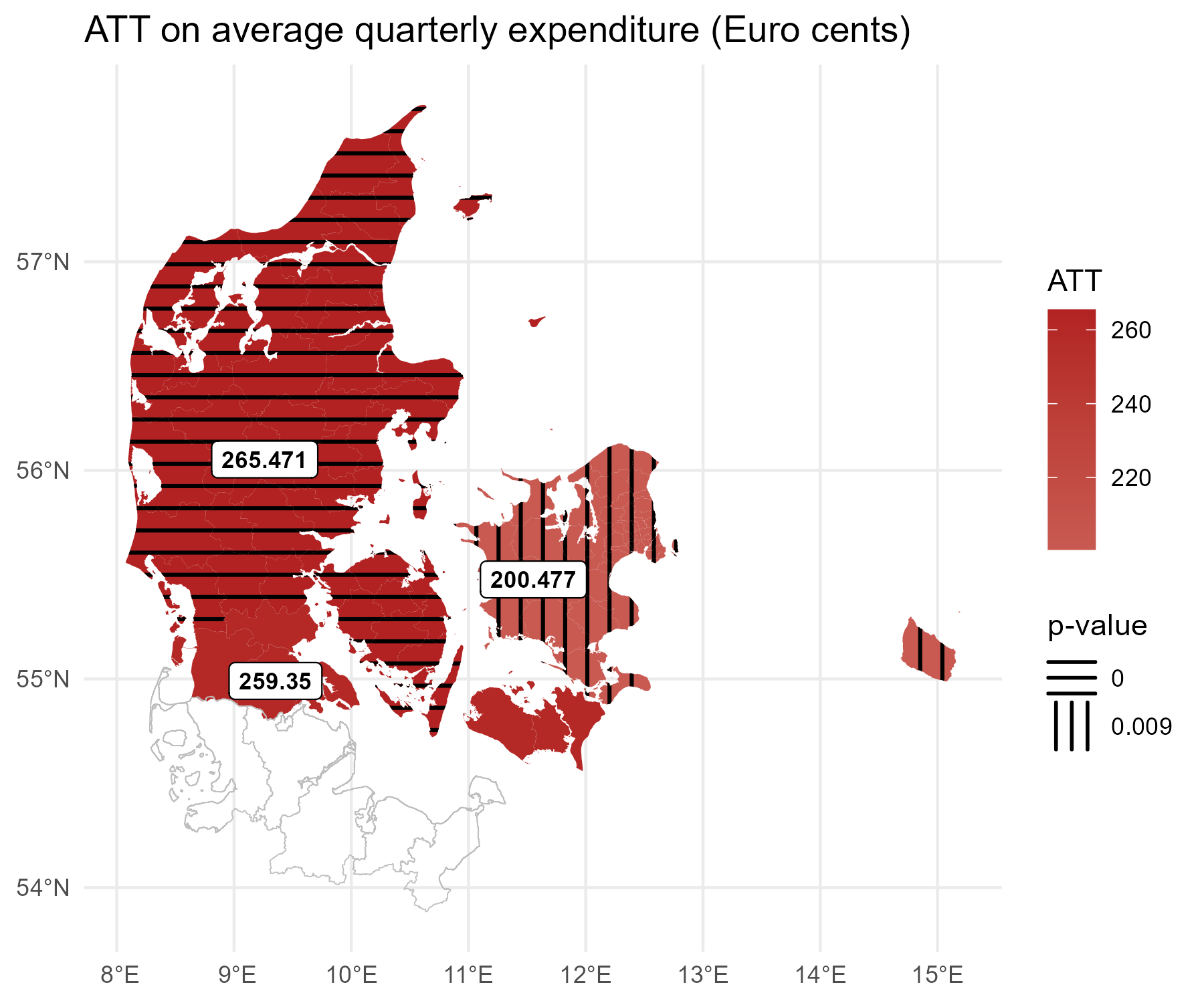}
    \caption{Map with ATT value per region for expenditure in  butter. Values that are significant at 5\% level are highlighted with black stripes.}
    \label{fig:map_att_total_expenditure_std_per_region}
\end{figure}

\begin{figure}[H]
    \centering
    \includegraphics[width=0.8\textwidth]{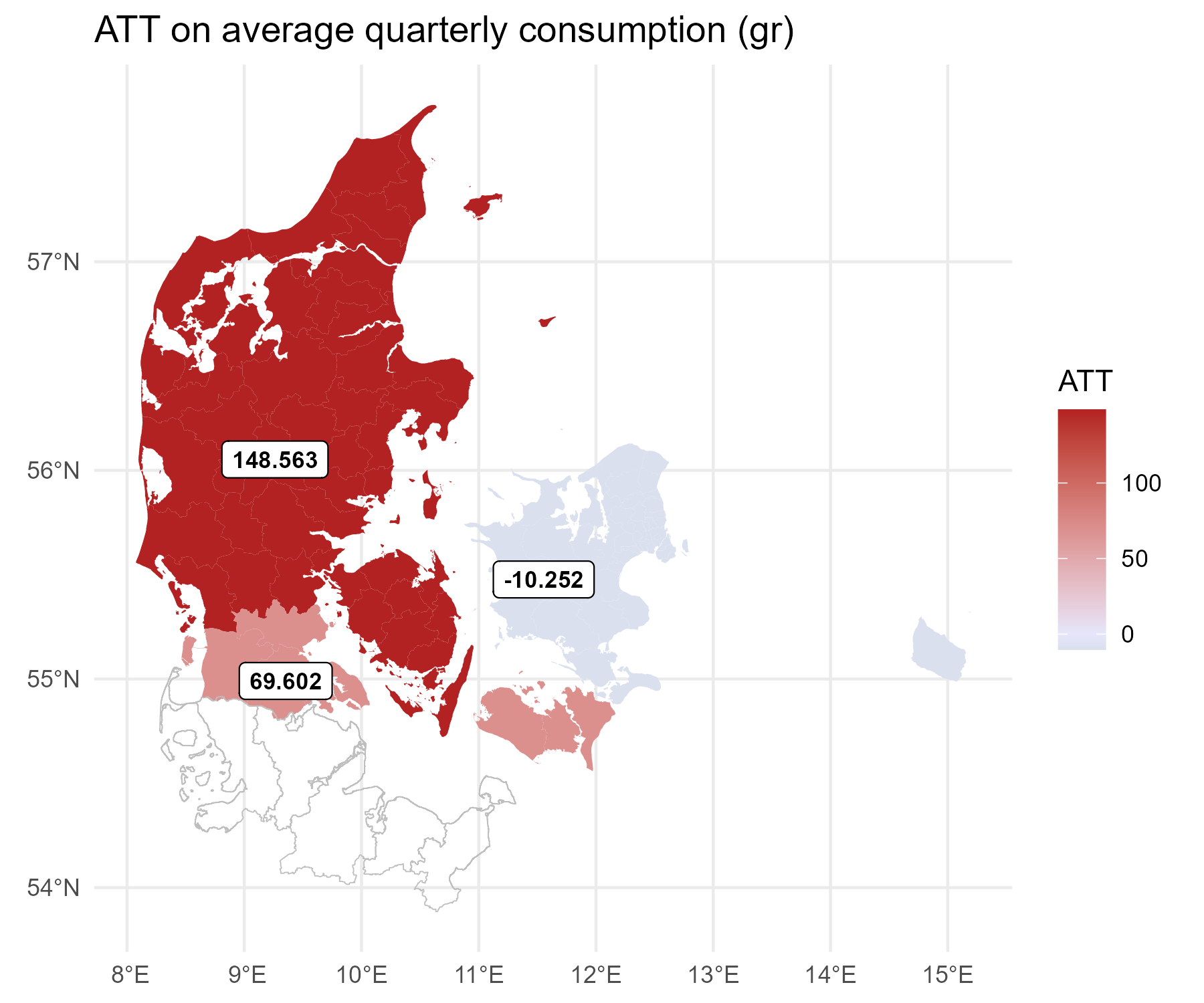}
    \caption{Map with ATT value per region for weight consumed of  butter. Values that are significant at 5\% level are highlighted with black stripes.}
    \label{fig:map_att_weight_std_per_region}
\end{figure}

\begin{figure}[H]
    \centering
    \includegraphics[width=0.8\textwidth]{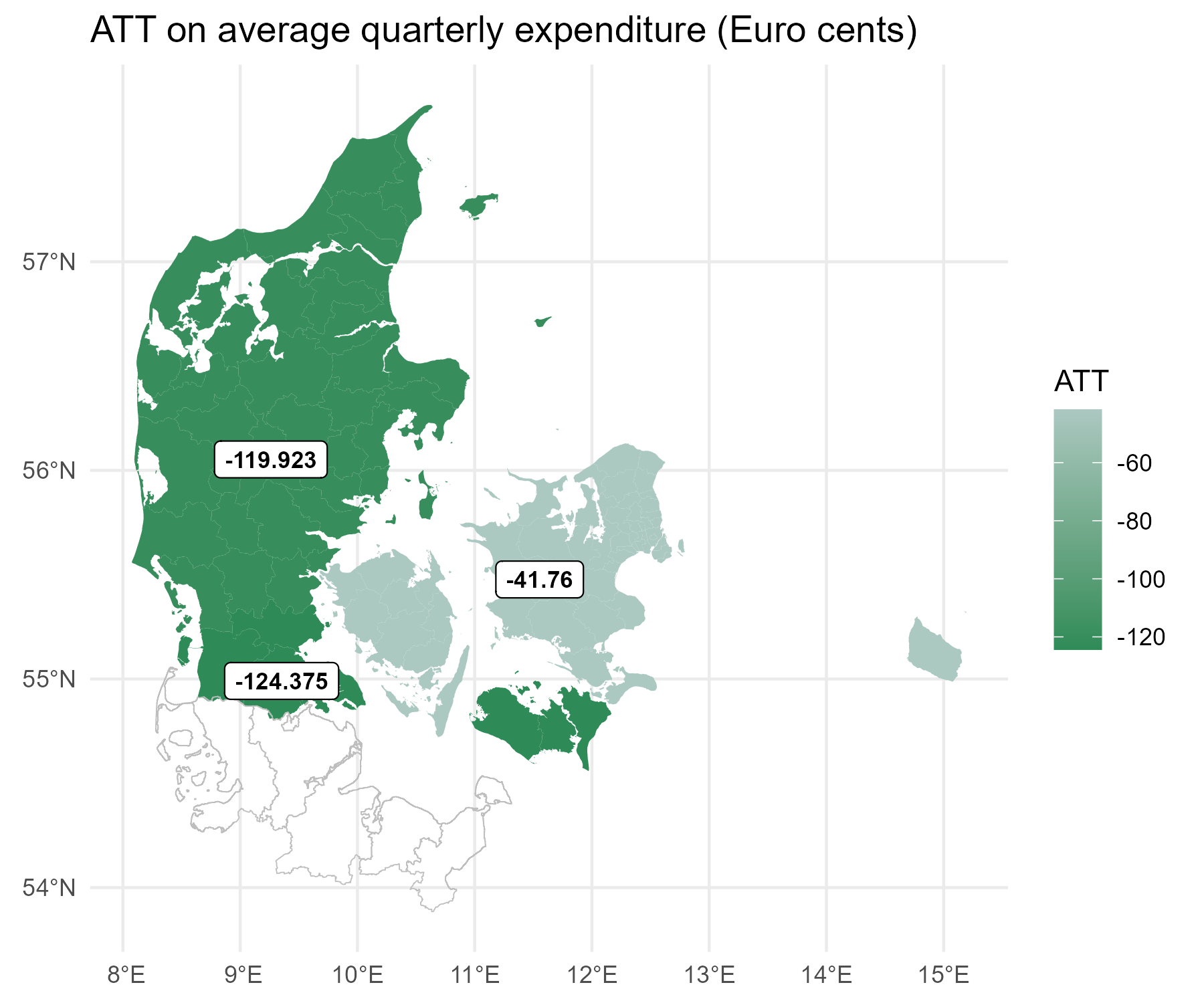}
    \caption{Map with ATT value per region for expenditure in  liver sausage. Values that are significant at 5\% level are highlighted with black stripes.}
    \label{fig:map_att_total_expenditure_std_per_region_ls}
\end{figure}

\begin{figure}[H]
    \centering
    \includegraphics[width=0.8\textwidth]{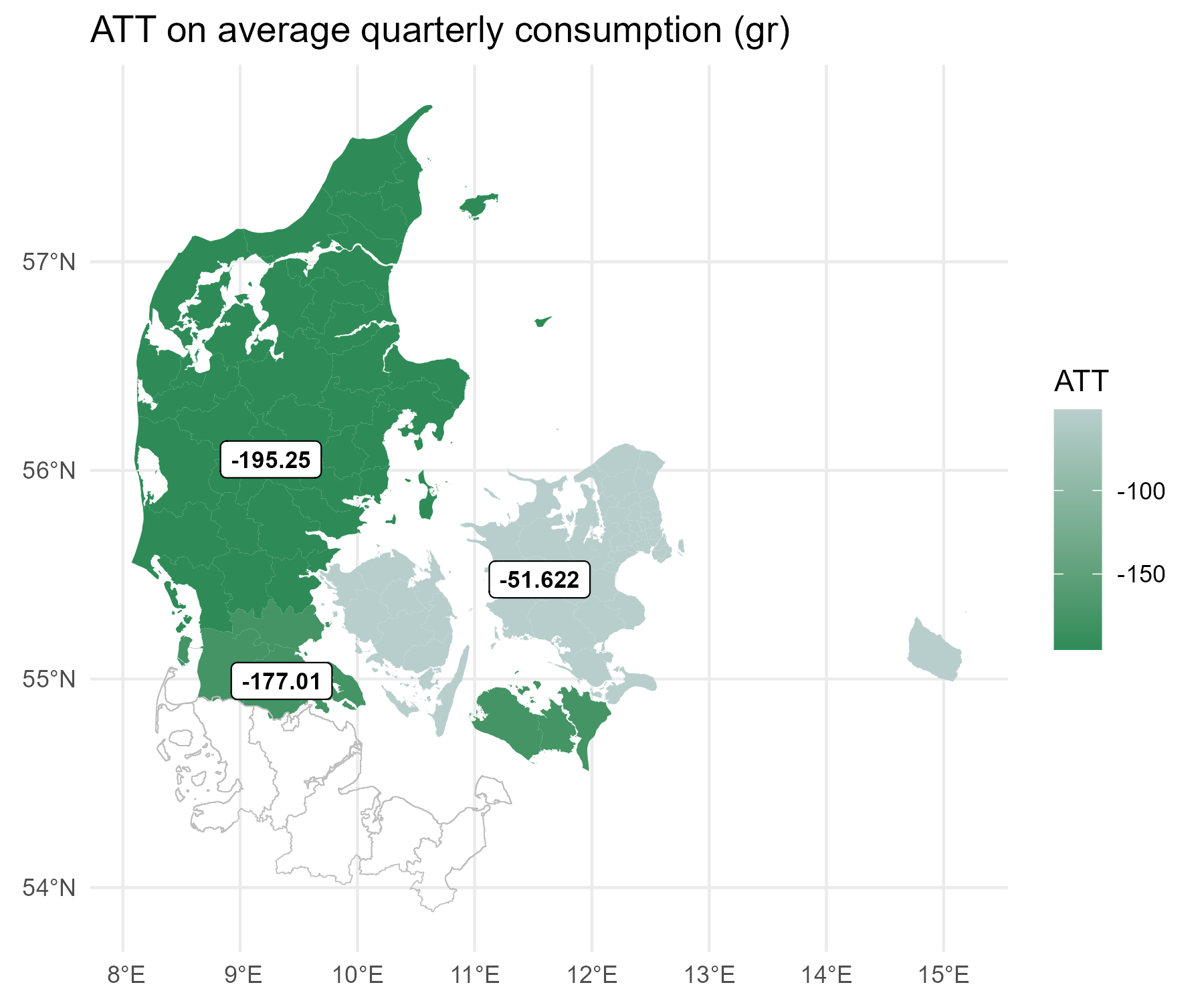}
    \caption{Map with ATT value per region for weight consumed of  liver sausage. Values that are significant at 5\% level are highlighted with black stripes.}
    \label{fig:map_att_weight_std_per_region_ls}
\end{figure}

In this section, we examine whether the causal effects of the Danish fat tax varied across different regions of Denmark, particularly focusing on whether households closer to the German border reacted differently compared to those further away. Our analysis did not reveal any significant evidence of heterogeneity in either expenditure or consumption for both products, indicating that the tax's impact was relatively uniform across regions as shown in Figures \ref{fig:map_att_total_expenditure_std_per_region}, \ref{fig:map_att_total_expenditure_std_per_region_ls}, \ref{fig:map_att_weight_std_per_region}, and \ref{fig:map_att_weight_std_per_region_ls}.

\subsection{The average treatment effect on the neighbors (ATN): German households}

In this section, we examine the potential spillover effects of the Danish fat tax on butter and liver sausage prices in Northern Germany.

 We estimate the effect of the Danish fat tax on the average prices paid by German households living close to the Danish border, using households living farther away as a control group. As shown in Table \ref{tab:spillover_butter_prices}, we find no evidence of significant spillover effects from the tax on the prices of butter and liver sausage paid by these German households. This suggests that German shops did not adjust the prices of these products in response to the Danish fat tax.

\begin{table}[H]
\centering
\caption{This table shows the ATN for Northern-German households.}
\label{tab:spillover_butter_prices}
\begin{tabular}{lccc}
\hline
\textbf{Variable} & \textbf{Estimate} & \textbf{Std. Error} & \textbf{P-value} \\
\hline
\textit{ATT on bordering Germans: butter} & 0.173 & 0.799 & 0.557\\  
\hline
\textit{ATT on bordering Germans: liver sausage} & -6.052 & 5.45 & 0.890\\
\hline
\end{tabular}
\end{table}


\section{Conclusion}
\label{sec:conclusion}
This study is the first to causally estimate the impact of Denmark's saturated fat tax on consumer behavior and expenditure, utilizing a culturally similar control group of Northern German households from Schleswig-Holstein and Hamburg. Employing the doubly robust difference-in-differences estimator by \citet{callaway2021difference}, we flexibly incorporated observed covariates and accounted for multiple hypothesis testing, enhancing the validity of our results. The longitudinal nature of our data provided valuable insights into the evolution of the tax's effects over time. We ensured the robustness of our findings by thoroughly examining the plausibility of identifying assumptions, particularly the parallel trends and no anticipation assumptions, and by conducting robustness checks related to cross-border shopping.

Our analysis offers a comprehensive examination of the policy's implications, including expenditure patterns, the effects of tax abolishment, and post-tax habit formation—areas previously unexplored in the literature. We observed that the fat tax effectively reduced consumption of certain high-fat products, such as bacon, cheese, liver sausage, and salami, with reductions ranging between 11\% and 13\%. Notably, the reduction in consumption of bacon, cheese, and liver sausage was primarily driven by low-income consumers, indicating that the tax may have effectively targeted the population most in need of dietary improvements. However, it remains unclear whether consumers substituted these products with healthier alternatives, highlighting an important avenue for future research. Importantly, the consumption reductions were not sustained after the tax was repealed, suggesting a lack of long-term behavioral change.

Conversely, we found that consumption of products like margarine, butter, sour cream, and cream remained largely unchanged during the tax period, despite significant increases in expenditure ranging from 14\% to 25\%. This suggests inelastic demand for these items, with consumers, regardless of income level, absorbing the higher costs without reducing consumption. This uniform increase across income groups indicates a regressive effect of the tax, disproportionately affecting lower-income households. Understanding why low-income consumers did not substitute away from these products is critical for refining the design of such taxes. Future research could delve deeper into consumer behavior mechanisms, potentially through more granular data on individual consumer characteristics.

An important finding of our study is the significant increase in cross-border shopping of butter by Danish households living close to the German border during the tax period. Despite this increase in cross-border purchases, the overall consumption of butter did not decrease significantly, suggesting that cross-border shopping mitigated the intended effect of the tax on reducing saturated fat intake. The ability of consumers to avoid the tax by purchasing goods abroad underscores a critical limitation of the policy, particularly for small countries with accessible borders. Our analysis indicates that the tax's effectiveness was compromised in regions where cross-border shopping was feasible, highlighting the need for policymakers to consider geographical factors and potential avoidance behaviors when designing such interventions.

Moreover, we observed that the increase in cross-border shopping persisted even after the tax was abolished, indicating a lasting change in consumer behavior. This suggests that once consumers became accustomed to purchasing goods across the border, they continued to do so, further diminishing the potential long-term impact of the tax. Understanding the factors that encourage or discourage cross-border shopping is essential for developing more effective tax policies that achieve public health objectives without unintended consequences.

Our investigation into post-tax dynamics revealed that the abolishment of the fat tax led to varied effects across different product categories. For some products, such as butter, margarine, and sour cream, consumption not only rebounded but exceeded pre-tax levels, particularly among high-income households. This raises concerns that the temporary nature of the tax may have inadvertently encouraged higher consumption of saturated fats after its repeal, possibly due to changes in consumer habits or perceptions. Further research extending beyond 2014 would be valuable to assess the long-term impact of the tax's removal.

Analysis of price dynamics indicated that the tax was not uniformly passed through to consumers across all products. While prices increased for butter, cream, margarine, and sour cream during the tax period, other products showed no significant price changes. After the tax was abolished, prices decreased for some items like margarine and liver sausage but remained elevated for butter and sour cream. These findings suggest complexities in how taxes are transmitted through the supply chain and underscore the need for policymakers to consider market structures and pricing behaviors when designing such taxes.

Overall, our findings suggest that a broad tax on saturated fats may not uniformly achieve desired public health outcomes, given the varied consumer responses across different products and the mitigating effect of cross-border shopping. The tax's regressive impact on low-income households further complicates its efficacy as a policy tool. To optimize the design of fat taxes, it is essential to understand the underlying mechanisms driving consumer choices, consider geographical and cross-border factors, and to implement targeted approaches that minimize adverse effects on vulnerable populations. Policymakers should also contemplate complementary measures, such as international cooperation on tax policies, subsidies for healthier alternatives, or educational campaigns, to encourage lasting dietary changes.

Several limitations of our study point to areas for future research. Our analysis was constrained by the lack of detailed data on the fat content of individual products, limiting our ability to assess substitution dynamics within product categories or to compare observed price changes with theoretical expectations of full tax pass-through. Additionally, we did not examine the direct health outcomes associated with the tax, such as changes in body mass index (BMI) or cardiovascular health indicators. Investigating these health-related consequences would provide a more comprehensive understanding of the tax's long-term implications. Lastly, exploring the role of health messaging and public perception during the tax's implementation and repeal could shed light on the sociocultural factors influencing consumer behavior.

In conclusion, while the Danish fat tax had some success in reducing consumption of specific high-fat products among certain consumer groups, its effectiveness was undermined by cross-border shopping and varied responses across products and income levels. The ability of consumers to avoid the tax by purchasing goods abroad highlights a significant challenge in implementing such policies, especially in countries with accessible borders. Future policies should consider these nuances to design more effective strategies that promote healthier eating habits without disproportionately impacting low-income households or being circumvented through cross-border purchases.

\bigskip


\bibliography{references}

\newpage
\begin{appendices}
    \section{Data processing steps and descriptives}
    
\subsection{Data processing steps}
\label{app:flowcharts}

\FloatBarrier

\begin{figure}[htbp]
    \centering
    \includegraphics[width=\textwidth]{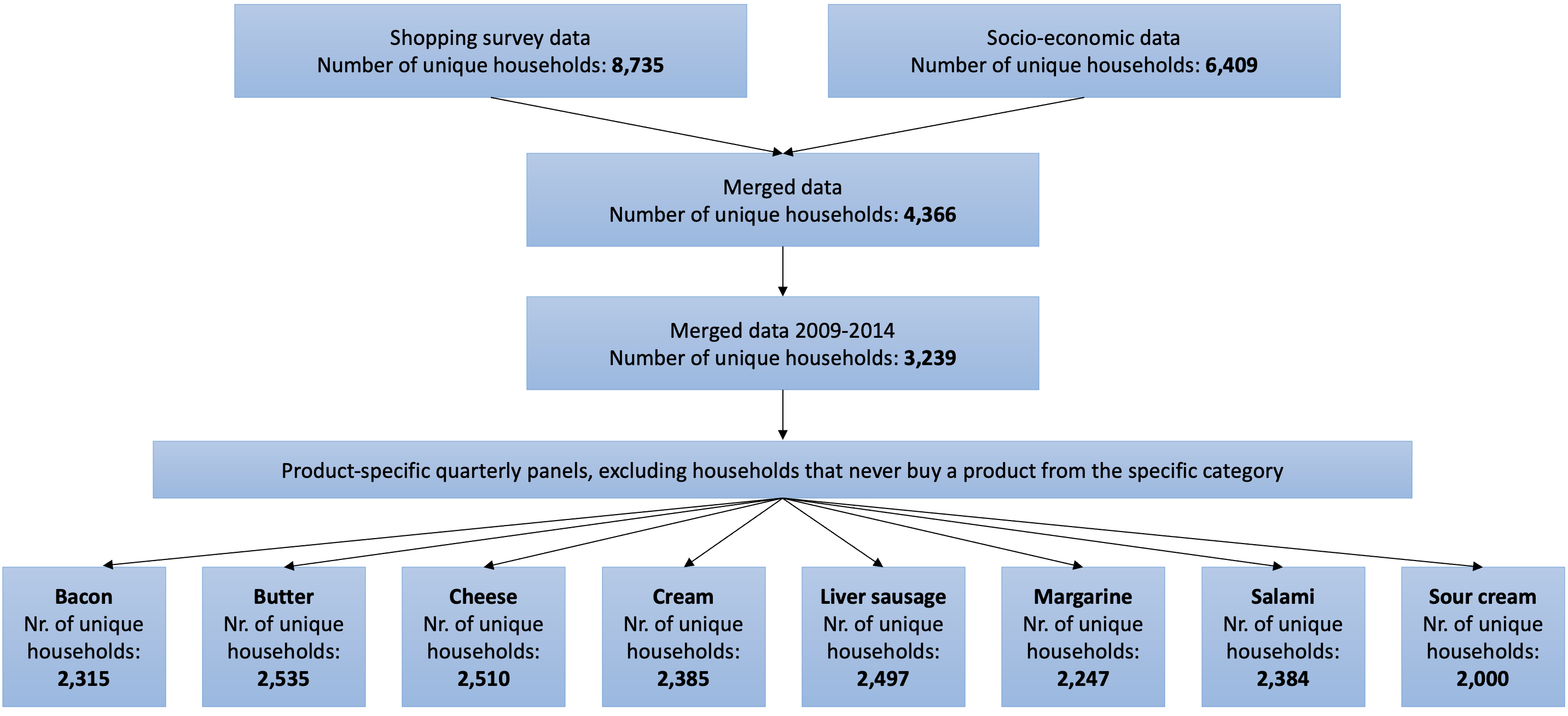}
    \caption{Data flow chart with the number of observations per step for constructing the Danish data set covering January 2009 to December 2014. For each product category, a quarterly panel is created which includes households that purchase some product from that category at least once between January 2009 and December 2014.}
    \label{fig:data_chart_dk}
\end{figure}

\newpage

\FloatBarrier

\hphantom{some text such that figure appears at top of page}

\begin{figure}[!t]
    \centering
    \includegraphics[width=\textwidth]{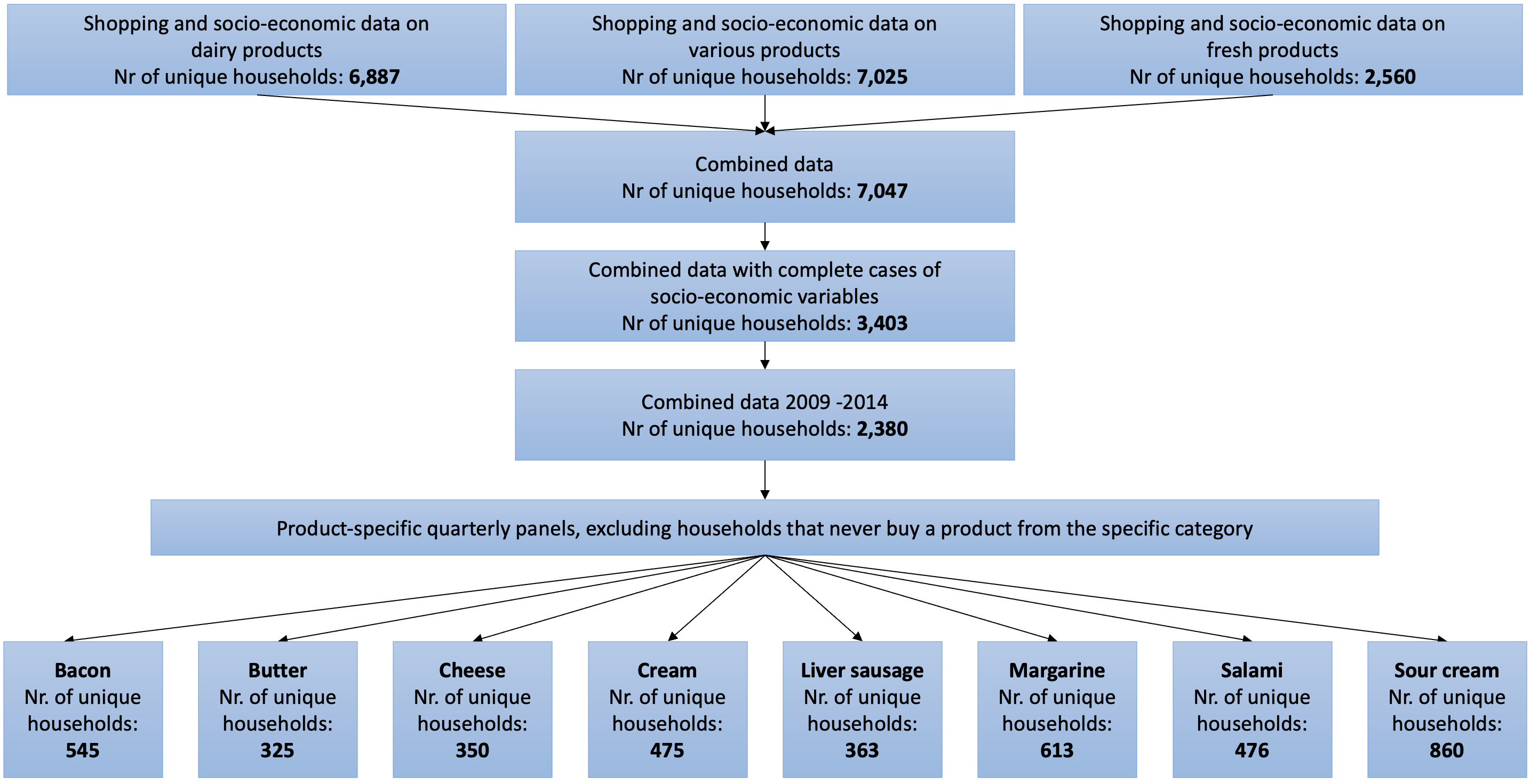}
    \caption{Data flow chart with the number of observations per step for constructing the Northern-German data set covering January 2009 to December 2014. For each product category, a quarterly panel is created which includes households that purchase some product from that category at least once between January 2009 and December 2014.}
    \label{fig:data_chart_de}
\end{figure}

\clearpage
    \newpage
    
\subsection{Summary statistics of pre-tax household characteristics}
\label{app:summary_characteristics}

\begingroup\footnotesize
\begin{longtable}{ll|rr|rr|rr}
  \caption{Summary statistics of pre-tax socio-economic household characteristics for Danish and Northern-German households used in the main analysis in Table \ref{tab:main_results}.} 
  \label{tab:sum_stat} \\ \hline \hline
 \textbf{Variable} & \textbf{Levels} & $\mathbf{n_{\mathrm{Denmark}}}$ & $\mathbf{\%_{\mathrm{Denmark}}}$ & $\mathbf{n_{\mathrm{Germany}}}$ & $\mathbf{\%_{\mathrm{Germany}}}$ & $\mathbf{n_{\mathrm{all}}}$ & $\mathbf{\%_{\mathrm{all}}}$ \\ 
  \hline
Geographic region & Capital region  & 583 & 21.5 & 0 & 0.0 & 583 & 12.9 \\ 
   & Zealand & 606 & 22.3 & 0 & 0.0 & 606 & 13.5 \\ 
   & Funen & 276 & 10.2 & 0 & 0.0 & 276 & 6.1 \\ 
   & North Jutland & 385 & 14.2 & 0 & 0.0 & 385 & 8.6 \\ 
   & East Jutland & 461 & 17.0 & 0 & 0.0 & 461 & 10.2 \\ 
   & South/West Jutland & 404 & 14.9 & 0 & 0.0 & 404 & 9.0 \\ 
   & Hamburg & 0 & 0.0 & 661 & 37.0 & 661 & 14.7 \\ 
   & Schleswig-Holstein & 0 & 0.0 & 1,125 & 63.0 & 1,125 & 25.0 \\ 
\hline
Age (head of hh.) & Under 30 & 578 & 21.3 & 203 & 11.4 & 781 & 17.4 \\ 
     & 30-39 & 717 & 26.4 & 362 & 20.3 & 1,079 & 24.0 \\ 
   & 40-49 & 614 & 22.6 & 469 & 26.3 & 1,083 & 24.1 \\ 
   & 50-59 & 514 & 18.9 & 333 & 18.6 & 847 & 18.8 \\ 
   & 60-69 & 137 & 5.0 & 304 & 17.0 & 441 & 9.8 \\ 
   & 70 or over & 155 & 5.7 & 115 & 6.4 & 270 & 6.0 \\ 
\hline
Children under 15 & 2,145 & 79.0 & 1,302 & 72.9 & 3,447 & 76.6 \\ 
   & 2 & 254 & 9.4 & 280 & 15.7 & 534 & 11.9 \\ 
   & 3 & 227 & 8.4 & 154 & 8.6 & 381 & 8.5 \\ 
   & 4 & 82 & 3.0 & 45 & 2.5 & 127 & 2.8 \\ 
   & 5 & 7 & 0.3 & 4 & 0.2 & 11 & 0.2 \\ 
   & 7 & 0 & 0.0 & 1 & 0.1 & 1 & 0.0 \\ 
   \hline
Gender (diary keeper)  & Male & 606 & 22.3 & 738 & 41.3 & 1344 & 29.9 \\
    & Female & 2,109 & 77.7 & 1,048 & 58.7 & 3,157 & 70.1\\ 
   \hline
Education  & 2 & 547 & 20.1 & 413 & 23.1 & 960 & 21.3 \\ 
   (ISCED level) & 3 & 1,015 & 37.4 & 666 & 37.3 & 1,681 & 37.4 \\ 
   & 5 & 396 & 14.6 & 176 & 9.8 & 572 & 12.7 \\ 
   & 6 & 583 & 21.5 & 211 & 11.8 & 794 & 17.6 \\ 
   & 7 & 174 & 6.4 & 320 & 17.9 & 494 & 11.0 \\ 
   \hline
Income (head of hh.) & Very low income& 843 & 31.1 & 610 & 34.1 & 1453 & 32.3 \\
    & Low income & 552 & 20.3 & 344 & 19.3 & 896 & 19.9 \\
   & Medium income & 362 & 13.3 & 298 & 16.7 & 660 & 14.7 \\ 
   & High income & 455 & 16.8 & 214 & 12.0 & 669 & 14.9 \\
   & Very high income  & 503 & 18.5 & 320 & 17.9 & 823 & 18.3 \\ 
   \hline
Household size & 1 & 990 & 36.5 & 457 & 25.6 & 1447 & 32.1 \\ 
   & 2 & 1,051 & 38.7 & 689 & 38.6 & 1,740 & 38.7 \\ 
   & 3 & 275 & 10.1 & 327 & 18.3 & 602 & 13.4 \\ 
   & 4 & 288 & 10.6 & 226 & 12.6 & 514 & 11.4 \\ 
   & 5 & 92 & 3.4 & 62 & 3.5 & 154 & 3.4 \\ 
   & 6 & 18 & 0.7 & 20 & 1.1 & 38 & 0.8 \\ 
   & 7 & 1 & 0.0 & 4 & 0.2 & 5 & 0.1 \\ 
   & 8 & 0 & 0.0 & 1 & 0.1 & 1 & 0.0 \\  
   \hline
Country & Denmark & 2,715 & 100.0 & 0 & 0.0 & 2,715 & 60.3 \\ 
   & Germany & 0 & 0.0 & 1,786 & 100.0 & 1,786 & 39.7 \\ 
   \hline
 & all & 2,715 & 100.0 & 1,786 & 100.0 & 4,501 & 100.0 \\ 
\hline
\hline
\end{longtable}
\endgroup
    \newpage
    
\subsection{Summary statistics for the consumption and price outcomes during the pre-tax and tax period}
\label{app:pretax_tax_dep}

\FloatBarrier

\begin{table}[!ht]
\centering
\caption{This table contains the average total quarterly consumption per household measured by the average weight and the average amount (counted by number of packages) purchased before and after the tax introduction. It further contains the average quarterly expenditure in Euro cents per household and the average price paid per 100 units. The pre-tax period covers January 2009 to September 2011, and the tax period covers October 2011 to December 2012.}
\label{tab:dep_quants}
\begin{adjustbox}{max width=\textwidth}
\footnotesize
\begin{tabular}{llcccccc}
  \hline \hline
 Product \hspace{1.5cm} & Outcome \hspace{1cm} & Pre-tax DK & Tax DK & $\Delta_{\text{DK}}$ & Pre-tax DE & Tax DE & $\Delta_{\text{DE}}$ \\ \hline
Bacon & Weight & 562.67 & 486.51 & -76.16 & 237.36 & 240.64 & 3.28 \\ 
   & Amount & 1.89 & 1.80 & -0.09 & 1.07 & 1.11 & 0.04 \\ 
   & Expenditure & 359.84 & 331.98 & -27.86 & 157.18 & 169 & 11.82 \\ 
   & Price & 67.74 & 71.00 & 3.26 & 77.45 & 82.93 & 5.48 \\
   \hline
Butter & Weight & 1,607.62 & 1,554.70 & -52.92 & 1,507.32 & 1,483.01 & -24.31 \\ 
   & Amount & 6.27 & 6.22 & -0.05 & 6.13 & 6.00 & -0.13 \\ 
   & Expenditure & 913.97 & 1,068.94 & 154.97 & 595.71 & 625.36 & 29.65 \\ 
   & Price & 59.65 & 72.36 & 12.71 & 41.06 & 43.41 & 2.35 \\
   \hline
Cheese & Weight & 2,615.39 & 2,059.62 & -555.77 & 2,678.16 & 2,569.77 & -108.39  \\ 
   & Amount & 3.96 & 3.52 & -0.44 & 9.44 & 9.30 & -0.14 \\ 
   & Expenditure & 2,236.91 & 1,845.75 & -391.16 & 1,942.41 & 1,948.03 & 5.62 \\ 
   & Price & 91.83 & 97.02 & 5.19 & 74.37 & 78.13 & 3.76 \\ 
   \hline
Cream & Weight & 1,465.81 & 1,548.64 & 82.83 & 1,099.53 & 1,108.92 & 9.39 \\ 
   & Amount & 3.47 & 3.53 & 0.06 & 5.10 & 5.14 & 0.04 \\ 
   & Expenditure & 481.91 & 569.35 & 87.44 & 242.29 & 267.57 & 25.28 \\ 
   & Price & 34.95 & 39.24 & 4.29 & 23.03 & 25.14 & 2.11 \\
   \hline
Liver sausage & Weight & 1,140.82 & 980.60 & -160.22 & 95.21 & 88.36 & -6.85 \\ 
   & Amount & 3.32 & 2.84 & -0.48 & 0.66 & 0.59 & -0.07  \\ 
   & Expenditure & 615.14 & 532.67 & -82.47 & 78.02 & 74.96 & -3.06 \\ 
   & Price & 61.48 & 63.19 & 1.71 & 86.28 & 91.05 & 4.77 \\ 
   \hline
Margarine & Weight & 1,842.00 & 1,693.05 & -148.95 & 1,969.10 & 1,760.18 & -208.92 \\ 
   & Amount & 3.87 & 3.63 & -0.24 & 4.12 & 3.58 & -0.54 \\ 
   & Expenditure & 511.17 & 570.55 & 59.38 & 368.39 & 350.34 & -18.05 \\ 
   & Price & 29.35 & 34.78 & 5.43 & 19.87 & 21.29 & 1.42 \\ 
   \hline
Salami & Weight & 398.79 & 338.49 & -60.3 & 474.31 & 463.43 & -10.88 \\ 
   & Amount & 1.93 & 1.81 & -0.12 & 3.05 & 2.97 & -0.08 \\ 
   & Expenditure & 478.57 & 405.15 & -73.42 & 449.1 & 452.91 & 3.81 \\ 
   & Price & 132.91 & 133.01 & 0.1 & 111.46 & 114.04 & 2.58 \\ 
   \hline
Sour cream & Weight & 613.76 & 604.32 & -9.44 & 506.46 & 524.83 & 18.37 \\ 
   & Amount & 1.33 & 1.30 & -0.03 & 2.47 & 2.55 & 0.08 \\ 
   & Expenditure & 176.17 & 186.28 & 10.11 & 127.57 & 132.27 & 4.70 \\ 
   & Price & 30.57 & 32.43 & 1.86 & 26.01 & 25.96 & -0.05 \\  
   \hline \hline
    \multicolumn{8}{p{1\textwidth}}{\footnotesize \textit{DK} indicates Danish households and \textit{DE} indicates Northern-German households. $\Delta$ indicates the difference.}\\ 
\end{tabular}
\end{adjustbox}
\end{table}

\begin{table}[!ht]
\centering
\label{tab:dep_quants_updated_products}
\begin{adjustbox}{max width=\textwidth}
\footnotesize
\begin{tabular}{llcccccc}
  \hline \hline
 Product \hspace{1.5cm} & Outcome \hspace{1cm} & Pre-tax DK & Tax DK & $\Delta_{\text{DK}}$ & Pre-tax DE & Tax DE & $\Delta_{\text{DE}}$ \\ \hline
 Eggs & Weight & 42.67 & 41.14 & -1.53 & 41.95 & 39.66 & -2.29 \\ 
   & Amount & 3.70 & 3.65 & -0.05 & 4.41 & 4.20 & -0.21 \\ 
   & Expenditure & 960.49 & 963.13 & 2.64 & 697.3 & 641.03 & -56.27 \\ 
   & Price & 2,538.28 & 2,582.03 & 43.75 & 1,764.54 & 1,758.29 & -6.25 \\ \hline
   Fruits & Weight & 2,056.73 & 730.00 & -1,326.73 & 18,663.45 & 16,788.45 & -1,875.00 \\ 
   & Amount & 17.36 & 6.28 & -11.08 & 18.15 & 17.21 & -0.94 \\ 
   & Expenditure & 3,634.88 & 1,244.02 & -2,390.86 & 2,921.78 & 2,764.65 & -157.13 \\ 
   & Price & 1,125.04 & 1,203.36 & 78.32 & 17.77 & 17.49 & -0.28 \\ \hline
Milk & Weight & 27,584.82 & 26,717.73 & -867.09 & 21,434.25 & 18,603.68 & -2,830.57 \\ 
   & Amount & 27.58 & 26.83 & -0.75 & 21.10 & 18.39 & -2.71 \\ 
   & Expenditure & 2,230.4 & 2,218.49 & -11.91 & 1,200.08 & 1,065.29 & -134.79 \\ 
   & Price & 8.67 & 8.91 & 0.24 & 5.84 & 6.00 & 0.16 \\ \hline
  Roast beef & Weight & 138.53 & 69.99 & -68.54 & 31.20 & 46.35 & 15.15 \\ 
   & Amount & 0.89 & 0.80 & -0.09 & 0.31 & 0.43 & 0.12 \\ 
   & Expenditure & 251.28 & 149.48 & -101.80 & 98.14 & 126.28 & 28.14 \\ 
   & Price & 229.07 & 233.34 & 4.27 & 317.64 & 283.44 & -34.20 \\ \hline
  Toilet paper & weight & 19.57 & 18.53 & -1.04 & 18.61 & 18.44 & -0.17 \\ 
   & Amount & 2.26 & 2.27 & 0.01 & 1.90 & 1.90 & 0.00 \\ 
   & Expenditure & 580.54 & 564.47 & -16.07 & 503.12 & 503.69 & 0.57 \\ 
   & Price & 3,107.14 & 3,136.17 & 29.03 & 2,721.62 & 2,742.86 & 21.24 \\ \hline
  Yoghurt & Weight & 6,397.78 & 6,285.79 & -111.99 & 4,963.91 & 5,062.14 & 98.23 \\ 
   & Amount & 7.31 & 7.25 & -0.06 & 18.67 & 19.05 & 0.38 \\ 
   & Expenditure & 1,092.00 & 1,108.03 & 16.03 & 894.50 & 896.38 & 1.88 \\ 
   & Price & 18.53 & 19.14 & 0.61 & 18.93 & 19.08 & 0.15 \\ 
   
   \hline \hline
    \multicolumn{8}{p{1\textwidth}}{\footnotesize \textit{DK} indicates Danish households and \textit{DE} indicates Northern-German households. $\Delta$ indicates the difference.}\\ 
\end{tabular}
\end{adjustbox}
\end{table}

\clearpage
    \newpage
    \section{Fitted propensity scores for overlap verification}
    \label{app:fitted_prop_scores}
    
\begin{figure}[!htbp]
    \centering
    
    \begin{minipage}{0.45\textwidth}
        \centering
        \includegraphics[width=\textwidth]{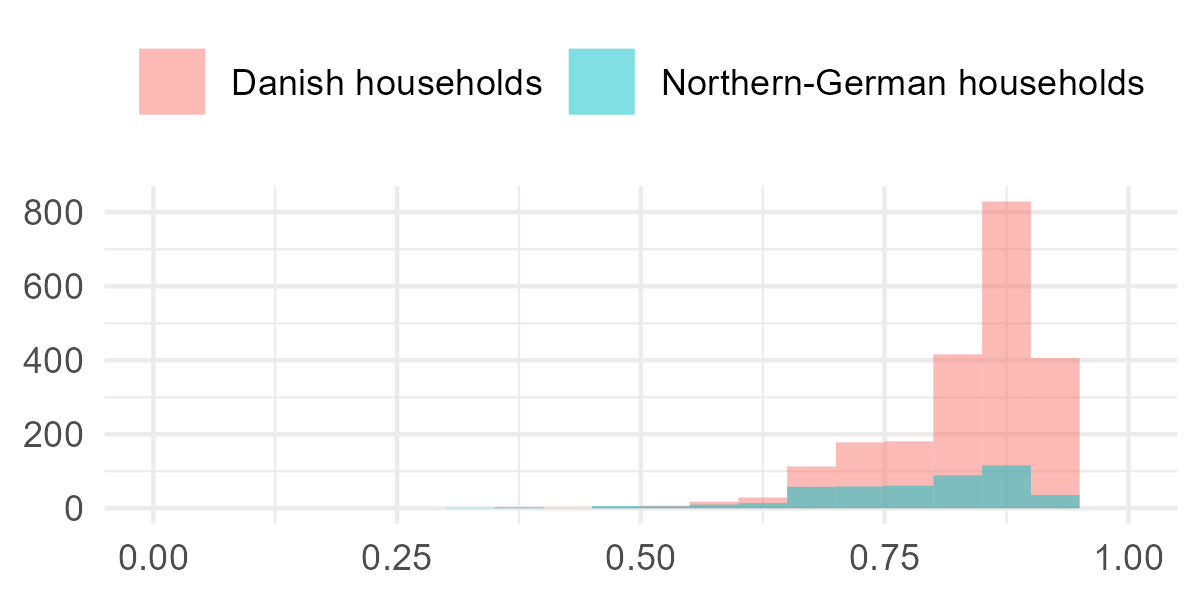}
        \caption{Bacon.}
        \label{fig:fitted_prop_bacon}
    \end{minipage}
    \hfill
    \begin{minipage}{0.45\textwidth}
        \centering
        \includegraphics[width=\textwidth]{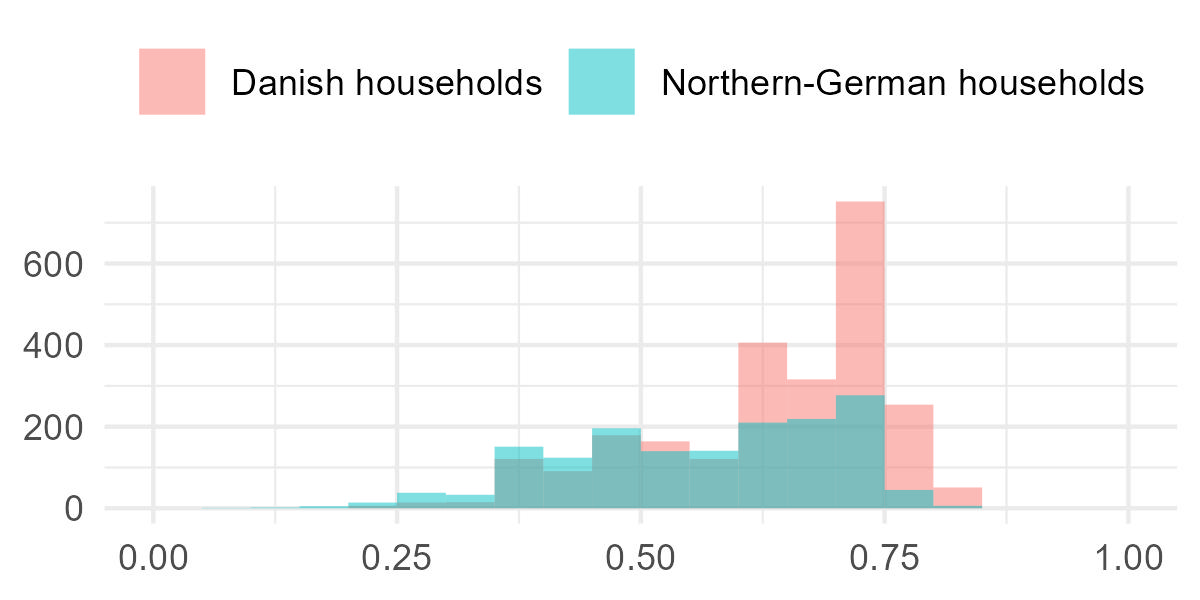}
        \caption{Butter.}
        \label{fig:fitted_prop_butter}
    \end{minipage}
    
    \begin{minipage}{0.45\textwidth}
        \centering
        \includegraphics[width=\textwidth]{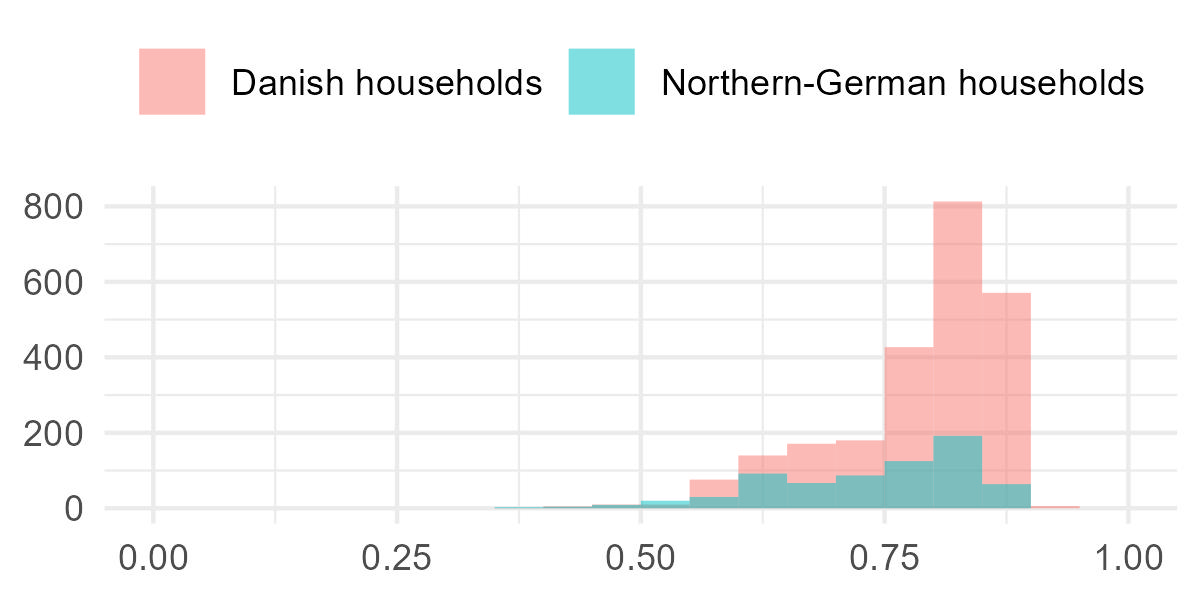}
        \caption{Cheese.}
        \label{fig:fitted_prop_cheese}
    \end{minipage}
    \hfill
    \begin{minipage}{0.45\textwidth}
        \centering
        \includegraphics[width=\textwidth]{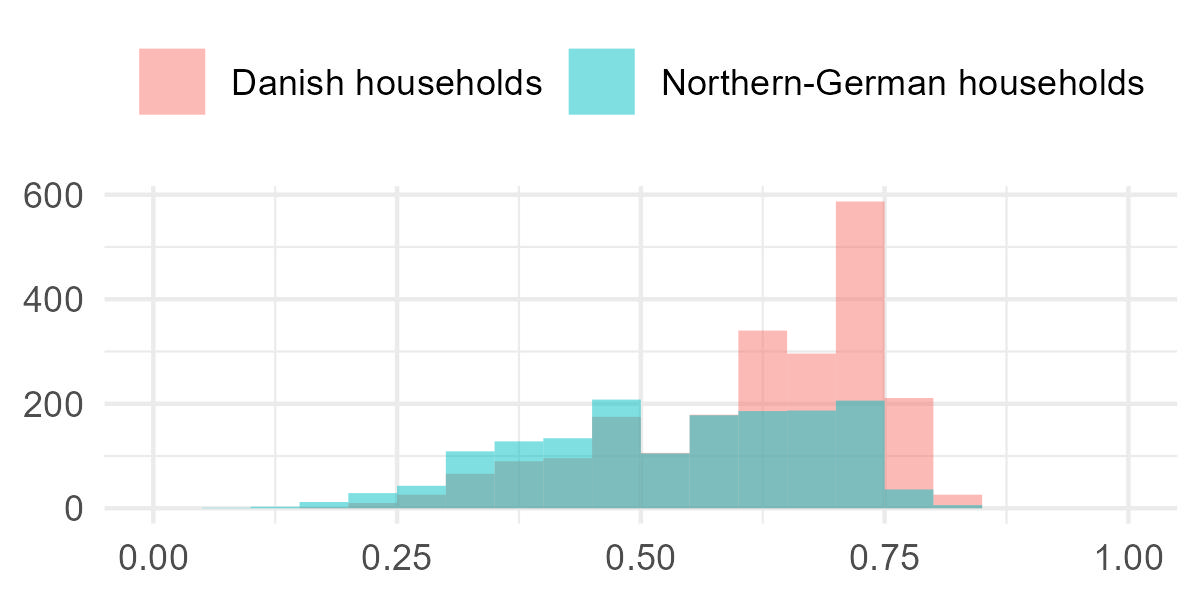}
        \caption{Cream.}
        \label{fig:fitted_prop_cream}
    \end{minipage}

     \begin{minipage}{0.45\textwidth}
        \centering
        \includegraphics[width=\textwidth]{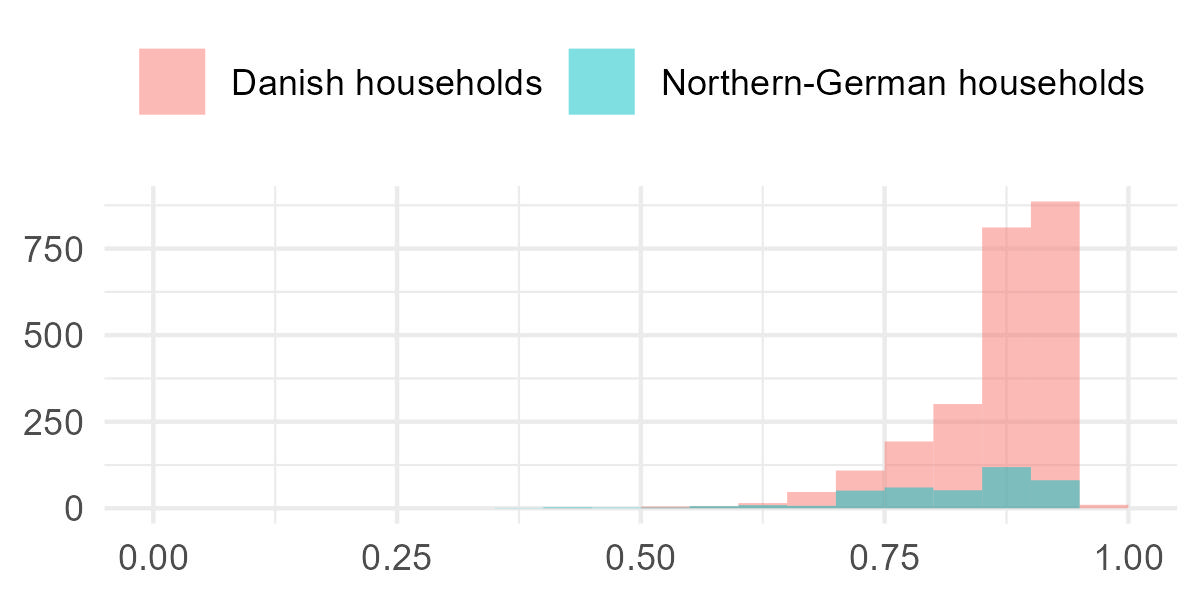}
        \caption{Liver sausage.}
        \label{fig:fitted_prop_liver}
    \end{minipage}
    \hfill
    \begin{minipage}{0.45\textwidth}
        \centering
        \includegraphics[width=\textwidth]{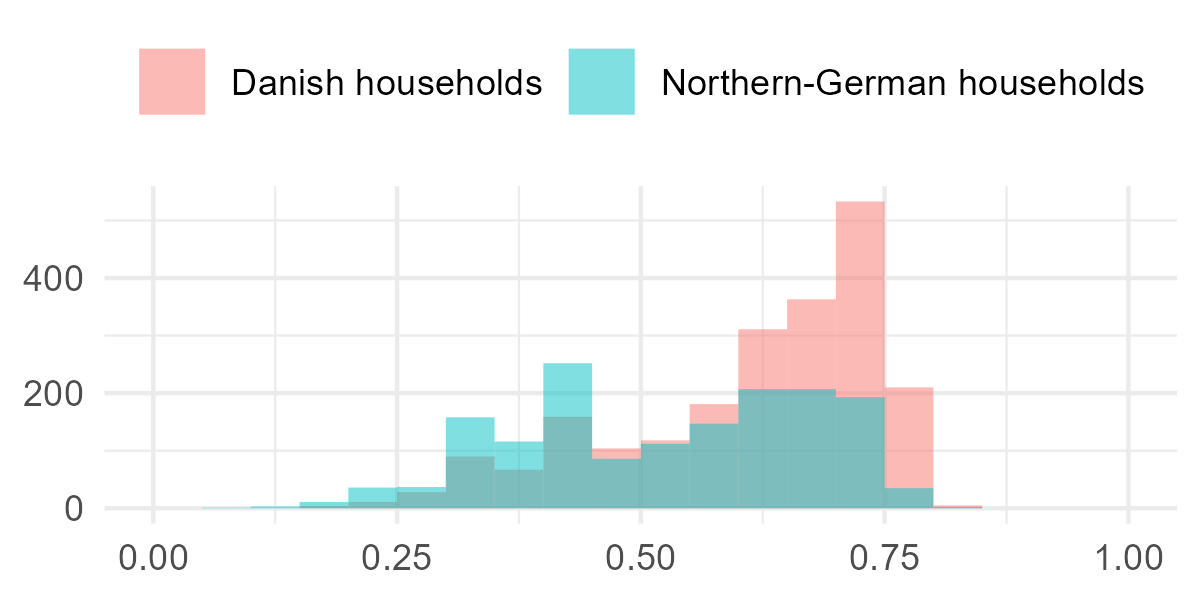}
        \caption{Margarine.}
        \label{fig:fitted_prop_margarine}
    \end{minipage}
    
    \begin{minipage}{0.45\textwidth}
        \centering
        \includegraphics[width=\textwidth]{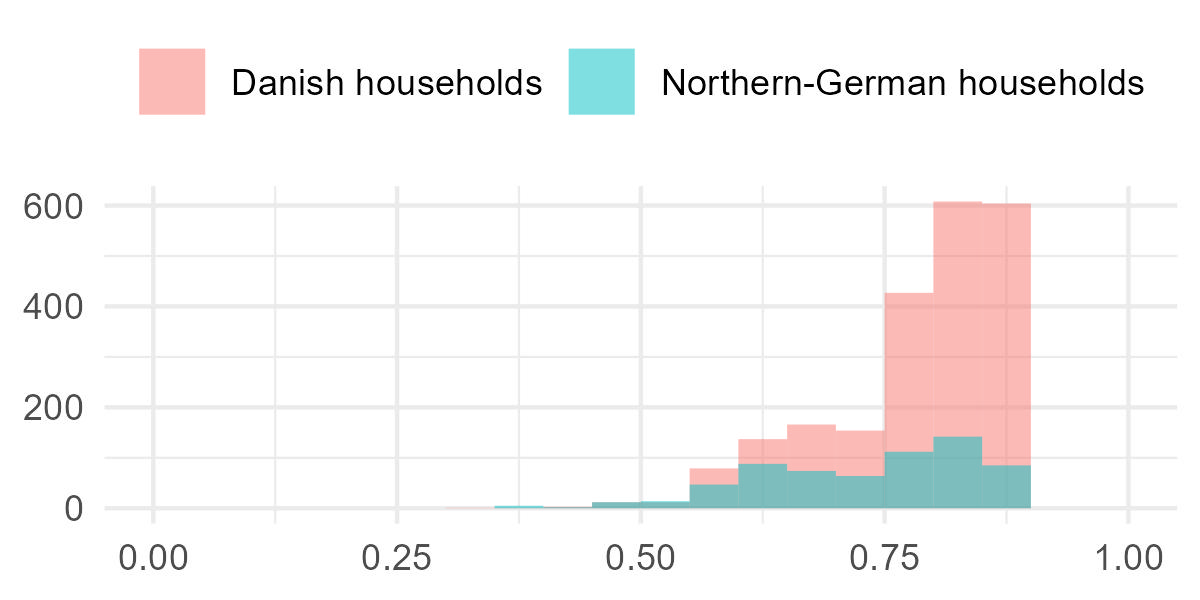}
        \caption{Salami.}
        \label{fig:fitted_prop_salami}
    \end{minipage}
    \hfill
    \begin{minipage}{0.45\textwidth}
        \centering
        \includegraphics[width=\textwidth]{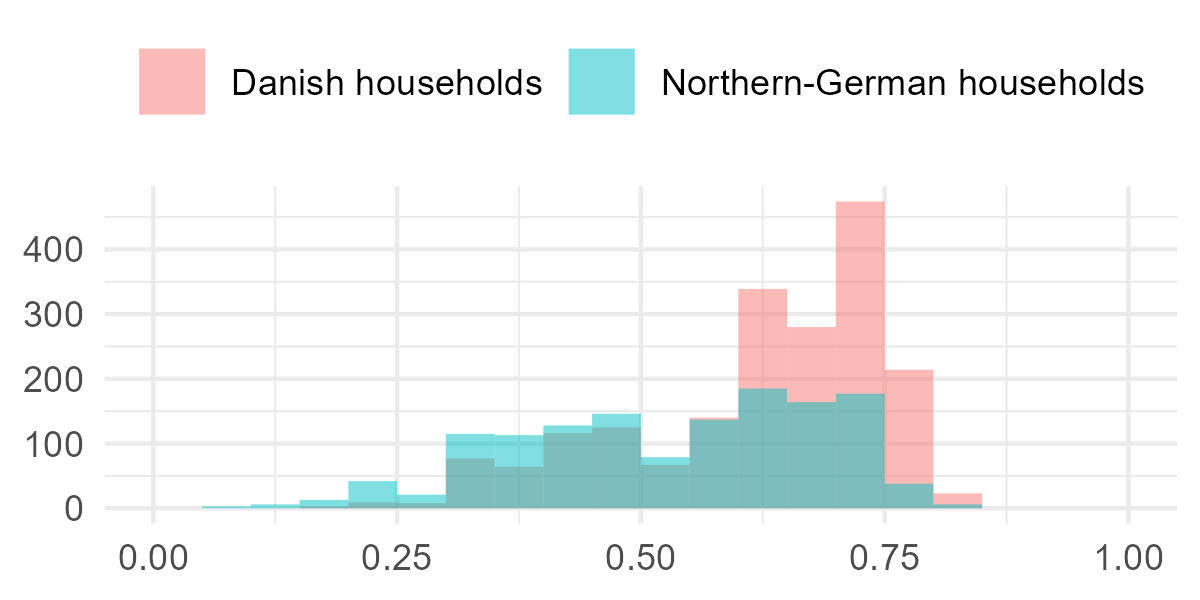}
        \caption{Sour cream.}
        \label{fig:fitted_prop_sour}
    \end{minipage}
    
    \caption{Fitted propensity scores for the Danish and Northern-German households obtained with a logistic regression controlling for the age of the head of the household, the number of children under 15, gender, educational attainment (ISCED), income level, and household size. The overlap assumption requires a positive probability of being treated at every covariate value, which seems to be valid for the cases depicted here.}
    \label{fig:hist_fitted_prop_scores}
\end{figure}

\section{Tables Anticipation Effects and Two-way Fixed Effects} \label{app:ae-twfe}
\begin{table}[H]
\centering
\caption{The causal effect of the Danish fat tax in the two quarters preceding the tax, using the first quarter of 2011 as reference period.}
\label{tab:anticipation_quarters}
\begin{adjustbox}{max width=\textwidth}
\small
\begin{tabular}{p{2.5cm} *{8}{p{1.6cm}<{\centering}}}
  \hline \hline
  & (1) & (2) & (3) & (4) & (5) & (6) & (7) & (8) \\
 & Bacon & Butter & Cheese & Cream & Liver sausage & Margarine & Salami & Sour cream \\ 
    \hline
  \multicolumn{9}{l}{\textbf{Panel A: Quarterly average weight consumed in gr/ml}} \\ \hline
  $ATT_{2011\textls{-}Q2}$  & -20.188 & -17.318 & -115.138 & -19.246 & 4.789 & -57.998 & 60.493 & 28.426 \\ 
  \textit{(se)} & (27.741) & (56.886) & (134.277) & (27.988) & (24.674) & (29.855) & (45.022) & (13.440) \\ 
  \textit{p-value} & 0.692 & 0.810 & 0.547 & 0.643 & 0.894 & 0.165 & 0.358 & 0.183\\
  $ATT_{2011\textls{-}Q3}$ & 18.208 & 35.762 & -78.247 & -10.580 & 27.649 & 53.792 & 91.394 & 20.983 \\ 
  \textit{(se)} & (42.933) & (100.254) & (143.327) & (87.195) & (63.777) & (135.263) & (49.936) & (48.075) \\ 
  \textit{p-value} & 0.656 & 0.667 & 0.691 & 0.799 & 0.457 & 0.223 & 0.170 & 0.301\\
    \hline
  \multicolumn{9}{l}{\textbf{Panel B: Quarterly average expenditure in Euro cents}} \\ \hline
  $ATT_{2011\textls{-}Q2}$& -0.034 & 0.220 & -0.133 & -0.335 & -0.005 & -0.149 & 0.311 & 0.274 \\ 
  \textit{(se)} & (0.211) & (0.411) & (0.495) & (0.348) & (0.150) & (0.261) & (0.269) & (0.207)  \\ 
  $ATT_{2011\textls{-}Q3}$ & 0.227 & 0.112 & -0.384 & -0.259 & 0.177 & 0.509 & 0.565 & 0.205 \\ 
  \textit{(se)} & (0.173) & (0.380) & (0.479) & (0.312) & (0.137) & (0.291) & (0.279) & (0.183) \\ 
  \textit{p-value} & 0.357 & 0.846 & 0.576 & 0.567 & 0.378 & 0.221 & 0.167 & 0.436\\
      \hline \hline
   \multicolumn{9}{p{1.15\textwidth}}{\footnotesize Significance level: * $p<0.1$ ** $p<0.05$ *** $p<0.01$. Standard errors are clustered at household level. The results are obtained controlling for the age of the head of the household, the number of children under 15 years, the gender of diary keeper, the educational level (ISCED), the income level, and the household size.}\\ 
\end{tabular}
\end{adjustbox}
\end{table}

\begin{table}[H]
\centering
\caption{TWFE estimates.}
\label{tab:twe}
\begin{adjustbox}{max width=\textwidth}
\begin{tabular}{B *{8}{C}}
  \hline \hline
  & (1) & (2) & (3) & (4) & (5) & (6) & (7) & (8) \\
 & Bacon & Butter & Cheese & Cream & Liver sausage & Margarine & Salami & Sour cream \\ 
  \hline
  \multicolumn{9}{l}{\textbf{Panel A: Quarterly average weight consumed in gr/ml}} \\ \hline
  \textit{Danish fat tax (ATT)} & $-80.801^{***}$ & $-31.697^{**}$ & $-446.679^{***}$ & $68.062^{**}$ & $-153.833^{***}$ & $57.257^{**}$ & $-49.589^{***}$ & $-28.427^{**}$ \\ 
  \textit{(se)} & (0.186) & (0.605) & (0.330) & (4.545) & (0.466) & (0.941) & (0.184) & (0.473) \\ 
  \textit{p-value} & 0.001 & 0.012 & 0.000 & 0.042 & 0.002 & 0.010 & 0.002 & 0.011 \\ 
  \textit{\% change} & -13.05\% & -2.04\% & -16.07\% & 5.03\% & -13.29\% & 2.98\% & -12.39\% & -5.16\%\\
    \hline
  \multicolumn{9}{l}{\textbf{Panel B: Quarterly average expenditure in Euro cents}} \\ \hline
  \textit{Danish fat tax (ATT)} & $-40.623^{***}$ & $123.457^{***}$ & $-396.318^{***}$ & $61.382^{***}$ & $-79.730^{***}$ & $77.253^{***}$ & $-77.510^{***}$ & $5.250^{**}$ \\ 
  \textit{(se)} & (0.148) & (0.176) & (0.066) & (0.758) & (0.215) & (0.030) & (0.125) & (0.097) \\ 
  \textit{p-value} & 0.002 & 0.001 & 0.000 & 0.008 & 0.002 & 0.000 & 0.001 & 0.012 \\ 
  \textit{\% change} & -10.39\% & 13.88\% & -16.71\% & 13.33\% & -12.72\% & 14.37\% & -16.29\% & 3.30\% \\
   \hline \hline
   \multicolumn{9}{p{1.3\textwidth}}{\normalsize Significance level: * $p<0.1$ ** $p<0.05$ *** $p<0.01$. Standard errors clustered at unit level. }\\ 
\end{tabular}
\end{adjustbox}
\end{table}
    \newpage
    \section{Evolution of tax effects over time}
    \label{app:dynamic_treatments}
    
\subsection{Bacon}
\FloatBarrier

\begin{figure}[!htbp]
    \centering
    \begin{subfigure}{0.75\textwidth} 
        \includegraphics[width=\textwidth]{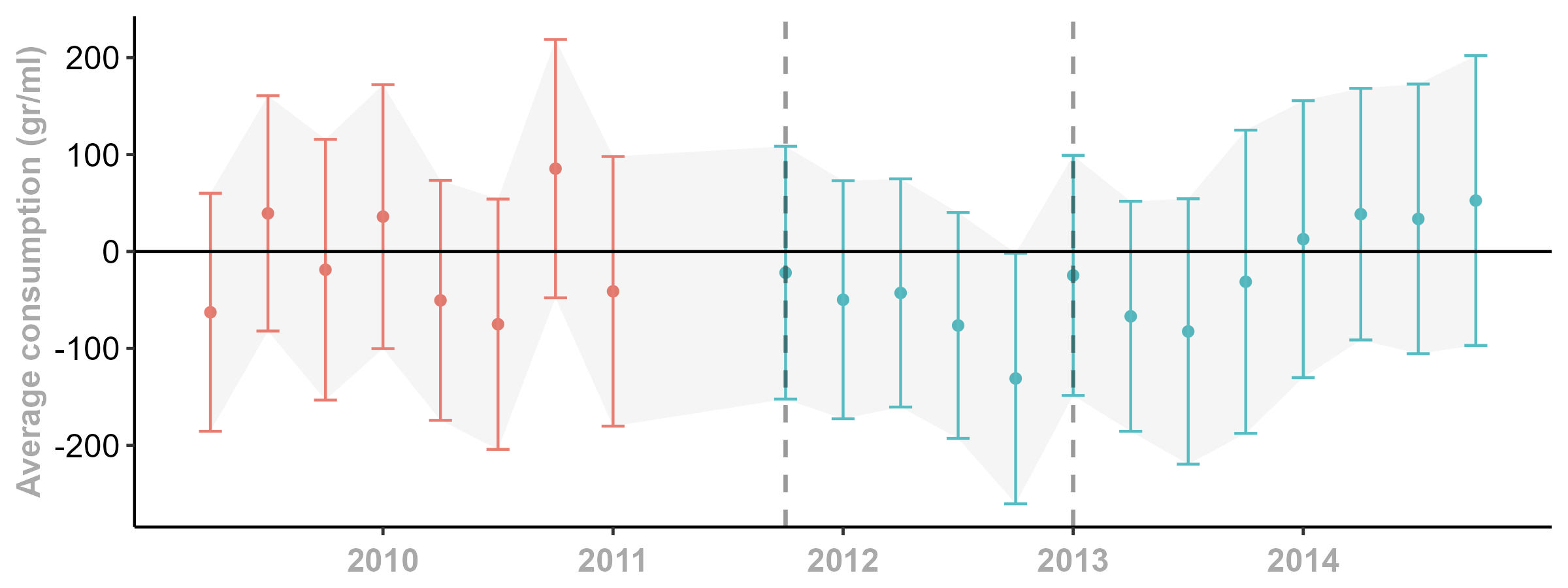}
        \caption{Average quarterly weight consumed (gr/ml) per household.}
    \end{subfigure}
    
    \begin{subfigure}{0.75\textwidth} 
        \includegraphics[width=\textwidth]{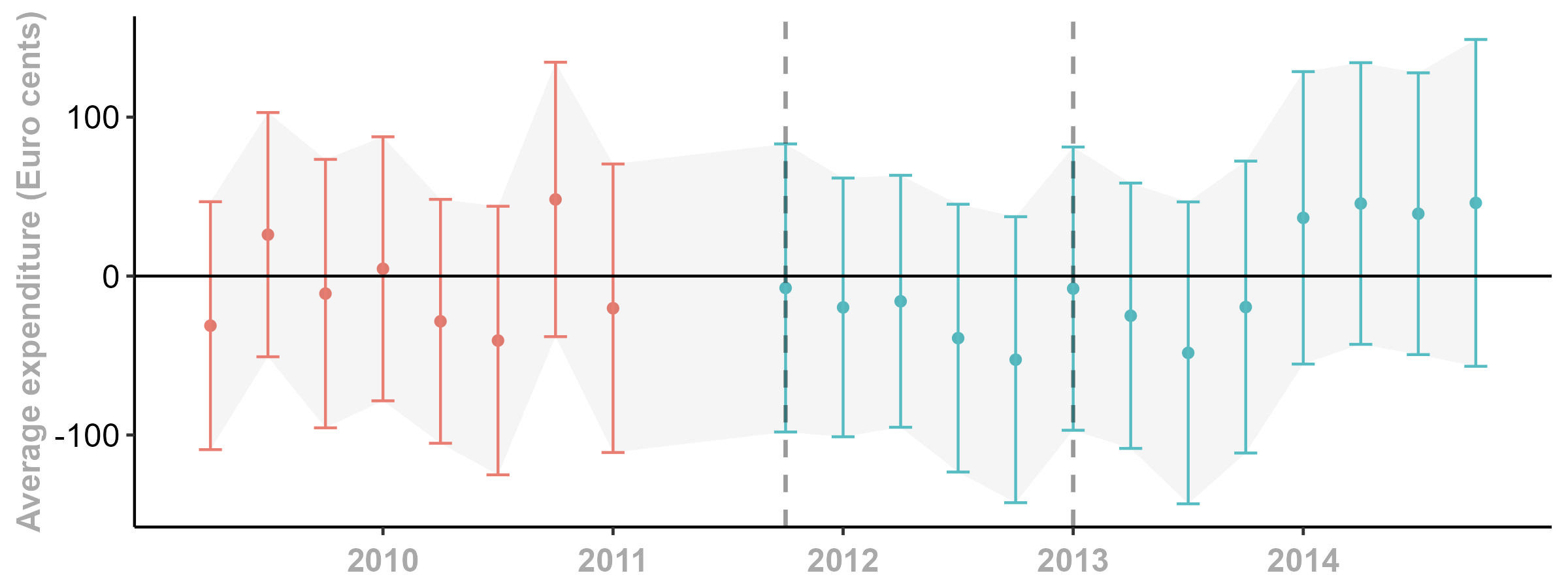}
        \caption{Average quarterly expenditure in Euro cents per household.}
    \end{subfigure}

    \begin{subfigure}{0.75\textwidth} 
        \includegraphics[width=\textwidth]{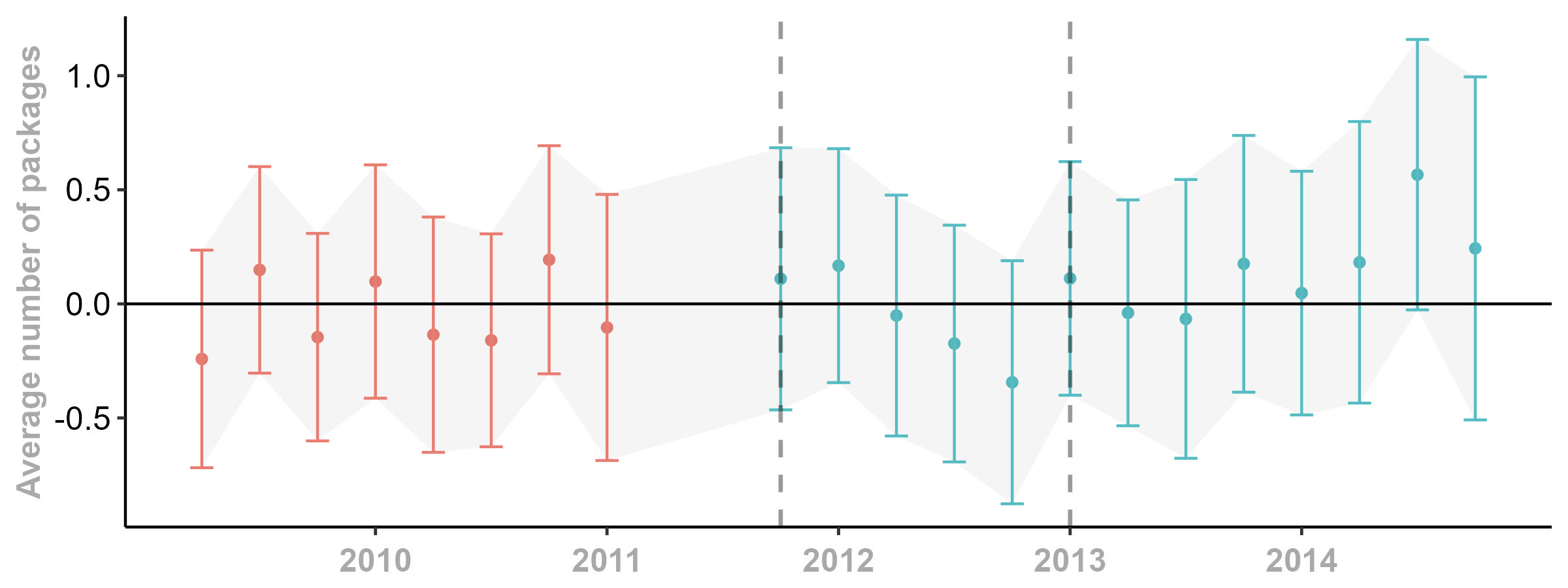}
        \caption{Average quarterly number of packages per household.}
    \end{subfigure}

    \caption{Time average treatment effects, $ATT(q)$. The two vertical dashed lines indicate the introduction and abolishment of the Danish fat tax, respectively. Red estimates and simultaneous 95\% confidence intervals for pre-tax periods, and blue estimates and simultaneous 95\% confidence intervals for tax and post-tax periods, clustered at the household level. Under conditional parallel trends, red estimates should be close to zero. Estimates are based on the doubly robust Equation (\ref{eq:dr_did}). Point estimates and standard errors are reported in Column (1) of Tables \ref{tab:att_t_weight} to \ref{tab:att_t_amount}.}
    \label{fig:main_graphs_bacon}
\end{figure}

\clearpage

\subsection{Butter}

\FloatBarrier

\begin{figure}[!htbp]
    \centering
    
    \begin{subfigure}{0.75\textwidth}
        \includegraphics[width=\textwidth]{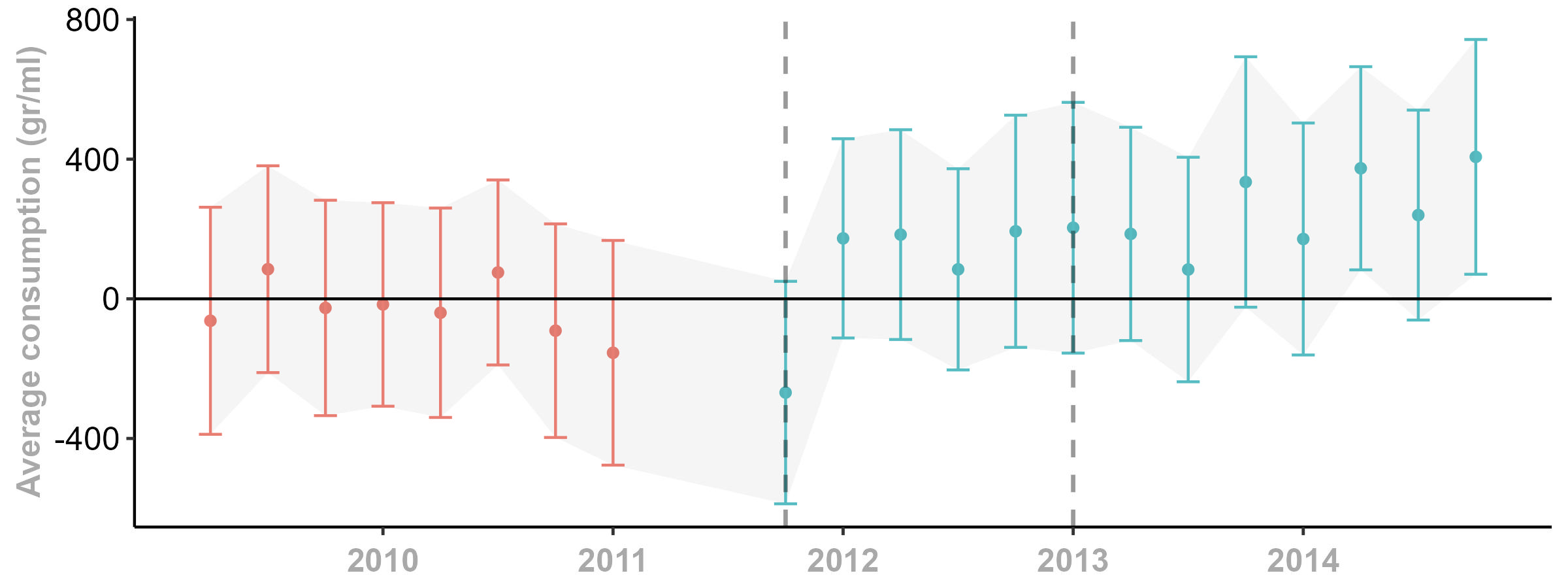}
        \caption{Average quarterly weight consumed (gr/ml) per household.}
        \label{fig:evolution_weight_butter}
    \end{subfigure}
    
    \begin{subfigure}{0.75\textwidth}
        \includegraphics[width=\textwidth]{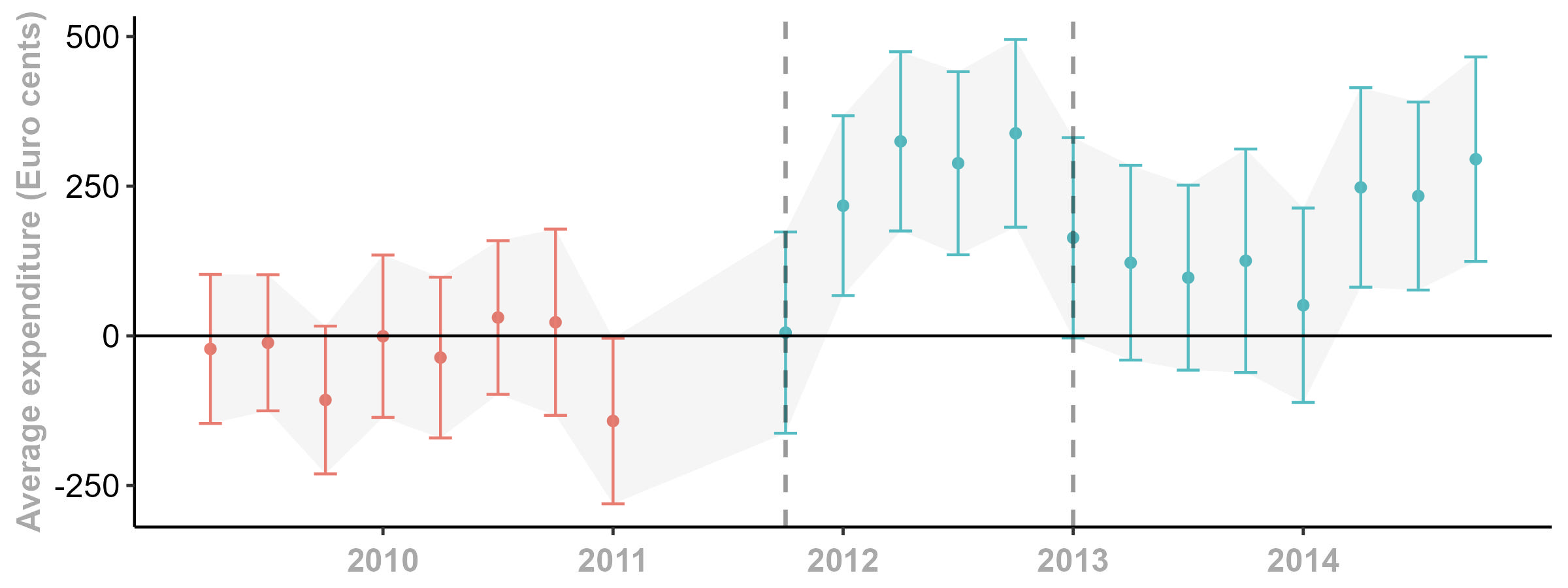}
        \caption{Average quarterly expenditure in Euro cents per household.}
        \label{fig:evolution_expenditure_butter}
    \end{subfigure}

    \begin{subfigure}{0.75\textwidth}
         \includegraphics[width=\textwidth]{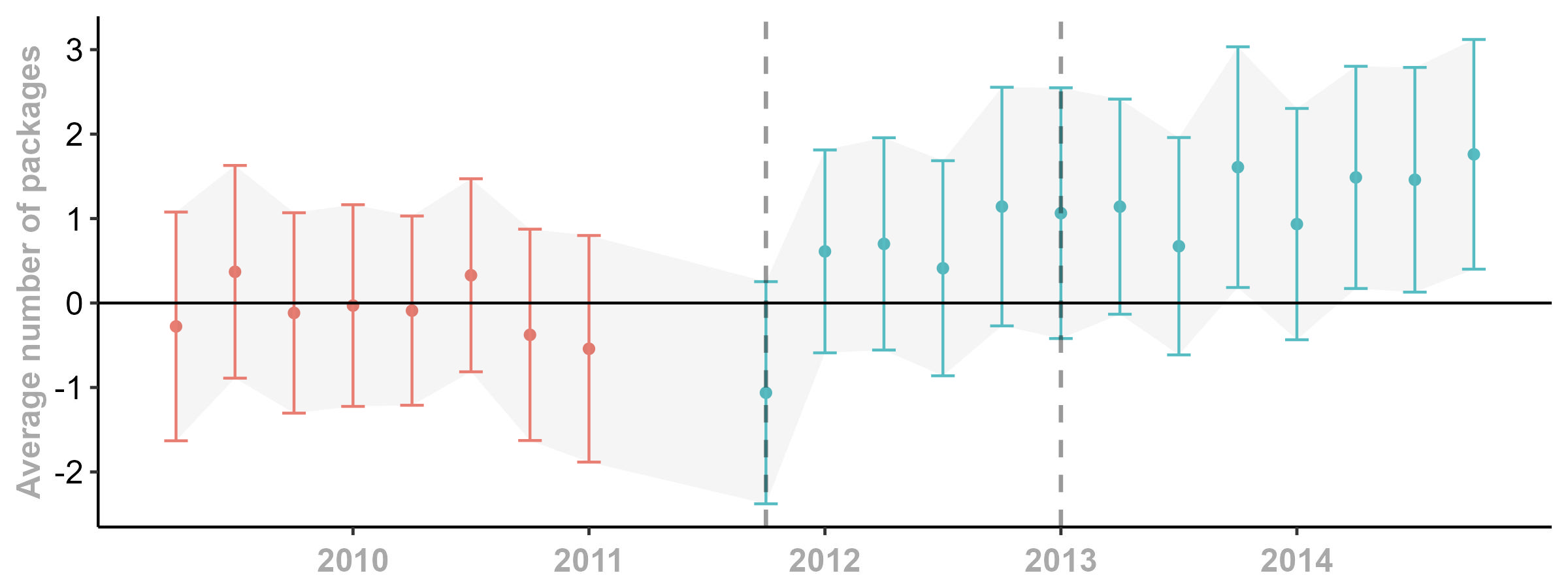}
        \caption{Average quarterly number of packages per household.}
        \label{fig:evolution_amount_butter}
    \end{subfigure}
    
    \caption{Time average treatment effects, $ATT(q)$. The two vertical dashed lines indicate the introduction and abolishment of the Danish fat tax, respectively. Red estimates and simultaneous 95\% confidence intervals for pre-tax periods, and blue estimates and simultaneous 95\% confidence intervals for tax and post-tax periods, clustered at the household level. Under conditional parallel trends, red estimates should be close to zero. Estimates are based on the doubly robust Equation (\ref{eq:dr_did}). Point estimates and standard errors are reported in Column (2) of Tables \ref{tab:att_t_weight} to \ref{tab:att_t_amount}.}
    \label{fig:main_graphs_butter}
\end{figure}

\clearpage

\subsection{Cheese}

\FloatBarrier

\begin{figure}[!htbp]
    \centering
    
    \begin{subfigure}{0.75\textwidth}
        \includegraphics[width=\textwidth]{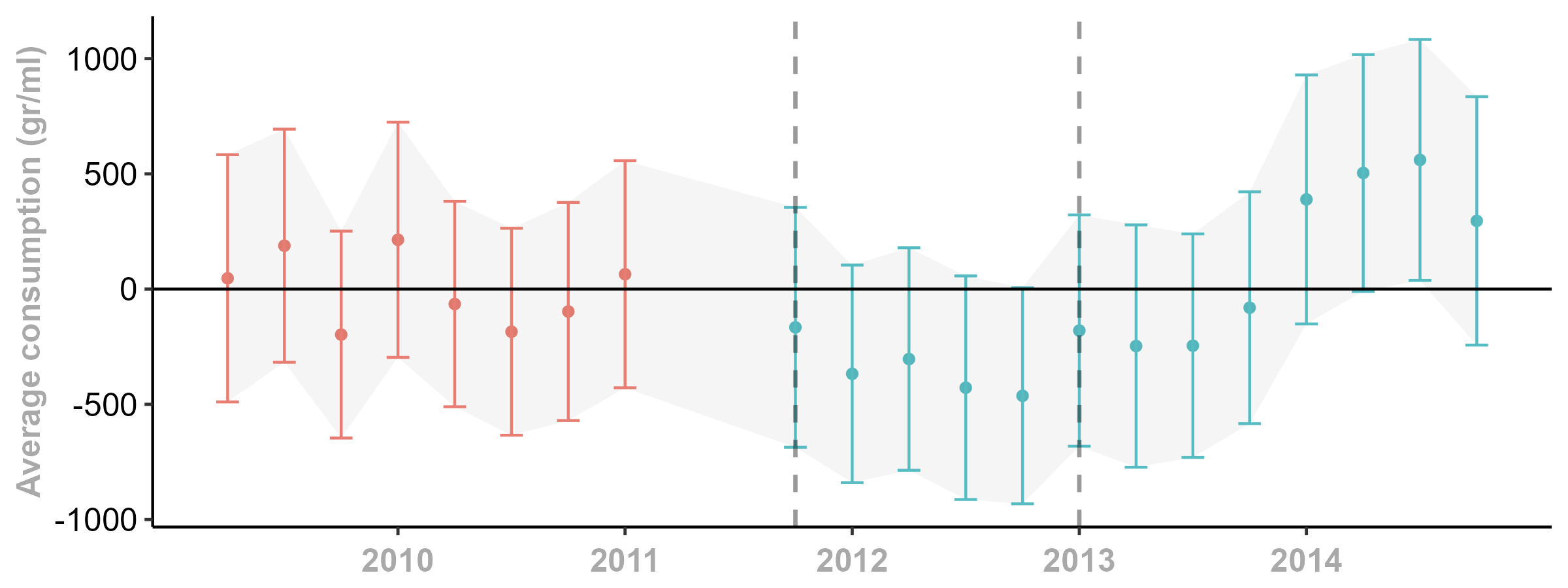}
        \caption{Average quarterly weight consumed (gr/ml) per household.}
        \label{fig:evolution_weight_cheese}
    \end{subfigure}
    
    \begin{subfigure}{0.75\textwidth}
        \includegraphics[width=\textwidth]{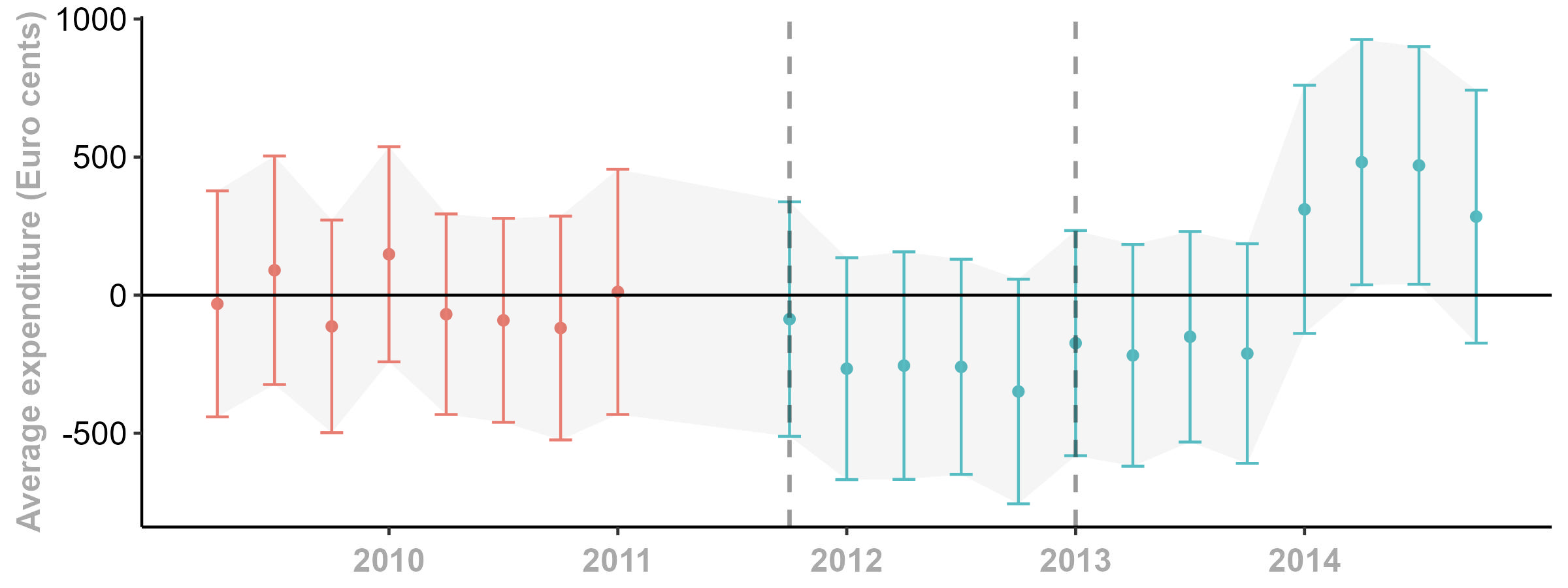}
        \caption{Average quarterly expenditure in Euro cents per household.}
        \label{fig:evolution_expenditure_cheese}
    \end{subfigure}

    \begin{subfigure}{0.75\textwidth}
         \includegraphics[width=\textwidth]{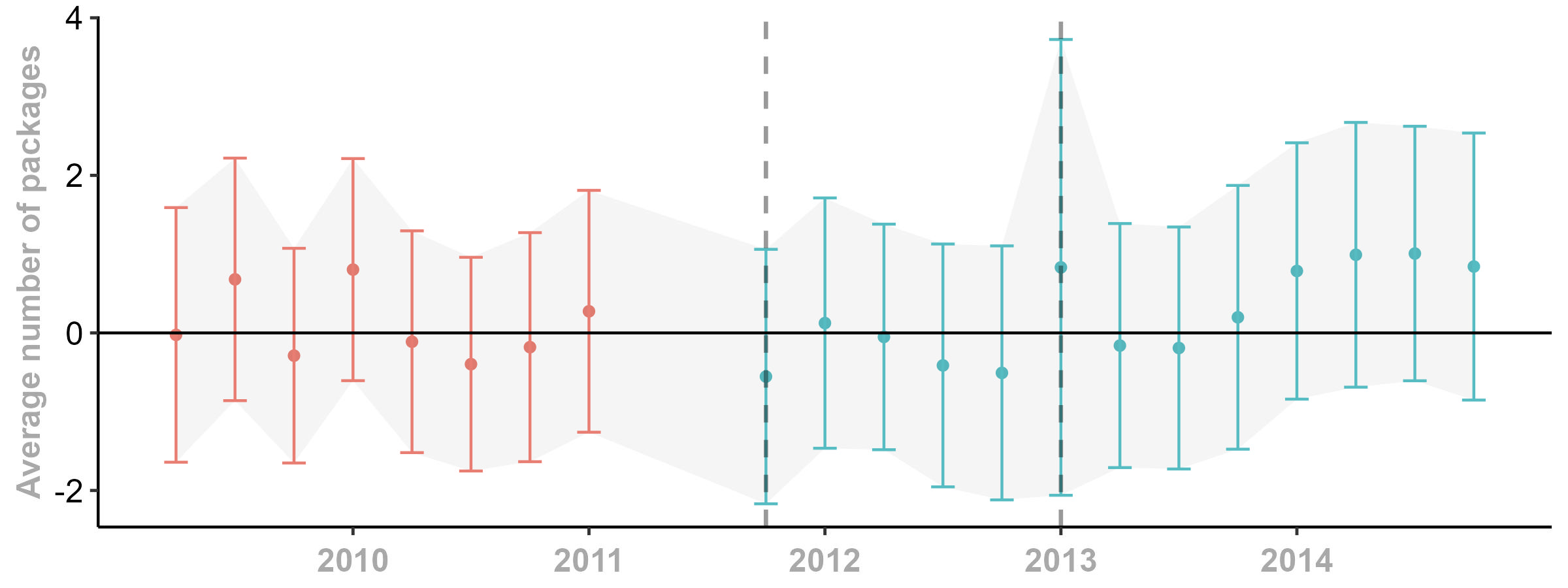}
        \caption{Average quarterly number of packages per household.}
        \label{fig:evolution_amount_cheese}
    \end{subfigure}
    
    \caption{Time average treatment effects, $ATT(q)$. The two vertical dashed lines indicate the introduction and abolishment of the Danish fat tax, respectively. Red estimates and simultaneous 95\% confidence intervals for pre-tax periods, and blue estimates and simultaneous 95\% confidence intervals for tax and post-tax periods, clustered at the household level. Under conditional parallel trends, red estimates should be close to zero. Estimates are based on the doubly robust Equation (\ref{eq:dr_did}). Point estimates and standard errors are reported in Column (3) of Tables \ref{tab:att_t_weight} to \ref{tab:att_t_amount}.}
    \label{fig:main_graphs_cheese}
\end{figure}

\clearpage

\subsection{Cream}

\FloatBarrier

\begin{figure}[!htbp]
    \centering
    
    \begin{subfigure}{0.75\textwidth}
        \includegraphics[width=\textwidth]{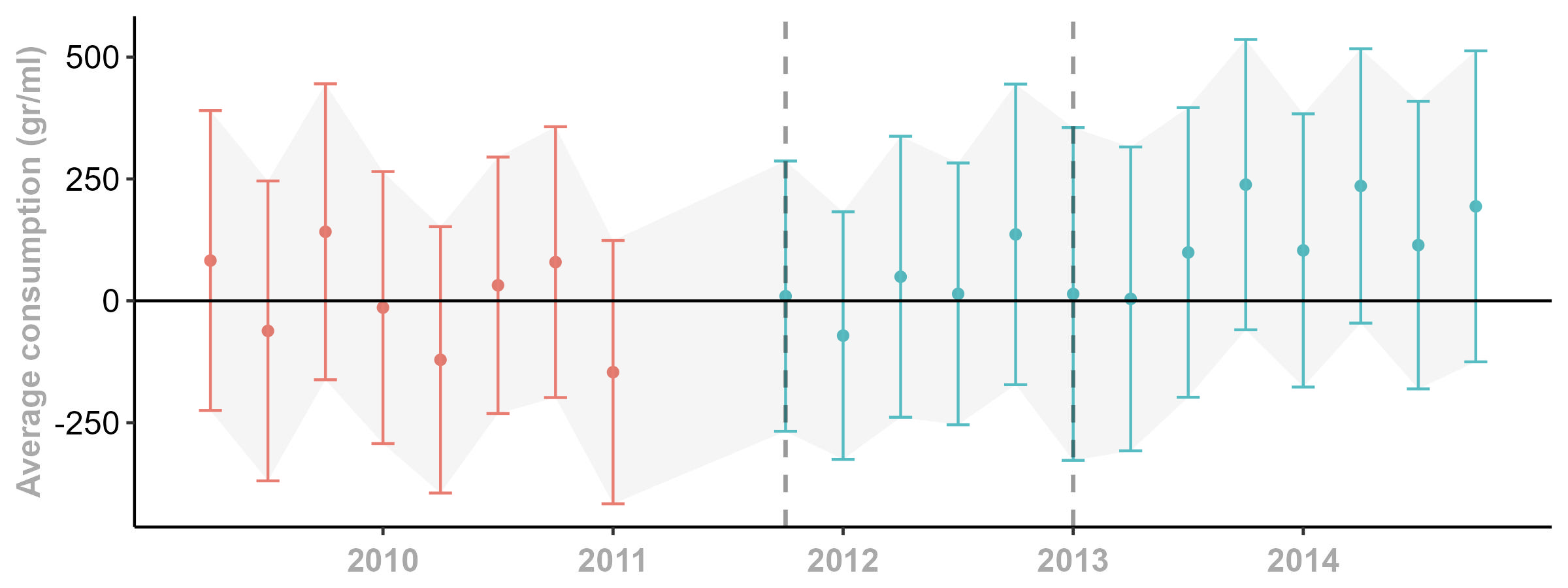}
        \caption{Average quarterly weight consumed (gr/ml) per household.}
        \label{fig:evolution_weight_cream}
    \end{subfigure}
    
    \begin{subfigure}{0.75\textwidth}
         \includegraphics[width=\textwidth]{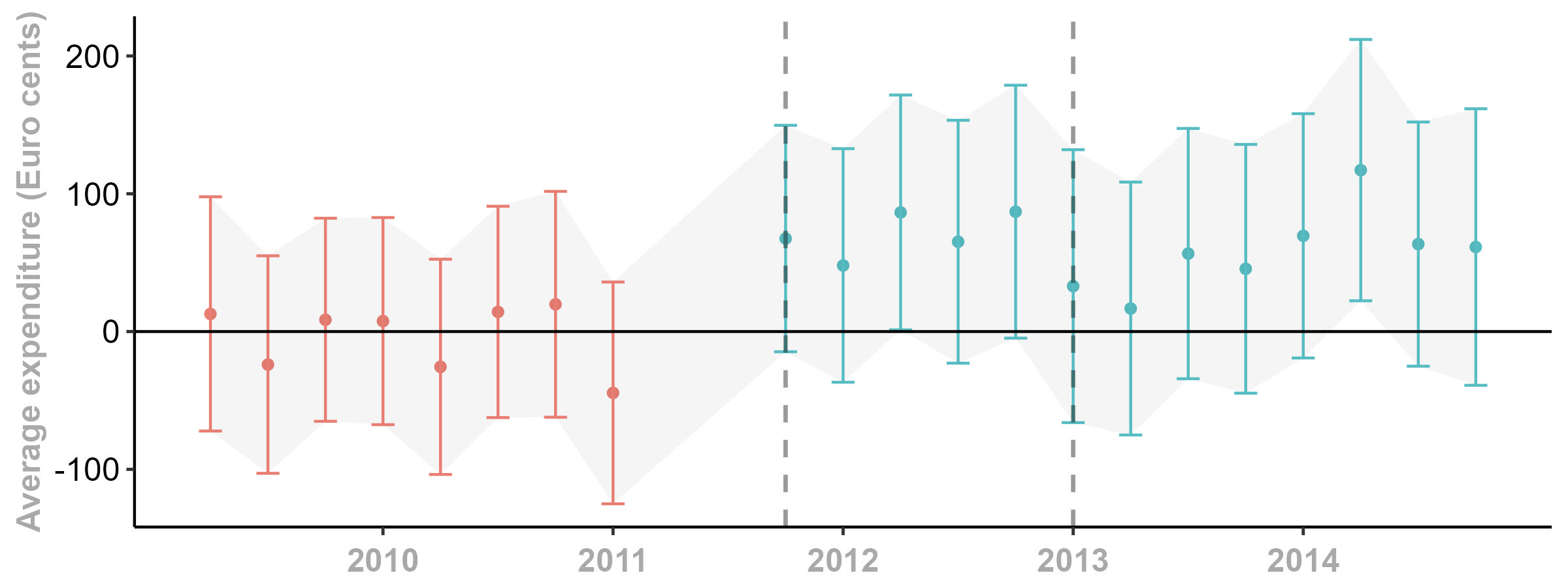}
        \caption{Average quarterly expenditure in Euro cents per household.}
        \label{fig:evolution_expenditure_cream}
    \end{subfigure}

    \begin{subfigure}{0.75\textwidth}
        \includegraphics[width=\textwidth]{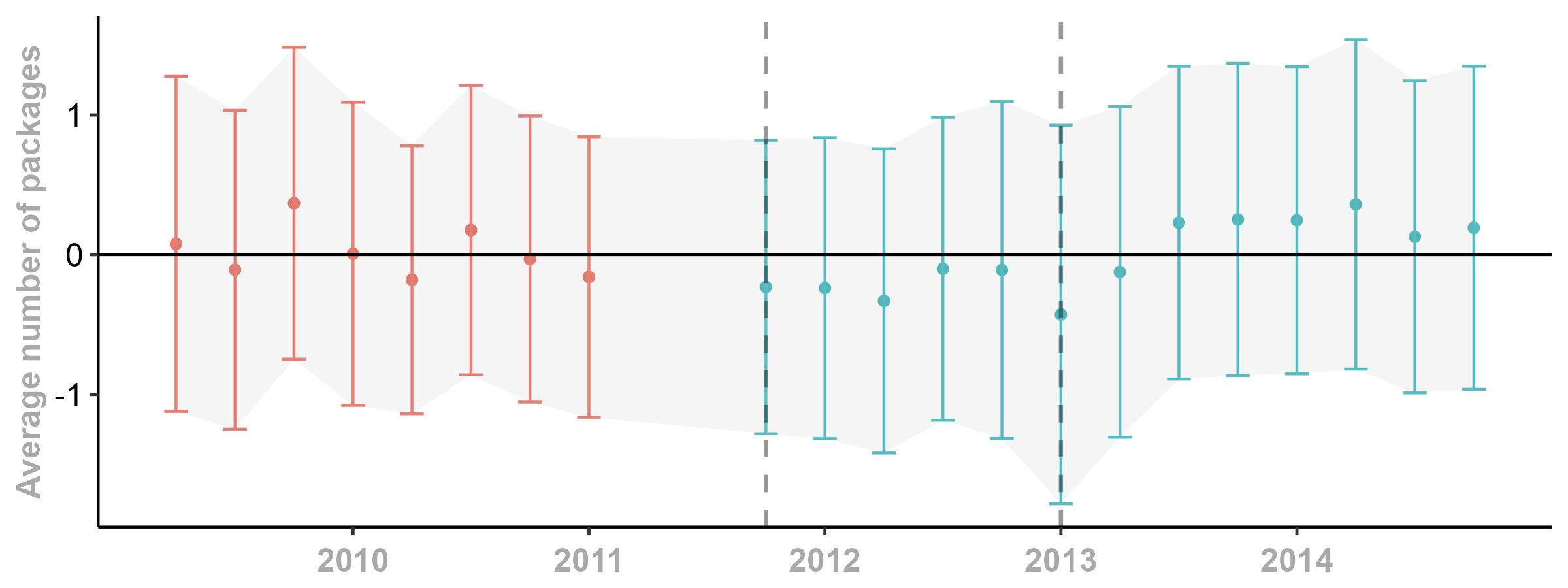}
        \caption{Average quarterly number of packages per household.}
        \label{fig:evolution_amount_cream}
    \end{subfigure}
    
    \caption{Time average treatment effects, $ATT(q)$. The two vertical dashed lines indicate the introduction and abolishment of the Danish fat tax, respectively. Red estimates and simultaneous 95\% confidence intervals for pre-tax periods, and blue estimates and simultaneous 95\% confidence intervals for tax and post-tax periods, clustered at the household level. Under conditional parallel trends, red estimates should be close to zero. Estimates are based on the doubly robust Equation (\ref{eq:dr_did}). Point estimates and standard errors are reported in Column (4) of Tables \ref{tab:att_t_weight} to \ref{tab:att_t_amount}.}
    \label{fig:main_graphs_cream}
\end{figure}

\clearpage

\subsection{Liver sausage}

\FloatBarrier

\begin{figure}[!htbp]
    \centering
    
    \begin{subfigure}{0.75\textwidth}
         \includegraphics[width=\textwidth]{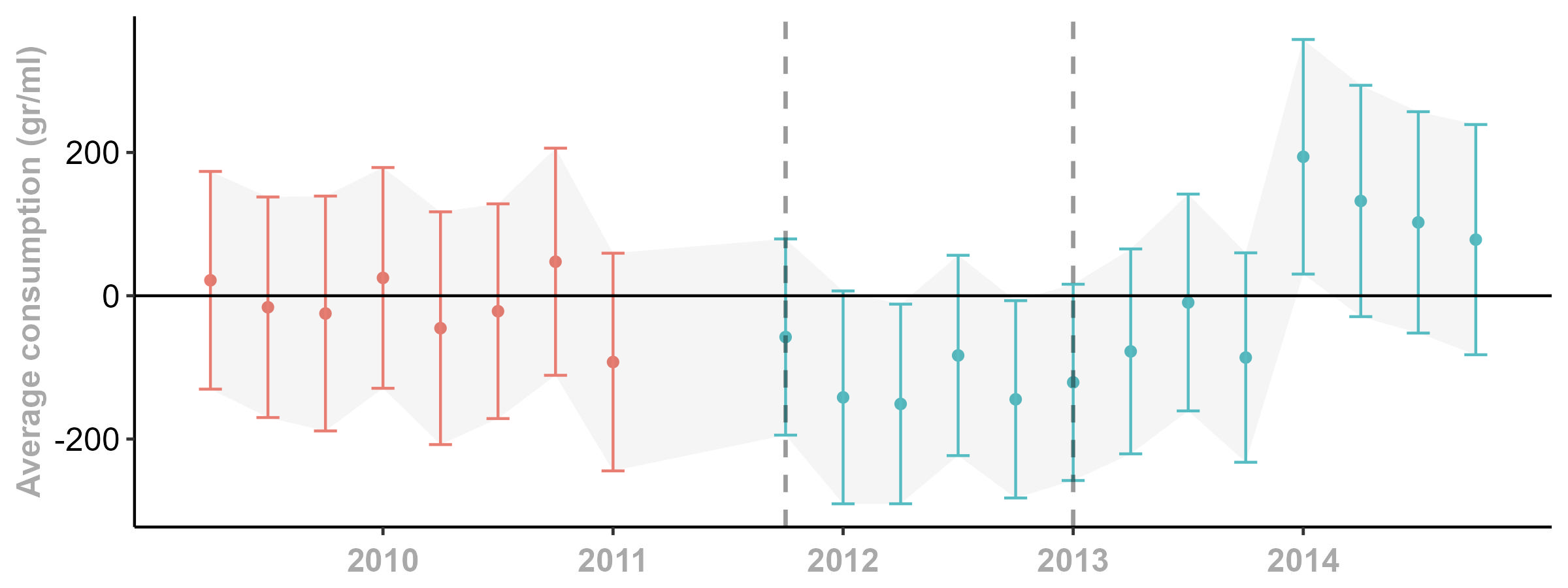}
        \caption{Average quarterly weight consumed (gr/ml) per household.}
        \label{fig:evolution_weight_liver}
    \end{subfigure}
    
    \begin{subfigure}{0.75\textwidth}
        \includegraphics[width=\textwidth]{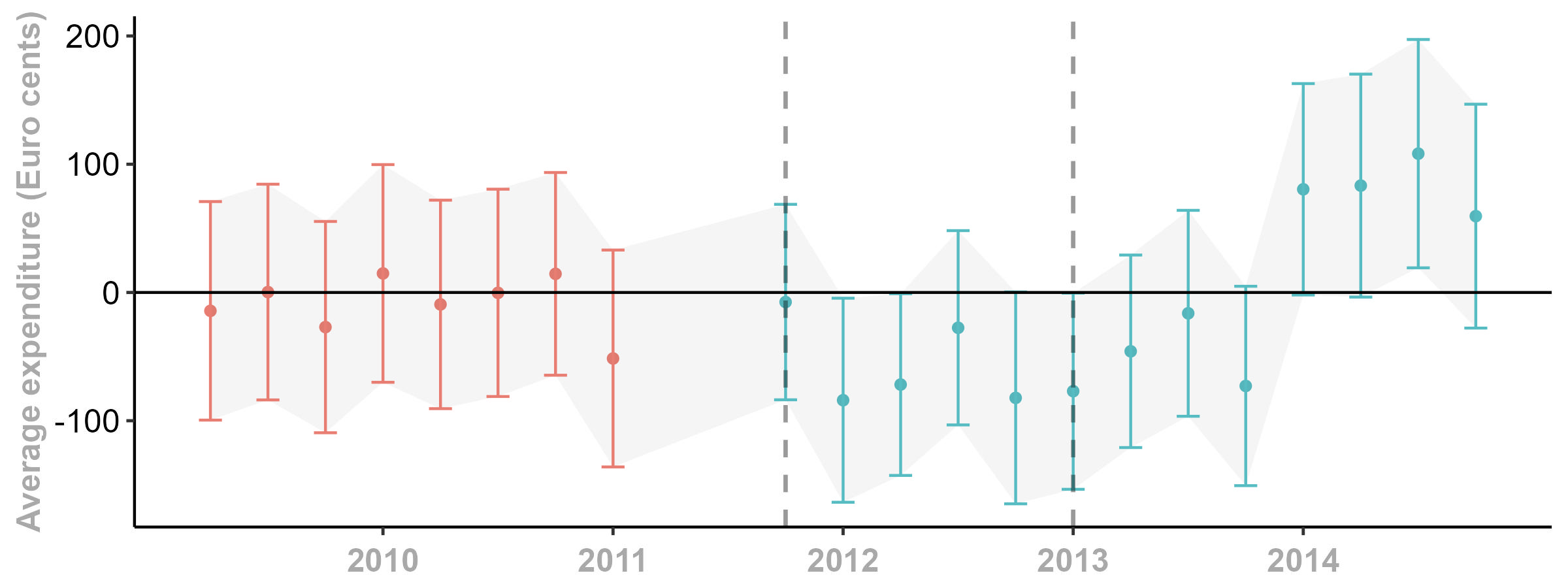}
        \caption{Average quarterly expenditure in Euro cents per household.}
        \label{fig:evolution_expenditure_liver}
    \end{subfigure}

    \begin{subfigure}{0.75\textwidth}
         \includegraphics[width=\textwidth]{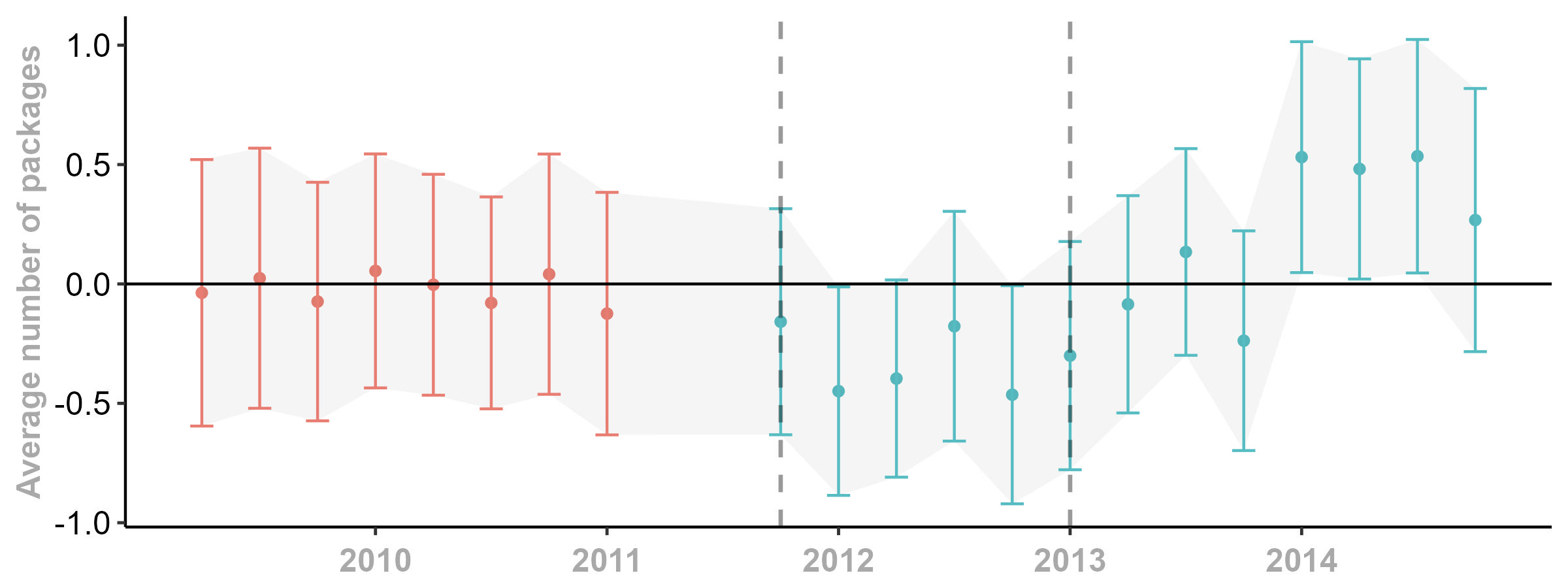}
        \caption{Average quarterly number of packages per household.}
        \label{fig:evolution_amount_liver}
    \end{subfigure}
    
    \caption{Time average treatment effects, $ATT(q)$. The two vertical dashed lines indicate the introduction and abolishment of the Danish fat tax, respectively. Red estimates and simultaneous 95\% confidence intervals for pre-tax periods, and blue estimates and simultaneous 95\% confidence intervals for tax and post-tax periods, clustered at the household level. Under conditional parallel trends, red estimates should be close to zero. Estimates are based on the doubly robust Equation (\ref{eq:dr_did}). Point estimates and standard errors are reported in Column (5) of Tables \ref{tab:att_t_weight} to \ref{tab:att_t_amount}.}
    \label{fig:main_graphs_liver}
\end{figure}

\clearpage

\subsection{Margarine}

\FloatBarrier

\begin{figure}[!htbp]
    \centering
    
    \begin{subfigure}{0.75\textwidth}
        \includegraphics[width=\textwidth]{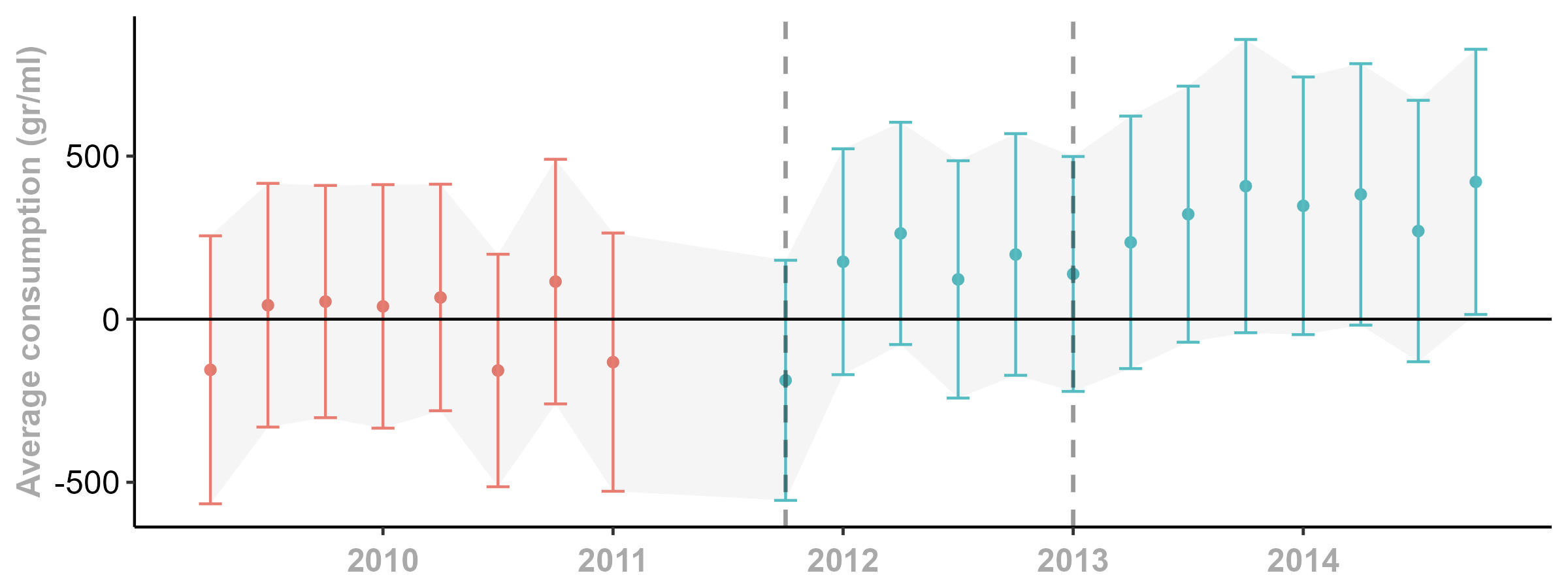}
        \caption{Average quarterly weight consumed (gr/ml) per household.}
        \label{fig:evolution_weight_margarine}
    \end{subfigure}

    \begin{subfigure}{0.75\textwidth}
         \includegraphics[width=\textwidth]{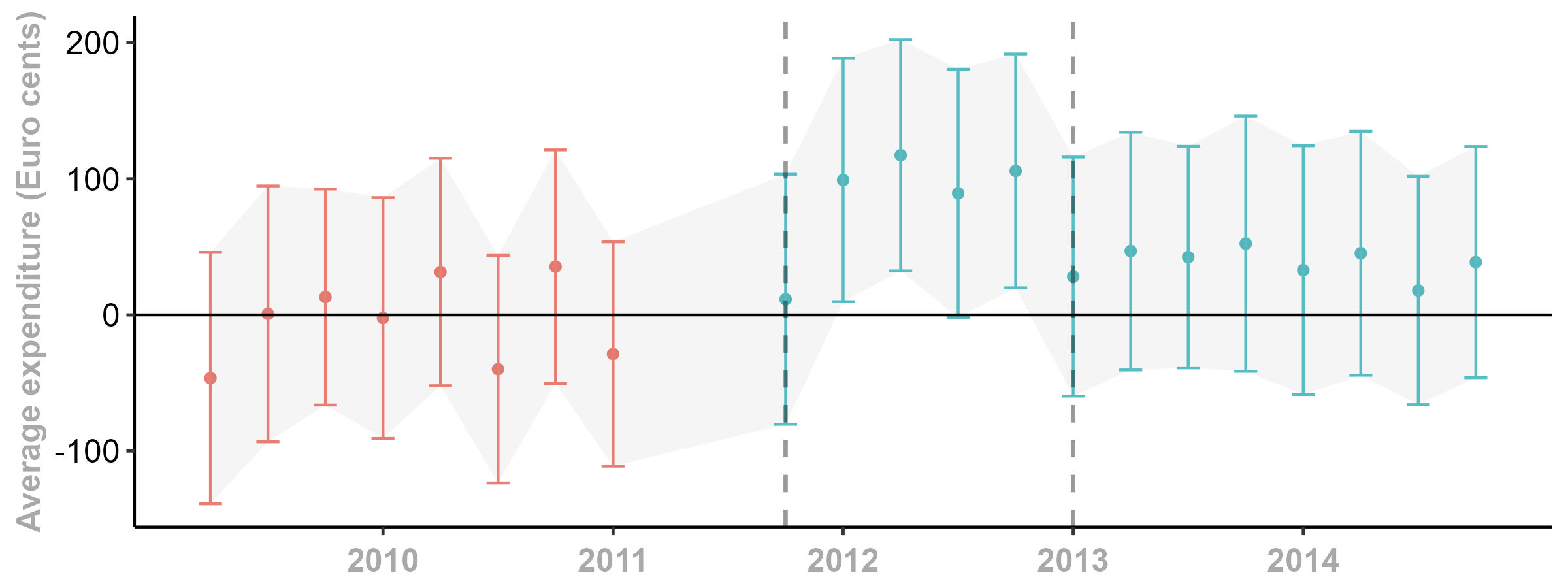}
        \caption{Average quarterly expenditure in Euro cents per household.}
        \label{fig:evolution_expenditure_margarine}
    \end{subfigure}

    \begin{subfigure}{0.75\textwidth}
        \includegraphics[width=\textwidth]{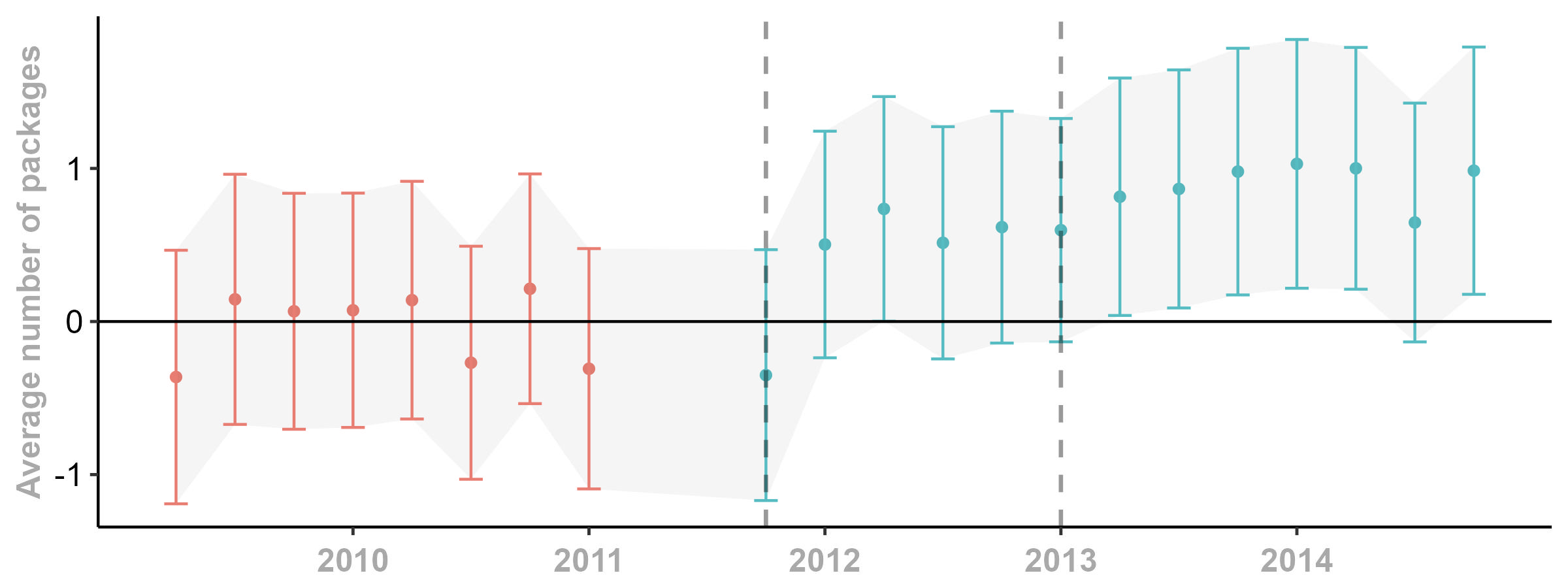}
        \caption{Average quarterly number of packages per household.}
        \label{fig:evolution_amount_margarine}
    \end{subfigure}
    
    \caption{Time average treatment effects, $ATT(q)$. The two vertical dashed lines indicate the introduction and abolishment of the Danish fat tax, respectively. Red estimates and simultaneous 95\% confidence intervals for pre-tax periods, and blue estimates and simultaneous 95\% confidence intervals for tax and post-tax periods, clustered at the household level. Under conditional parallel trends, red estimates should be close to zero. Estimates are based on the doubly robust Equation (\ref{eq:dr_did}). Point estimates and standard errors are reported in Column (6) of Tables \ref{tab:att_t_weight} to \ref{tab:att_t_amount}.}
    \label{fig:main_graphs_margarine}
\end{figure}

\clearpage

\subsection{Salami}

\FloatBarrier

\begin{figure}[!htbp]
    \centering
    
    \begin{subfigure}{0.75\textwidth}
        \includegraphics[width=\textwidth]{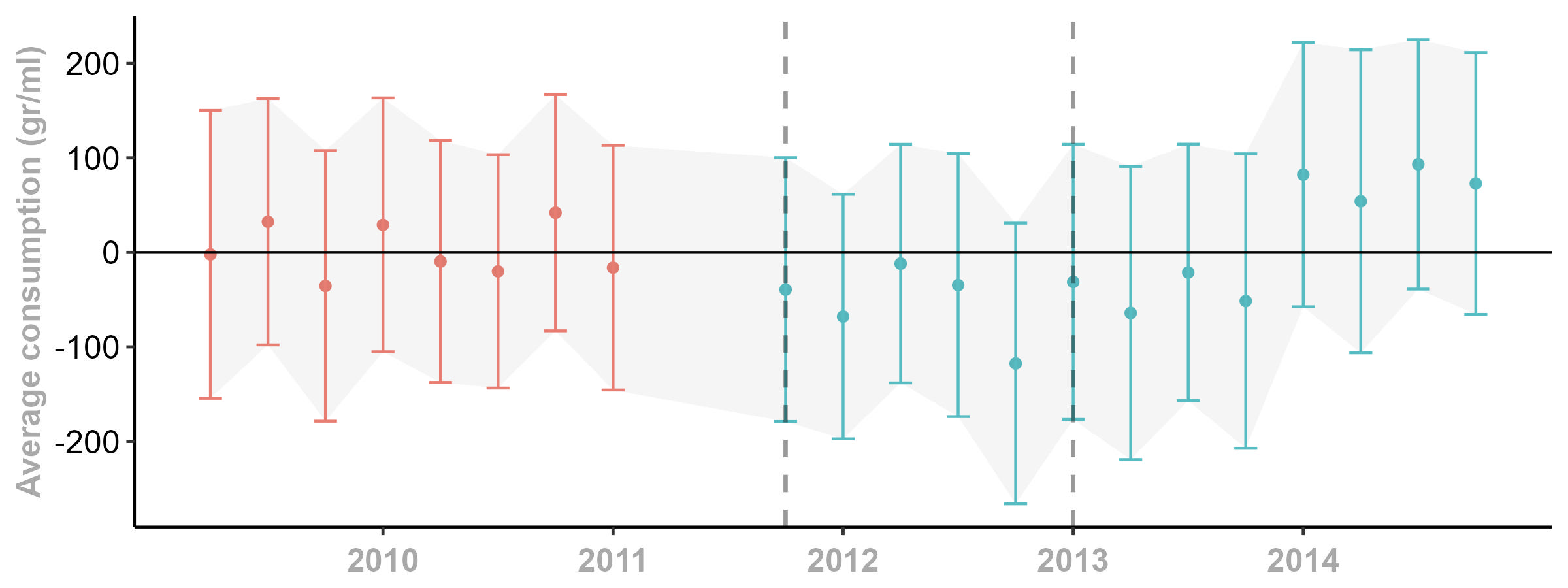}
        \caption{Average quarterly weight consumed (gr/ml) per household.}
        \label{fig:evolution_weight_salami}
    \end{subfigure}
    
    \begin{subfigure}{0.75\textwidth}
        \includegraphics[width=\textwidth]{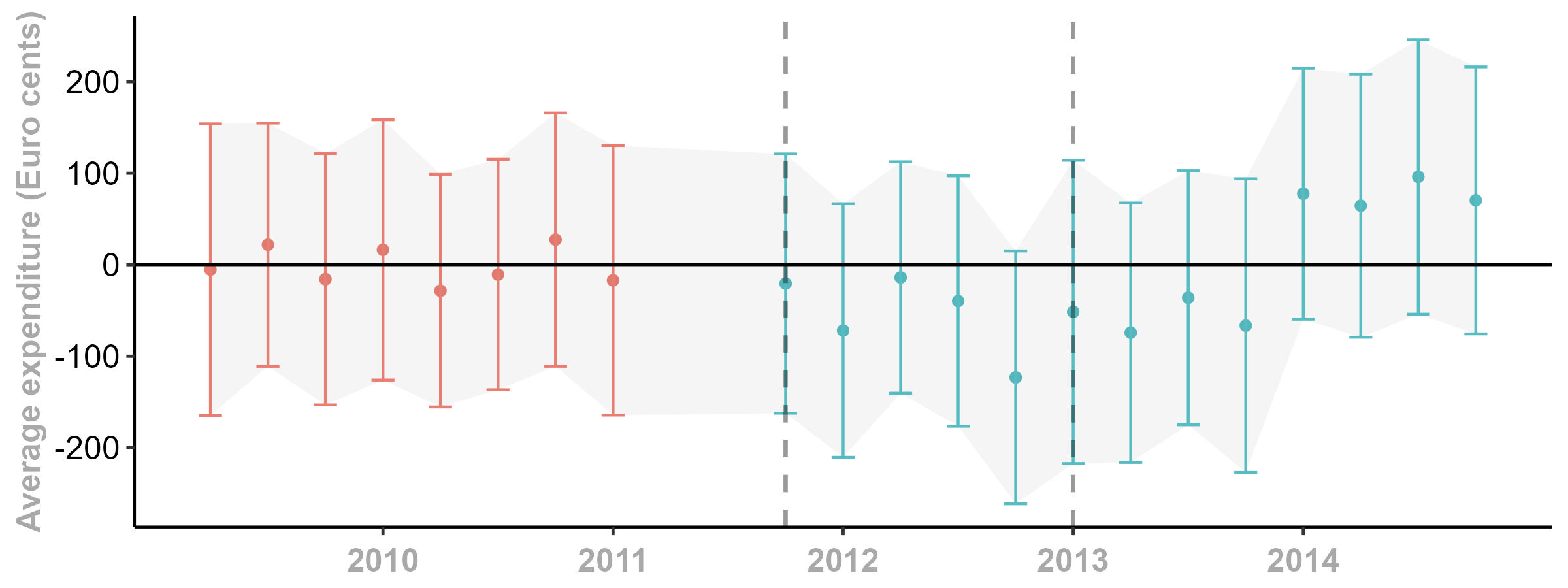}
        \caption{Average quarterly expenditure in Euro cents per household.}
        \label{fig:evolution_expenditure_salami}
    \end{subfigure}

    \begin{subfigure}{0.75\textwidth}
        \includegraphics[width=\textwidth]{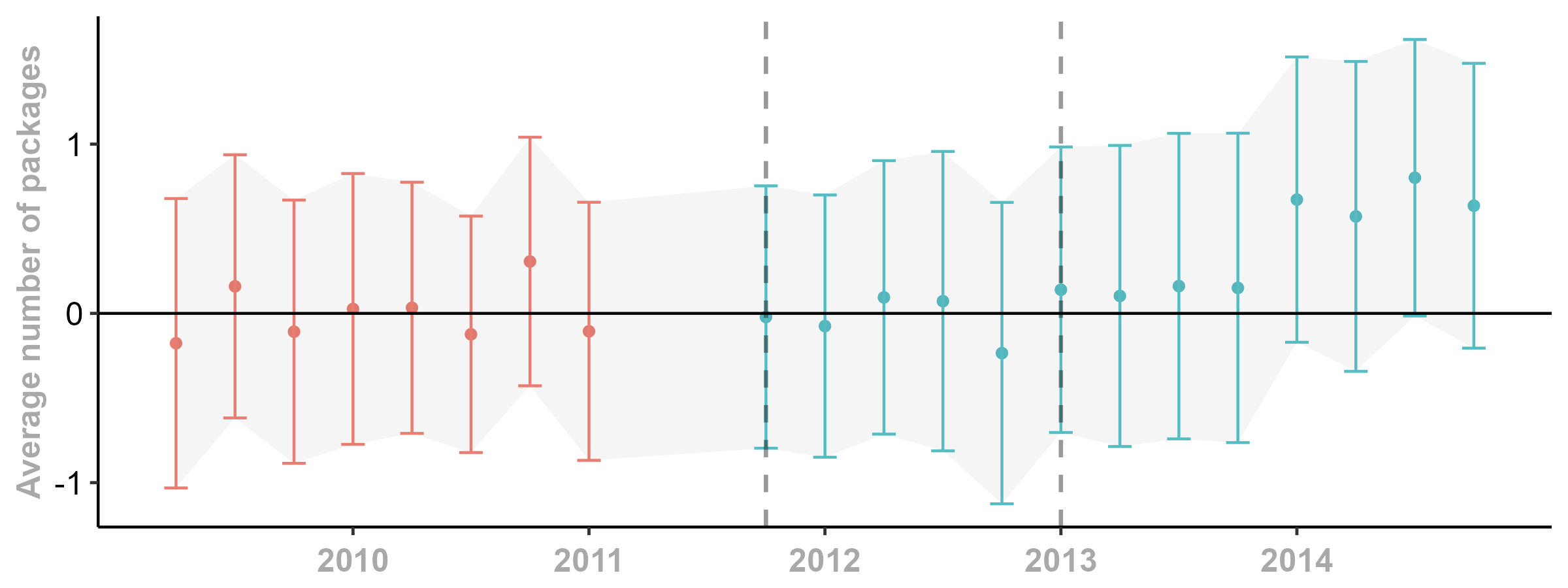}
        \caption{Average quarterly number of packages per household.}
        \label{fig:evolution_amount_salami}
    \end{subfigure}
    
    \caption{Time average treatment effects, $ATT(q)$. The two vertical dashed lines indicate the introduction and abolishment of the Danish fat tax, respectively. Red estimates and simultaneous 95\% confidence intervals for pre-tax periods, and blue estimates and simultaneous 95\% confidence intervals for tax and post-tax periods, clustered at the household level. Under conditional parallel trends, red estimates should be close to zero. Estimates are based on the doubly robust Equation (\ref{eq:dr_did}). Point estimates and standard errors are reported in Column (7) of Tables \ref{tab:att_t_weight} to \ref{tab:att_t_amount}.}
    \label{fig:main_graphs_salami}
\end{figure}

\clearpage

\subsection{Sour cream}

\FloatBarrier

\begin{figure}[!htbp]
    \centering
    
    \begin{subfigure}{0.75\textwidth}
        \includegraphics[width=\textwidth]{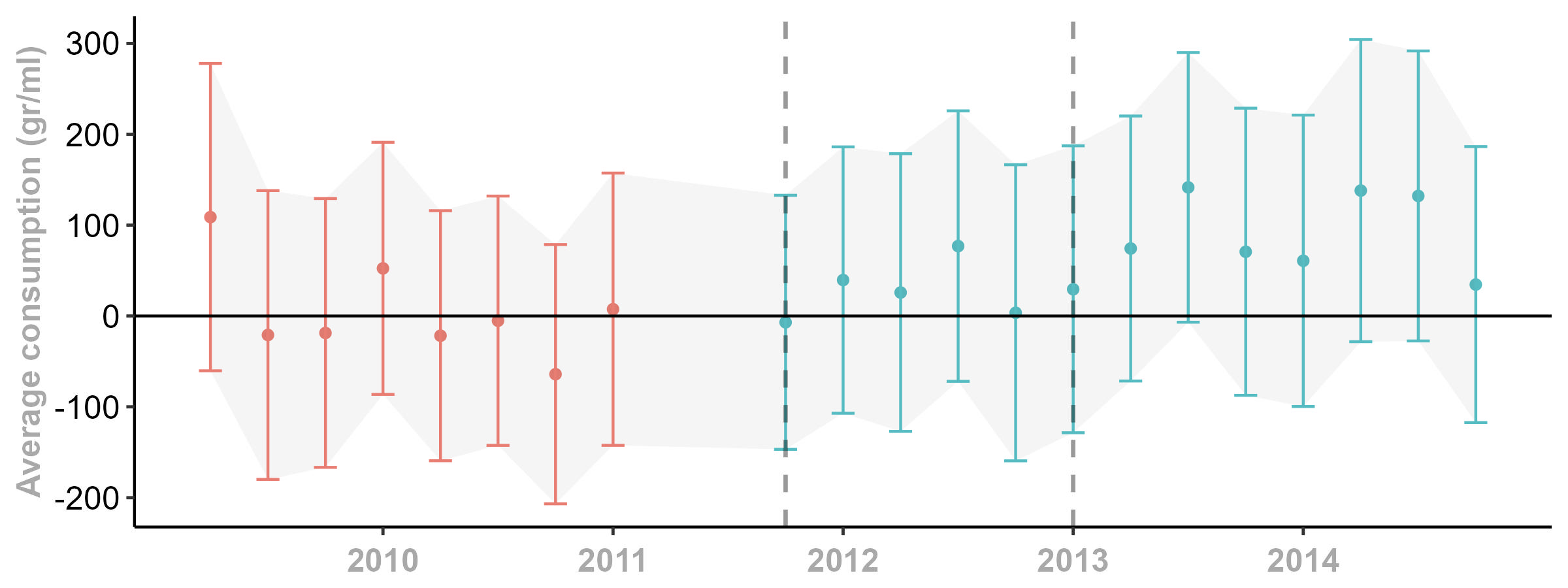}
        \caption{Average quarterly weight consumed (gr/ml) per household.}
        \label{fig:evolution_weight_sour}
    \end{subfigure}
    
    \begin{subfigure}{0.75\textwidth}
        \includegraphics[width=\textwidth]{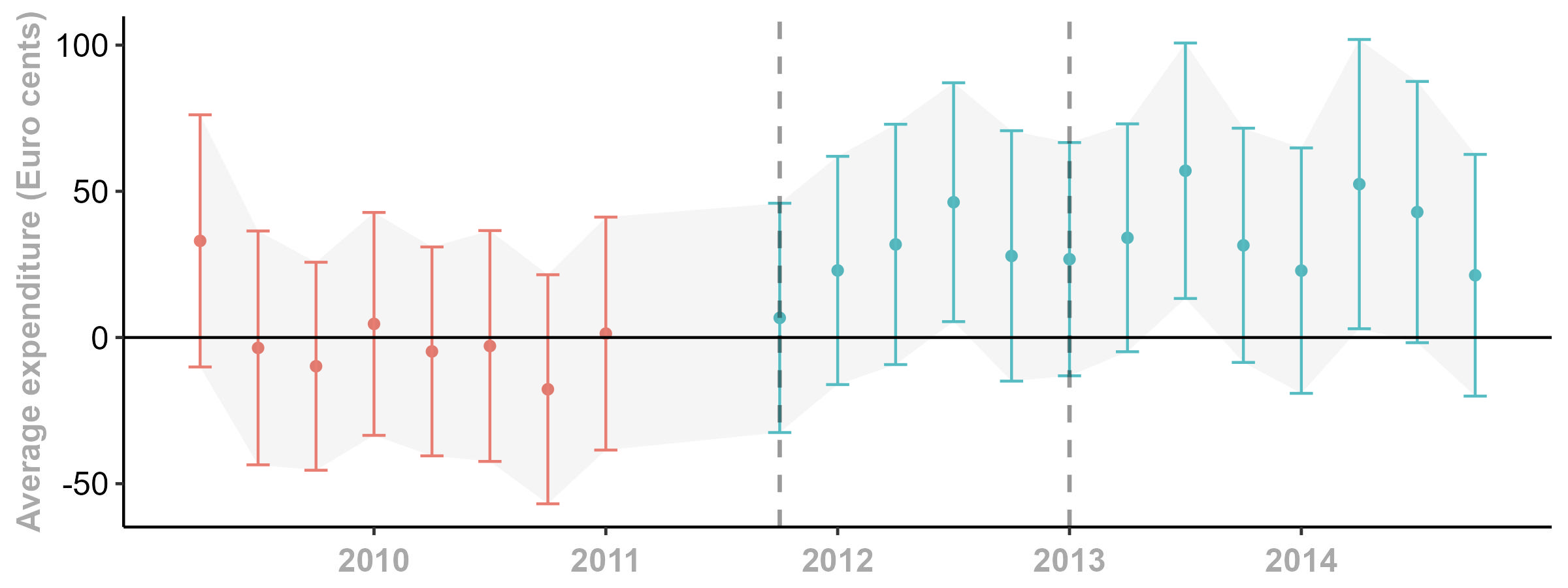}

        \caption{Average quarterly expenditure in Euro cents per household.}
        \label{fig:evolution_expenditure_sour}
    \end{subfigure}
    
    \begin{subfigure}{0.75\textwidth}
        \includegraphics[width=\textwidth]{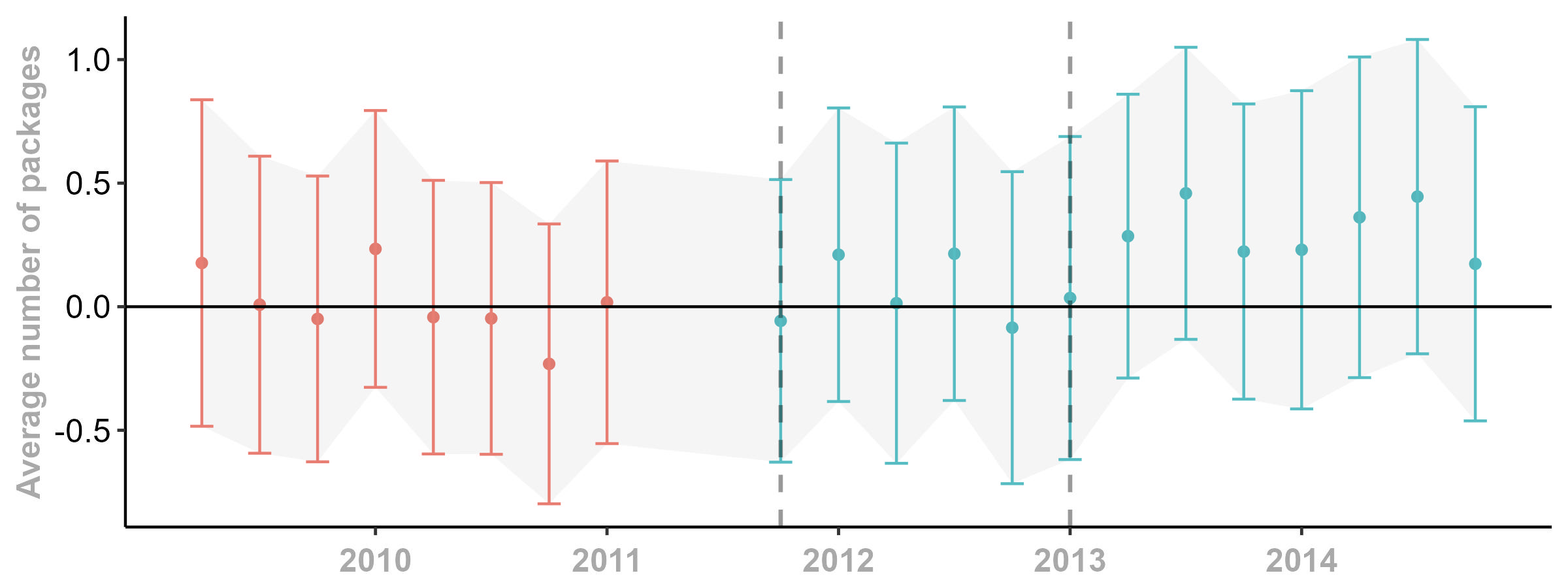}
        \caption{Average quarterly number of packages per household.}
        \label{fig:evolution_amount_sour}
    \end{subfigure}
    
    \caption{Time average treatment effects, $ATT(q)$. The two vertical dashed lines indicate the introduction and abolishment of the Danish fat tax, respectively. Red estimates and simultaneous 95\% confidence intervals for pre-tax periods, and blue estimates and simultaneous 95\% confidence intervals for tax and post-tax periods, clustered at the household level. Under conditional parallel trends, red estimates should be close to zero. Estimates are based on the doubly robust Equation (\ref{eq:dr_did}). Point estimates and standard errors are reported in Column (8) of Tables \ref{tab:att_t_weight} to \ref{tab:att_t_amount}.}
    \label{fig:main_graphs_sour}
\end{figure}

\clearpage

    \newpage
    
\normalsize

\begin{table}[ht]
\centering
\caption{Dynamic treatment effects, $ATT(q)$, for quarterly average weight consumed in gr/ml.}
\label{tab:att_t_weight}
\begin{adjustbox}{max width=0.95\textwidth}
\footnotesize
\begin{tabular}{A *{8}{D}}
  \hline \hline
  & (1) & (2) & (3) & (4) & (5) & (6) & (7) & (8) \\
 & Bacon & Butter & Cheese & Cream & Liver sausage & Margarine & Salami & Sour cream \\ 
  \hline
    $ATT_{2009\textls{-}Q2}$ & -69.13 & -62.88 & 46.73 & 82.69 & 18.71 & -155.22 & -40.79 & 108.83 \\ [-0.35em] 
  (se) & (49.88) & (108.28) & (176.68) & (99.88) & (65.42) & (132.20) & (59.06) & (54.68) \\ 
  $ATT_{2009\textls{-}Q3}$ & 54.29 & 84.87 & 183.78 & -61.67 & 30.27 & 42.89 & 44.76 & -20.94 \\ [-0.35em] 
  (se) & (52.08) & (107.77) & (167.77) & (98.53) & (64.20) & (130.56) & (48.70) & (53.75) \\ 
  $ATT_{2009\textls{-}Q4}$ & -22.12 & -26.10 & -205.94 & 141.64 & -45.28 & 54.04 & -39.73 & -18.69 \\ [-0.35em] 
  (se) & (51.48) & (95.67) & (174.33) & (98.05) & (66.10) & (127.49) & (48.99) & (47.99) \\ 
  $ATT_{2010\textls{-}Q1}$  & 60.76 & -16.14 & 207.32 & -13.82 & 19.83 & 39.28 & 25.31 & 52.41 \\ [-0.35em] 
  (se)  & (55.35) & (102.61) & (163.09) & (91.72) & (65.27) & (120.48) & (48.77) & (46.31) \\ 
  $ATT_{2010\textls{-}Q2}$ & -51.84 & -39.90 & -53.03 & -120.96 & -38.56 & 66.58 & -11.71 & -21.71 \\ [-0.35em] 
  (se) & (49.27) & (93.27) & (159.11) & (87.73) & (62.45) & (125.17) & (45.25) & (43.43) \\
  $ATT_{2010\textls{-}Q3}$ & -79.11 & 75.50 & -183.07 & 31.91 & -2.71 & -157.30 & -32.46 & -5.21 \\ [-0.35em] 
  (se) & (47.49) & (98.59) & (162.64) & (93.02) & (59.56) & (119.92) & (50.48) & (44.93) \\ 
  $ATT_{2010\textls{-}Q4}$ & 98.44 & -91.26 & -102.94 & 79.42 & 45.10 & 115.30 & 50.27 & -64.08 \\ [-0.35em] 
  (se) & (51.84) & (109.75) & (159.96) & (98.80) & (58.94) & (129.49) & (51.06) & (45.00) \\ 
  $ATT_{2011\textls{-}Q1}$ & -51.77 & -154.55 & 53.94 & -146.26 & -93.69 & -131.65 & -23.13 & 7.46 \\ [-0.35em] 
  (se)& (49.12) & (111.97) & (164.07) & (89.22) & (65.03) & (141.44) & (47.87) & (52.78) \\ 
  $ATT_{2011\textls{-}Q4}$ & -38.20 & $-268.51^{*}$ & -166.35 & 9.60 & -69.14 & -187.42 & -50.96 & -6.96 \\ [-0.35em] 
  (se) & (47.50) & (104.34) & (168.12) & (87.73) & (62.90) & (133.87) & (51.35) & (48.24) \\ 
  $ATT_{2012\textls{-}Q1}$ & -37.38 & 173.17  & -374.49 & -71.29 & -118.45 & 176.18 & -78.86 & 39.55 \\[-0.35em] 
  (se) & (43.02) & (99.00) & (163.40) & (87.94) & (62.85) & (119.92) & (51.04) & (47.33) \\ 
  $ATT_{2012\textls{-}Q2}$ & -10.37 & 183.92 & -305.58  & 49.40 & $-151.23^{*}$ & 263.02 & -10.87 & 25.81 \\ [-0.35em] 
  (se) & (42.52) & (104.80) & (170.58) & (98.59) & (59.75) & (114.91) & (48.00) & (51.86) \\ 
  $ATT_{2012\textls{-}Q3}$ & -67.53 & 84.42 & $-432.40^{*}$ & 14.28 & -72.74 & 121.99 & -61.36 & 76.91 \\[-0.35em] 
  (se) & (47.41) & (97.27) & (151.62) & (93.62) & (60.68) & (126.53) & (48.25) & (53.78) \\  
  $ATT_{2012\textls{-}Q4}$ & $-130.98^{*}$ & 193.46 & $-476.73^{*}$ & 136.32 & $-193.07^{*}$ & 198.39 & -131.82 & 3.52 \\ [-0.35em] 
  (se) & (50.01) & (101.65) & (161.99) & (105.67) & (64.65) & (125.44) & (53.77) & (56.03) \\
  $ATT_{2013\textls{-}Q1}$ & -35.86 & 203.63 & -190.38 & 14.09 & -117.06 & 138.68 & -66.08 & 29.40 \\[-0.35em] 
  (se) & (46.82) & (113.12) & (172.27) & (113.39) & (61.08) & (127.11) & (52.69) & (47.13) \\
  $ATT_{2013\textls{-}Q2}$ & -58.44 & 185.93 & -252.18 & 4.00 & -74.42 & 235.68 & -74.10 & 4.28 \\ [-0.35em] 
  (se) & (44.69) & (102.01) & (175.48) & (109.42) & (64.31) & (128.40) & (55.48) & (49.50) \\
  $ATT_{2013\textls{-}Q3}$ & -68.15 & 83.94 & -240.44 & 99.26 & -11.90 & $321.98^{*}$ & -32.42 & $141.56^{*}$ \\[-0.35em] 
  (se) & (47.15) & (107.25) & (166.99) & (109.50) & (65.00) & (122.13) & (47.74) & (53.04) \\ 
  $ATT_{2013\textls{-}Q4}$ & -4.98 & $334.66^{**}$ & -84.59 & 238.34 & -107.59 & $408.00^{**}$ & -69.05 & 70.68 \\  [-0.35em] 
  (se) & (53.13) & (122.59) & (153.75) & (108.01) & (66.53) & (132.07) & (53.49) & (52.88) \\ 
  $ATT_{2014\textls{-}Q1}$ & 35.33 & 171.35 & 409.49 & 103.39 & $199.76^{*}$ & $347.68^{*}$ & 66.88 & 60.76 \\[-0.35em] 
  (se) & (50.44) & (115.20) & (177.38) & (92.15) & (68.70) & (134.56) & (54.76) & (50.25) \\
  $ATT_{2014\textls{-}Q2}$ & 66.08 & $373.91^{**}$ & $532.98^{**}$ & 235.67 & $177.27^{*}$ & $382.61^{*}$ & 71.63 & 138.02 \\ [-0.35em] 
  (se) & (49.66) & (107.07) & (175.17) & (100.61) & (69.07) & (126.95v & (56.85) & (57.16) \\ 
  $ATT_{2014\textls{-}Q3}$ & 55.80 & 239.85 & $571.49^{**}$ & 114.29 & 91.23 & 270.31 & 108.76 & 132.13 \\[-0.35em] 
  (se) & (49.77) & (106.39) & (157.07) & (97.36) & (66.37) & (136.47) & (47.13) & (54.38) \\ 
  $ATT_{2014\textls{-}Q4}$ & 94.84 & $406.58^{**}$ & 310.35& 193.80 & 90.84 & $420.97^{**}$ & 61.84 & 34.57 \\[-0.35em] 
  (se) & (54.20) & (115.83) & (185.60) & (100.97) & (68.32) & (134.54) & (54.52) & (49.74) \\
   \hline \hline
\end{tabular}
\end{adjustbox}
\end{table}

\begin{table}[ht]
\centering
\caption{Dynamic treatment effects, $ATT(q)$, for quarterly average expenditure in Euro cents.}
\label{tab:att_t_expenditure}
\begin{adjustbox}{max width=0.95\textwidth}
\footnotesize
\begin{tabular}{A *{8}{D}}
  \hline \hline
  & (1) & (2) & (3) & (4) & (5) & (6) & (7) & (8) \\
 & Bacon & Butter & Cheese & Cream & Liver sausage & Margarine & Salami & Sour cream \\ 
  \hline
  $ATT_{2009\textls{-}Q2}$ & -31.04 & -21.88 & -30.48 & 12.80 & -1.18 & -46.41 & -33.34 & 33.05 \\ [-0.35em]
  (se) & (36.93) & (43.08) & (146.61) & (26.52) & (42.08) & (32.55) & (55.09) & (14.97) \\ 
  $ATT_{2009\textls{-}Q3}$ & 35.68 & -11.59 & 84.82 & -23.96 & 31.27 & 0.78 & 28.94 & -3.56 \\ [-0.35em]
  (se) & (36.72) & (40.63) & (131.02) & (25.57) & (38.95) & (30.84) & (49.14) & (13.45) \\ 
  $ATT_{2009\textls{-}Q4}$ & -9.44 & $-107.16^{*}$ & -120.58 & 8.55 & -44.16 & 13.18 & -24.89 & -9.83 \\ [-0.35em]
  (se) & (36.81) & (40.67) & (124.97) & (25.30) & (38.11) & (28.32) & (51.16) & (11.70) \\ 
  $ATT_{2010\textls{-}Q1}$ & 23.51 & -0.60 & 144.59 & 7.57 & 20.36 & -2.28 & 21.20 & 4.65 \\ [-0.35em]
  (se) & (35.88) & (45.53) & (122.55) & (27.11) & (43.61) & (28.83) & (50.98) & (12.32) \\ 
  $ATT_{2010\textls{-}Q2}$ & -32.24 & -36.30 & -59.73 & -25.60 & -9.85 & 31.54 & -32.31 & -4.75 \\ [-0.35em]
  (se) & (35.10) & (41.60) & (122.78) & (27.02) & (40.23) & (28.84) & (45.35) & (12.76) \\ 
  $ATT_{2010\textls{-}Q3}$ & -39.12 & 30.61 & -87.50 & 14.24 & 11.29 & -39.83 & -24.43 & -2.91 \\ [-0.35em]
  (se) & (31.80) & (43.56) & (134.71) & (26.56) & (38.32) & (27.40) & (47.54) & (13.14) \\ 
  $ATT_{2010\textls{-}Q4}$ & 48.00 & 22.71 & -121.89 & 19.77 & 16.34 & 35.52 & 38.51 & -17.71 \\ [-0.35em]
  (se) & (37.24) & (49.20) & (135.13) & (25.68) & (40.65) & (27.97) & (51.33) & (12.49) \\ 
  $ATT_{2011\textls{-}Q1}$ & -23.38 & $-142.21^{*}$ & 1.83 & -44.53 & -49.93 & -28.71 & -33.84 & 1.34 \\ [-0.35em]
  (se) & (35.96) & (51.37) & (142.75) & (28.18) & (43.26) & (31.01) & (48.12) & (13.06) \\ 
  $ATT_{2011\textls{-}Q4}$ & -25.36 & 5.41 & -83.51 & 67.51 & -25.99 & 11.52 & -16.28 & 6.71 \\ [-0.35em]
  (se) & (37.24) & (50.84) & (136.30) & (26.61) & (38.77) & (30.44) & (51.35) & (14.04) \\
  $ATT_{2012\textls{-}Q1}$ & -14.43 & $217.48^{***}$ & -264.10 & 47.97 & -87.25 & $99.13^{**}$ & -71.69 & 22.93 \\ [-0.35em]
  (se) & (31.82) & (48.47) & (128.00) & (28.60) & (39.82) & (28.47) & (47.25) & (13.76) \\ 
  $ATT_{2012\textls{-}Q2}$ & 12.74 & $324.86^{***}$ & -248.37 & $86.44^{*}$ & $-98.86^{*}$ & $117.37^{***}$ & -2.84 & 31.84 \\ [-0.35em]
  (se) & (29.84) & (50.69) & (130.35) & (29.09) & (38.18) & (28.61) & (46.73) & (13.39) \\ 
  $ATT_{2012\textls{-}Q3}$ & -38.24 & $288.38^{***}$ & -254.46 & 65.20 & -45.69 & $89.35^{**}$ & -42.75 & $46.27^{**}$ \\ [-0.35em]
  (se) & (34.55) & (49.14) & (127.70) & (25.62) & (40.62) & (29.98) & (50.34) & (13.97) \\ 
  $ATT_{2012\textls{-}Q4}$ & -60.10 & $338.27^{***}$ & $-350.39^{*}$ & $87.00^{*}$ & $-149.40^{**}$ & $105.84^{**}$ & -118.69 & 27.90 \\ [-0.35em]
  (se) & (37.80) & (56.32) & (136.13) & (30.84) & (43.71) & (28.86) & (51.54) & (14.39) \\ 
  $ATT_{2013\textls{-}Q1}$ & -32.24 & $163.85^{**}$ & -179.63 & 32.94 & $-101.29^{*}$ & 28.18 & -58.35 & 26.79 \\ [-0.35em]
  (se) & (35.70) & (53.27) & (142.15) & (29.79) & (38.01) & (29.36) & (57.03) & (13.87) \\ 
  $ATT_{2013\textls{-}Q2}$ & -27.56 & 122.16 & -219.92 & 16.71 & -79.78 & 46.91 & -57.49 & 34.10 \\ [-0.35em]
  (se) & (32.52) & (55.70) & (134.93) & (30.63) & (41.05) & (28.91) & (48.70) & (13.90) \\ 
  $ATT_{2013\textls{-}Q3}$ & -50.83 & 97.25 & -145.88 & 56.58 & -52.54 & 42.48 & -18.54 & $57.03^{***}$ \\ [-0.35em]
  (se) & (36.92) & (55.79) & (139.93) & (32.66) & (36.21) & (29.52) & (49.46) & (14.24) \\ 
  $ATT_{2013\textls{-}Q4}$ & -19.64 & 125.40 & -210.05& 45.55 & $-108.22^{*}$ & 52.37 & -61.48 & 31.53 \\ [-0.35em]
  (se) & (35.77) & (58.63) & (136.76) & (30.07) & (42.44) & (28.48) & (51.21) & (14.14) \\ 
  $ATT_{2014\textls{-}Q1}$ & 27.81 & 51.14 & 329.05 & 69.48 & 54.22 & 32.90 & 87.94 & 22.87 \\ [-0.35em]
  (se) & (37.11) & (57.51) & (144.13) & (29.96) & (42.00) & (25.59) & (52.09) & (15.57) \\ 
  $ATT_{2014\textls{-}Q2}$ & 52.16 & $247.94^{***}$ & $506.17^{**}$ & $117.15^{**}$ & 77.91 & 45.32 & 102.81 & $52.46^{**}$ \\ [-0.35em]
  (se) & (37.18) & (53.58) & (147.39) & (31.26) & (38.47) & (29.46) & (51.35) & (16.24) \\ 
  $ATT_{2014\textls{-}Q3}$ & 44.99 & $233.53^{***}$ & $480.43^{**}$ & 63.47 & 73.67 & 17.97 & $137.29^{*}$ & $42.90^{*}$ \\ [-0.35em]
  (se) & (34.80) & (52.20) & (139.03) & (30.35) & (42.48) & (29.35) & (50.23) & (15.62) \\ 
  $ATT_{2014\textls{-}Q4}$ & 63.37& $295.09^{***}$  & 297.82  & 61.34  & 34.85 & 38.80 & 84.19 & 21.29 \\ [-0.35em]
  (se) & (39.28) & (55.95) & (138.15) & (31.53) & (40.90) & (28.42) & (51.81) & (13.30) \\ 
   \hline \hline
\end{tabular}
\end{adjustbox}
\end{table}

\begin{table}[ht]
\centering
\caption{Dynamic treatment effects, $ATT(q)$, for quarterly average amount (in packages) consumed.}
\label{tab:att_t_amount}
\footnotesize
\begin{adjustbox}{max width=0.95\textwidth}
\begin{tabular}{A *{8}{D}}
  \hline \hline
  & (1) & (2) & (3) & (4) & (5) & (6) & (7) & (8) \\
 & Bacon & Butter & Cheese & Cream & Liver sausage & Margarine & Salami & Sour cream \\ 
  \hline
  $ATT_{2009\textls{-}Q2}$ & -0.21 & -0.28 & -0.03 & 0.08 & -0.02 & -0.36 & -0.33 & 0.18 \\ [-0.35em] 
  (se) & (0.26) & (0.46) & (0.52) & (0.40) & (0.28) & (0.27) & (0.31) & (0.21) \\ 
  $ATT_{2009\textls{-}Q3}$ & 0.15 & 0.37 & 0.66 & -0.11 & 0.29 & 0.14 & 0.21 & 0.01 \\ [-0.35em] 
  (se) & (0.23) & (0.40) & (0.51) & (0.38) & (0.27) & (0.27) & (0.29) & (0.19) \\ 
  $ATT_{2009\textls{-}Q4}$ & -0.11 & -0.12 & -0.30 & 0.37 & -0.23 & 0.07 & -0.08 & -0.05 \\ [-0.35em] 
  (se) & (0.25) & (0.38) & (0.48) & (0.35) & (0.27) & (0.24) & (0.27) & (0.19) \\ 
  $ATT_{2010\textls{-}Q1}$ & 0.16 & -0.03 & 0.79 & 0.01 & 0.05 & 0.07 & -0.02 & 0.23 \\ [-0.35em] 
  (se) & (0.27) & (0.41) & (0.46) & (0.35) & (0.27) & (0.26) & (0.28) & (0.18) \\ 
  $ATT_{2010\textls{-}Q2}$ & -0.11 & -0.09 & -0.08 & -0.18 & 0.05 & 0.14 & 0.01 & -0.04 \\ [-0.35em] 
  (se) & (0.25) & (0.36) & (0.48) & (0.35) & (0.26) & (0.26) & (0.28) & (0.18) \\ 
  $ATT_{2010\textls{-}Q3}$ & -0.06 & 0.33 & -0.39 & 0.18 & -0.04 & -0.27 & -0.12 & -0.05 \\ [-0.35em] 
  (se) & (0.22) & (0.36) & (0.44) & (0.34) & (0.27) & (0.25) & (0.24) & (0.19) \\ 
  $ATT_{2010\textls{-}Q4}$ & 0.13 & -0.38 & -0.19 & -0.03 & 0.13 & 0.21 & 0.23 & -0.23 \\ [-0.35em] 
  (se) & (0.20) & (0.41) & (0.47) & (0.33) & (0.29) & (0.26) & (0.26) & (0.19) \\ 
  $ATT_{2011\textls{-}Q1}$ & -0.11 & -0.54 & 0.25 & -0.16 & -0.10 & -0.31 & -0.11 & 0.02 \\ [-0.35em] 
  (se) & (0.23) & (0.42) & (0.51) & (0.35) & (0.29) & (0.28) & (0.27) & (0.18) \\ 
  $ATT_{2011\textls{-}Q4}$ & 0.07 & -1.06 & -0.56 & -0.23 & -0.33 & -0.35 & -0.02 & -0.06 \\ [-0.35em] 
  (se) & (0.24) & (0.44) & (0.58) & (0.32) & (0.27) & (0.28) & (0.32) & (0.19) \\ 
  $ATT_{2012\textls{-}Q1}$ & 0.22 & 0.61 & 0.10 & -0.24 & -0.40 & 0.50 & -0.12 & 0.21 \\ [-0.35em] 
  (se) & (0.19) & (0.39) & (0.50) & (0.33) & (0.28) & (0.26) & (0.30) & (0.19) \\ 
  $ATT_{2012\textls{-}Q2}$ & 0.07 & 0.70 & -0.07 & -0.33 & -0.50 & $0.74^{*}$ & 0.17 & 0.01 \\ [-0.35em] 
  (se) & (0.20) & (0.38) & (0.47) & (0.37) & (0.27) & (0.27) & (0.30) & (0.21) \\ 
  $ATT_{2012\textls{-}Q3}$ & -0.12 & 0.41 & -0.44 & -0.10 & -0.26 & 0.51 & 0.03 & 0.21 \\ [-0.35em] 
  (se) & (0.19) & (0.41) & (0.47) & (0.36) & (0.27) & (0.28) & (0.33) & (0.20) \\ 
  $ATT_{2012\textls{-}Q4}$ & -0.29 & $1.14^{*}$ & -0.55 & -0.11 & $-0.82^{*}$ & 0.62 & -0.10 & -0.08 \\ [-0.35em] 
  (se) & (0.21) & (0.42) & (0.50) & (0.37) & (0.30) & (0.26) & (0.30) & (0.23) \\ 
  $ATT_{2013\textls{-}Q1}$ & 0.02 & 1.06 & 0.79 & -0.43 & -0.30 & 0.60 & 0.15 & 0.04 \\ [-0.35em] 
  (se) & (0.22) & (0.48) & (0.92) & (0.44) & (0.29) & (0.26) & (0.30) & (0.20) \\ 
  $ATT_{2013\textls{-}Q2}$ & -0.10 & $1.14^{*}$ & -0.19 & -0.12 & -0.20 & $0.81^{**}$ & 0.22 & 0.29 \\ [-0.35em] 
  (se) & (0.20) & (0.39) & (0.49) & (0.46) & (0.29) & (0.26) & (0.31) & (0.19) \\ 
  $ATT_{2013\textls{-}Q3}$ & -0.02 & 0.67 & -0.20 & 0.23 & -0.09 & $0.87^{**}$ & 0.28 & 0.46 \\ [-0.35em] 
  (se) & (0.23) & (0.40) & (0.54) & (0.39) & (0.27) & (0.25) & (0.31) & (0.21) \\ 
  $ATT_{2013\textls{-}Q4}$ & 0.22 & $1.61^{**}$ & 0.18 & 0.25 & -0.44 & $0.98^{**}$ & 0.39 & 0.22 \\ [-0.35em] 
  (se) & (0.22) & (0.50) & (0.51) & (0.37) & (0.28) & (0.28) & (0.32) & (0.19) \\ 
  $ATT_{2014\textls{-}Q1}$ & 0.06 & 0.94 & 0.83 & 0.25 & 0.39 & $1.03^{**}$ & $0.83^{*}$ & 0.23 \\ [-0.35em] 
  (se) & (0.22) & (0.43) & (0.51) & (0.38) & (0.32) & (0.29) & (0.30) & (0.20) \\ 
  $ATT_{2014\textls{-}Q2}$ & 0.24 & $1.49^{**}$ & 1.05 & 0.36 & 0.63 & $1.00^{**}$ & 0.77 & 0.36 \\ [-0.35em] 
  (se) & (0.21) & (0.43) & (0.53) & (0.36) & (0.28) & (0.29) & (0.34) & (0.21) \\ 
  $ATT_{2014\textls{-}Q3}$ & $0.67^{**}$ & $1.46^{**}$ & 1.02 & 0.13 & 0.32 & 0.65 & $0.95^{**}$ & 0.44\\ [-0.35em] 
  (se) & (0.21) & (0.43) & (0.52) & (0.38) & (0.30) & (0.28) & (0.30) & (0.21) \\ 
  $ATT_{2014\textls{-}Q4}$ & 0.40 & $1.76^{**}$ & 0.86 & 0.19 & 0.16 & $0.98^{**}$ & 0.78 & 0.17 \\ [-0.35em] 
  (se) & (0.24) & (0.47) & (0.53) & (0.37) & (0.32) & (0.29) & (0.32) & (0.20) \\ 
   \hline \hline
\end{tabular}
\end{adjustbox}
\end{table}

    \newpage
    \section{Robustness checks}

\subsection{Results when excluding Southern-Danish households}
\label{app:robust_southern_dk}

\begin{table}[!htb]
\centering
\caption{The causal effect of the Danish fat tax on the average quarterly consumption and expenditure per household during the tax period (October 2011 to December 2012). Southern-Danish households located in South/West Jutland and Funen are excluded.}
\label{tab:robust_southern_dk}
\begin{adjustbox}{max width=\textwidth}
\begin{tabular}{B *{8}{C}}
  \hline \hline
  & (1) & (2) & (3) & (4) & (5) & (6) & (7) & (8) \\
 & Bacon & Butter & Cheese & Cream & Liver sausage & Margarine & Salami & Sour cream \\ 
  \hline
  \multicolumn{9}{l}{\textbf{Panel A: Quarterly total weight consumed in gr/ml}} \\ \hline
  \textit{Danish fat tax (ATT)} & -62.619 & 56.607 & $-411.433^{**}$ & 6.714 & $-108.259^{*}$ & 118.680 & -49.521 & 18.472 \\ 
  \textit{(se)} & (34.487) & (88.085) & (130.721) & (77.347) & (39.793) & (98.689) & (35.699) & (39.050) \\
  \textit{p-value} & 0.220 & 0.671 & 0.030 & 0.952 & 0.055 & 0.413 & 0.337 & 0.746\\
  \textit{\% change} & -11.34\% & 3.72\% & -15.78\% & 0.50\% & -10.23\% & 6.93\% & -13.64\% & 3.30\%  \\ 
  \textit{obs.} & 23,073 & 34,792 & 25,939 & 33,376 & 23,399 & 32,761 & 24,857 & 30,678 \\ 
  \textit{nr. of Danish hh.} & 1,690 & 1,892 & 1,824 & 1,700 & 1,813 & 1,647 & 1,706 & 1,520 \\ 
  \textit{nr. of North.-German hh.} & 480 & 1,616 & 698 & 1,596 & 428 & 1,614 & 658 & 1,409 \\ 
 \hline
  \multicolumn{9}{l}{\textbf{Panel B: Quarterly total expenditure in Euro cents}} \\ \hline
  \textit{Danish fat tax (ATT)} & -29.556 & $227.535^{***}$ & $-298.029^{*}$ & $63.645^{*}$ & $-48.729^{*}$ & $79.356^{**}$ & -45.785 & 22.480 \\  
  \textit{(se)} & (22.454) & (41.715) & (104.902) & (23.633) & (20.870) & (23.583) & (38.443) & (10.042) \\ 
  \textit{p-value} & 0.355 & 0.000 & 0.051 & 0.071 & 0.098 & 0.021 & 0.407 & 0.120\\
  \textit{\% change} & -8.22\% & 24.07\% & -13.07\% & 13.16\% & -8.44\% & 15.28\% & -10.44\% & 13.42\%  \\ 
  \textit{obs.} & 23,073 & 34,792 & 25,939 & 33,376 & 23,399 & 32,761 & 24,857 & 30,678 \\ 
  \textit{nr. of Danish hh.} & 1,690 & 1,892 & 1,824 & 1,700 & 1,813 & 1,647 & 1,706 & 1,520 \\ 
  \textit{nr. of North.-German hh.} & 480 & 1,616 & 698 & 1,596 & 428 & 1,614 & 658 & 1,409 \\ 
      \hline
  \multicolumn{8}{l}{\textbf{Panel C: Quarterly total amount (in packages) consumed}} \\ \hline
  \textit{Danish fat tax (ATT)} & -0.117 & 0.278 & -0.349 & -0.272 & $-0.321^{*}$ & 0.406 & -0.004 & 0.014 \\ 
  \textit{(se)} & (0.142) & (0.331) & (0.409) & (0.274) & (0.121) & (0.208) & (0.213) & (0.146) \\ 
  \textit{p-value} & 0.567 & 0.555 & 0.561 & 0.494 & 0.071 & 0.179 & 0.991 & 0.949\\
  \textit{\% change} & -6.17\% & 4.64\% & -8.45\% & -8.41\% & -10.23\% & 11.19\% & -0.20\% & 1.13\%  \\
   \textit{obs.} & 23,073 & 34,792 & 25,939 & 33,376 & 23,399 & 32,761 & 24,857 & 30,678 \\ 
  \textit{nr. of Danish hh.} & 1,690 & 1,892 & 1,824 & 1,700 & 1,813 & 1,647 & 1,706 & 1,520 \\ 
  \textit{nr. of North.-German hh.} & 480 & 1,616 & 698 & 1,596 & 428 & 1,614 & 658 & 1,409 \\ 
   \hline \hline
      \multicolumn{9}{p{1.3\textwidth}}{\normalsize Significance level: * $p<0.1$ ** $p<0.05$ *** $p<0.01$. Standard errors are clustered at the household level. \textit{\% change} compares the estimated tax average to the pre-tax sample mean for Danish households. The estimates are obtained using the doubly robust DID approach, controlling for age of head of household, number of children under 15, gender, ISCED, income level, and household size.}\\ 
\end{tabular}
\end{adjustbox}
\end{table}

    \newpage

\subsection{Results under unconditional parallel trends for the weight outcome}
\label{app:unconditional_pta_weight}

\FloatBarrier
\begin{figure}[!htbp]
\centering
\begin{minipage}{0.49\textwidth}
        \centering
        \includegraphics[width=\textwidth]{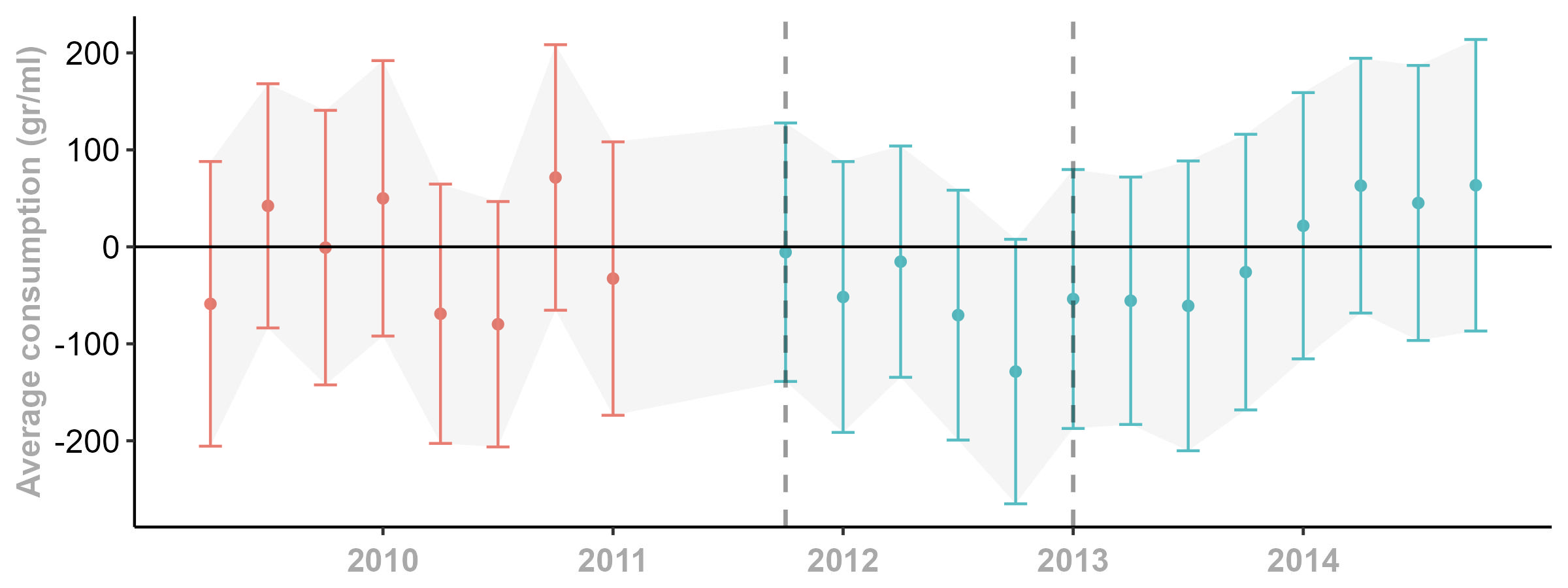}
        \caption{Bacon.}
    \end{minipage}
    \hfill
    \begin{minipage}{0.49\textwidth}
        \centering
        \includegraphics[width=\textwidth]{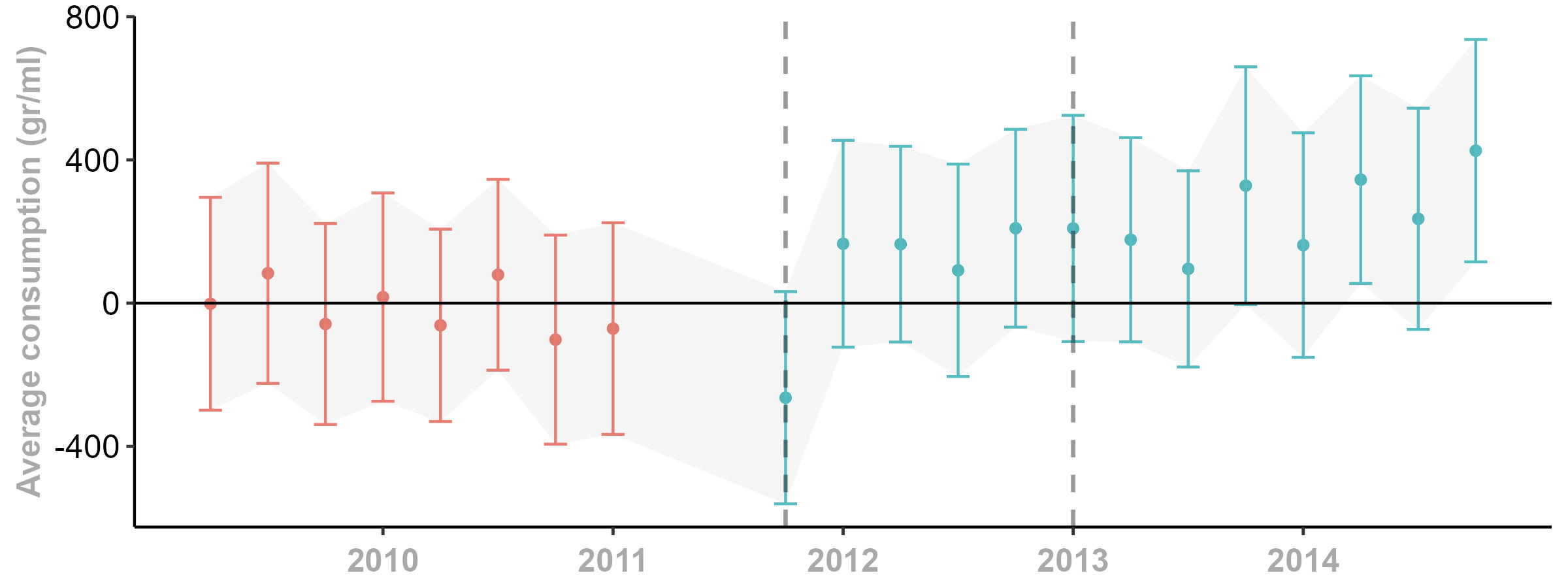}
        \caption{Butter.}
    \end{minipage}
    
    \begin{minipage}{0.45\textwidth}
        \centering
        \includegraphics[width=\textwidth]{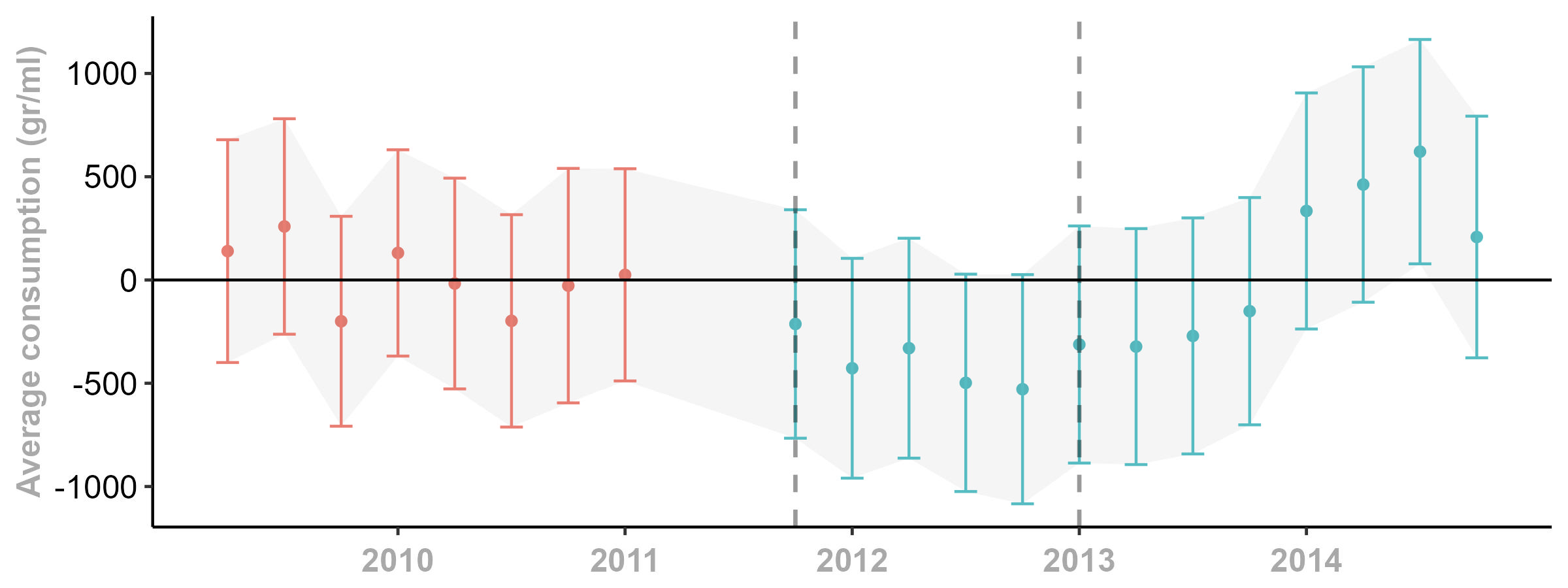}
        \caption{Cheese.}
    \end{minipage}
    \hfill
    \begin{minipage}{0.45\textwidth}
        \centering
        \includegraphics[width=\textwidth]{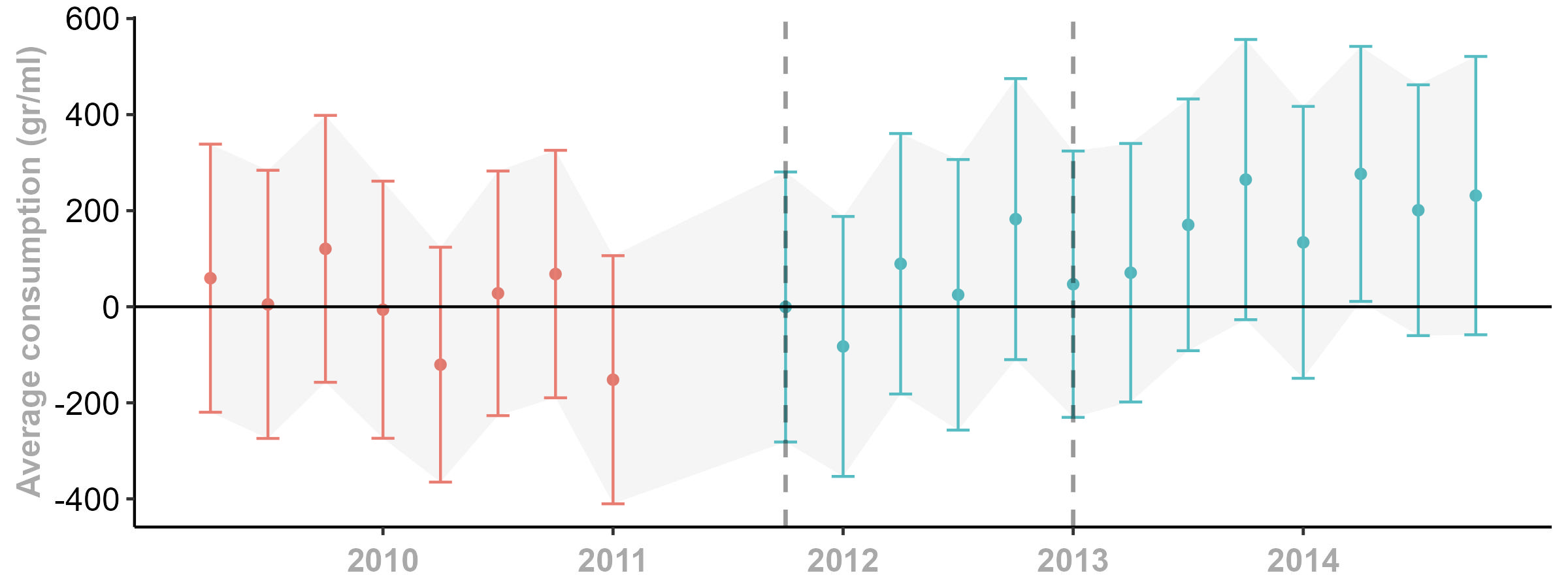}
        \caption{Cream.}
    \end{minipage}

     \hfill
     \begin{minipage}{0.45\textwidth}
        \centering
        \includegraphics[width=\textwidth]{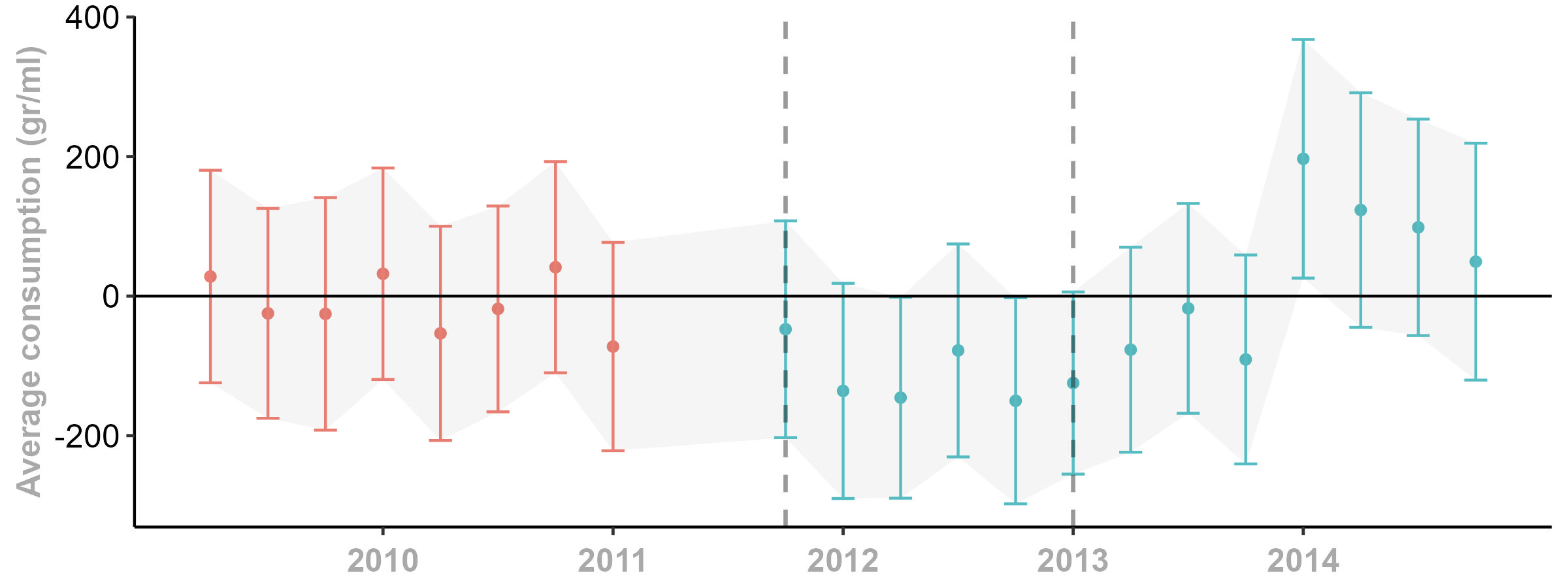}
        \caption{Liver sausage.}
    \end{minipage}
    \hfill
    \begin{minipage}{0.45\textwidth}
        \centering
        \includegraphics[width=\textwidth]{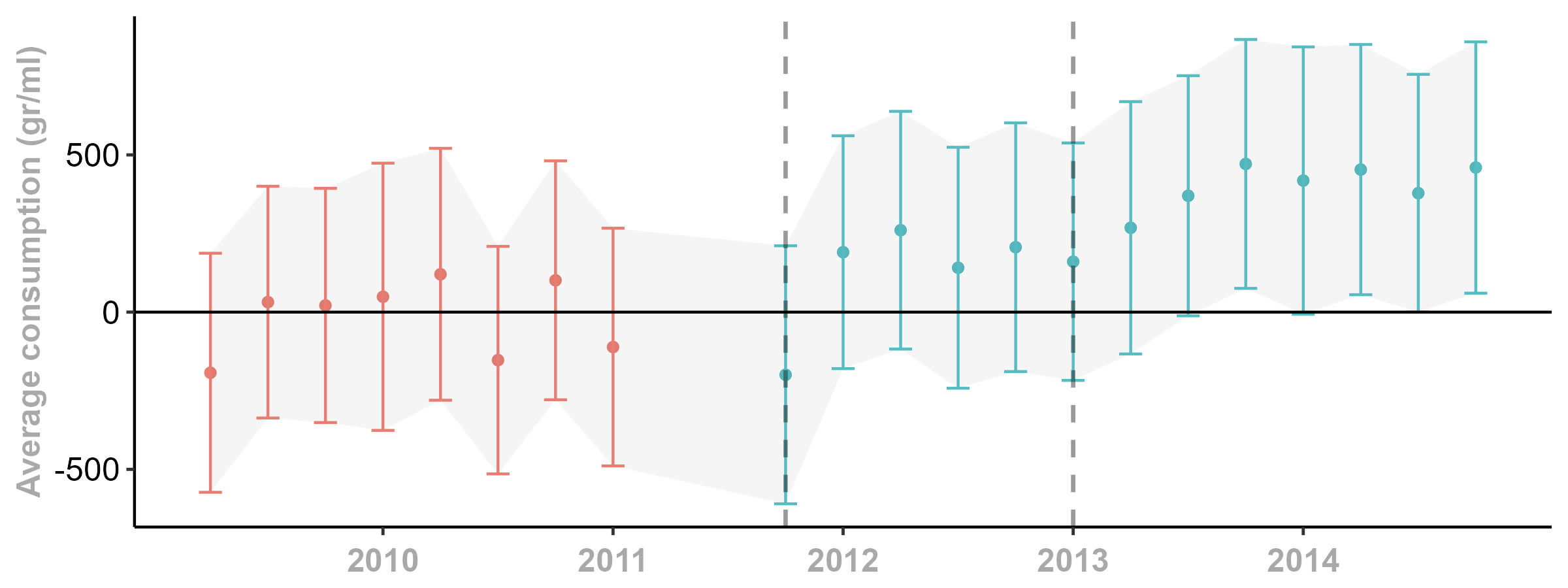}
        \caption{Margarine.}
    \end{minipage}
    \hfill

    \begin{minipage}{0.45\textwidth}
        \centering
        \includegraphics[width=\textwidth]{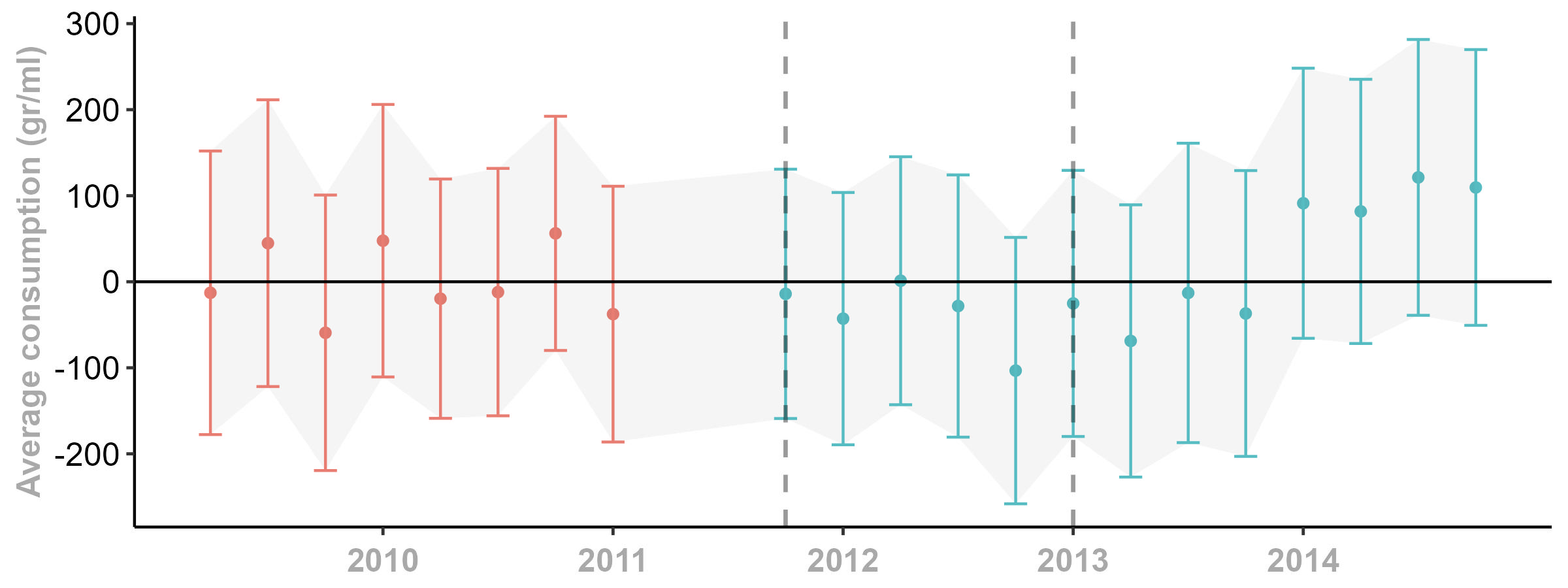}
        \caption{Salami.}
    \end{minipage}
    \hfill
    \begin{minipage}{0.45\textwidth}
        \centering
        \includegraphics[width=\textwidth]{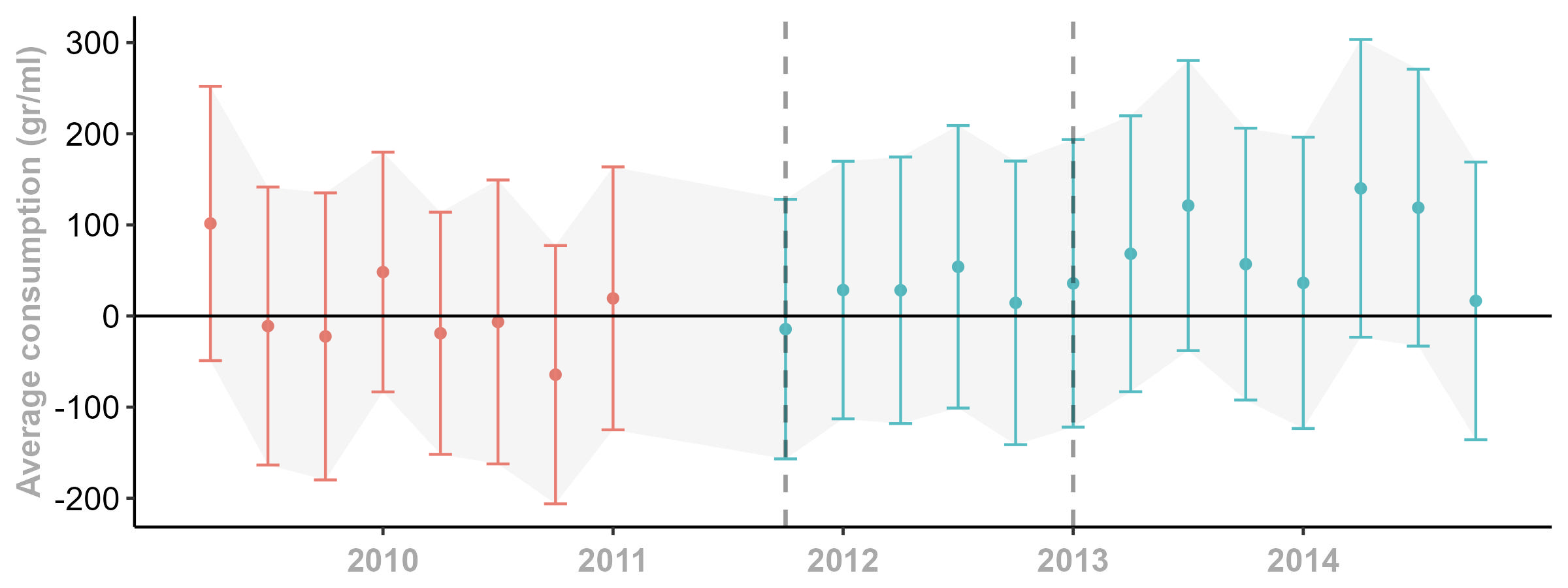}
        \caption{Sour cream.}
    \end{minipage}
    \caption{Average quarterly weight consumed in gr/ml for the different product categories. The figures depict the trends when unconditional parallel trends is assumed and provide insights into the strength of the influence of the household characteristics on the differences in trends between Northern-German and Danish households. The trends are similar to those observed under conditional parallel trends (see Appendix \ref{app:dynamic_treatments}), indicating that the household characteristics are not strongly influencing differences in consumption trends between Danish and Northern-German households.}
    \label{fig:unconditional_pta_weight}
\end{figure}

\clearpage

\subsection{Results under unconditional parallel trends for the total expenditure outcome}
\label{app:unconditional_pta_expenditure}
\FloatBarrier
\begin{figure}[!htbp]
\centering
\begin{minipage}{0.49\textwidth}
        \centering
        \includegraphics[width=\textwidth]{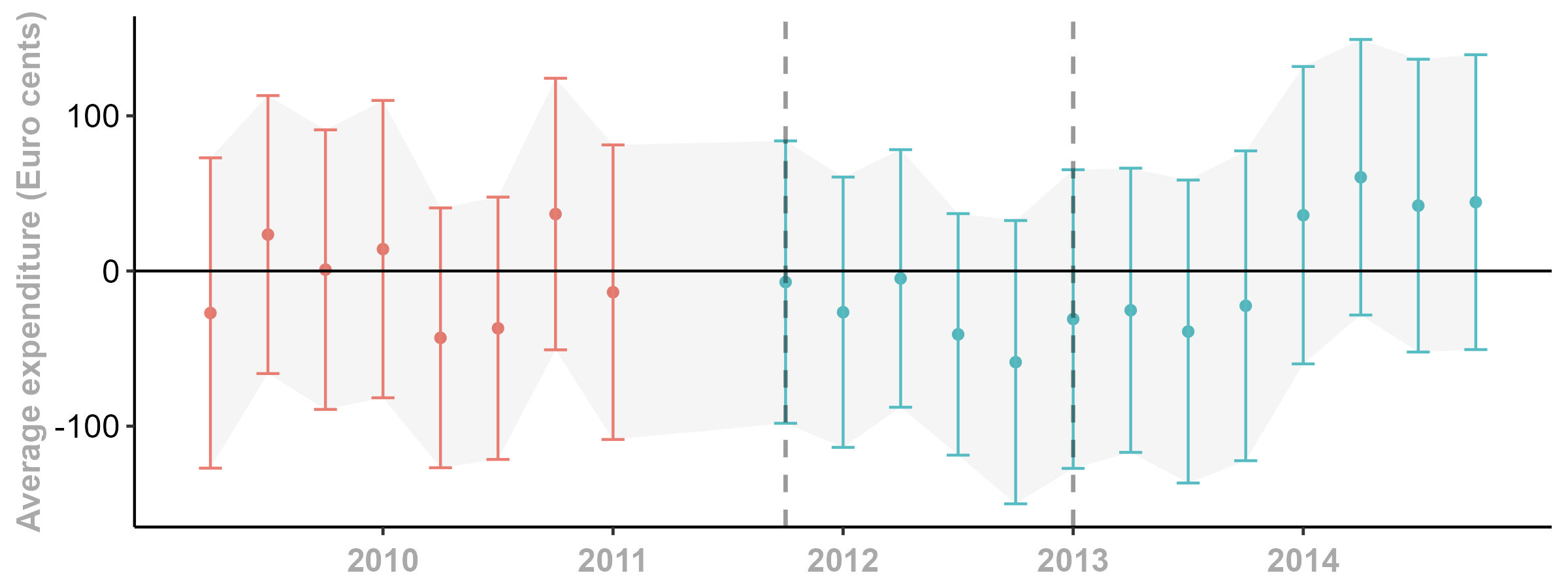}
        \caption{Bacon.}
    \end{minipage}
    \hfill
    \begin{minipage}{0.49\textwidth}
        \centering
        \includegraphics[width=\textwidth]{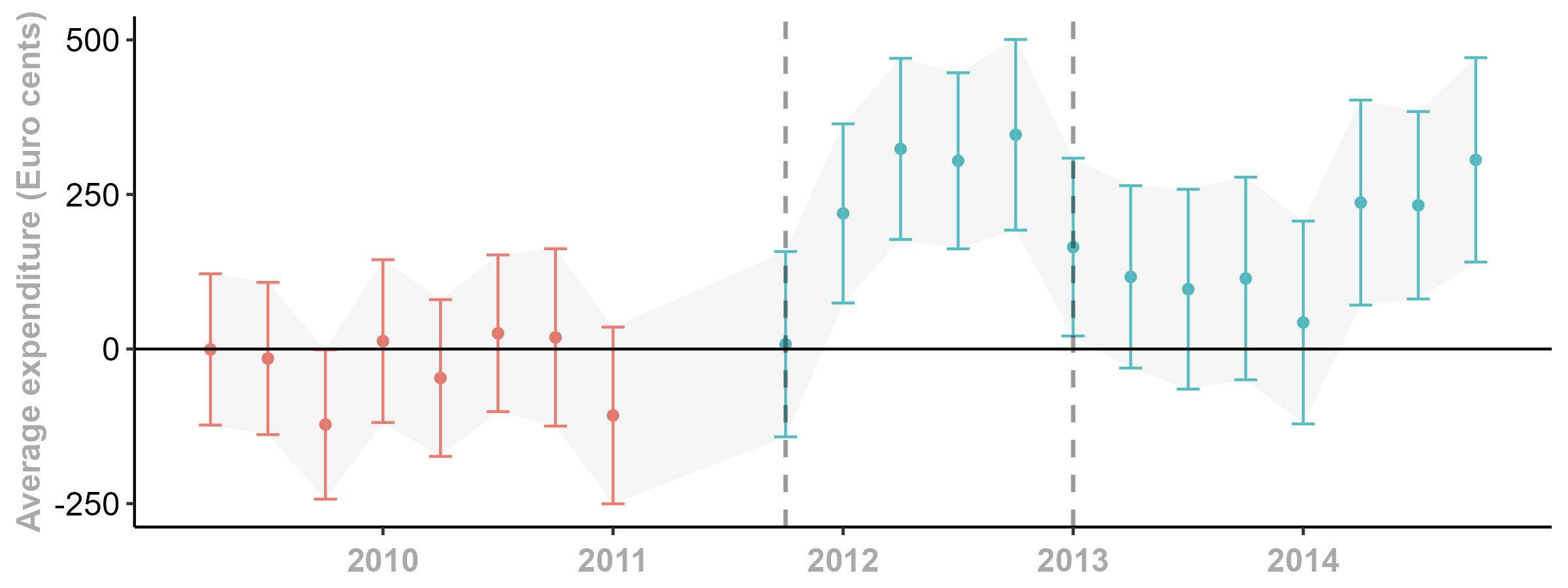}
        \caption{Butter.}
    \end{minipage}
    
    \begin{minipage}{0.45\textwidth}
        \centering
        \includegraphics[width=\textwidth]{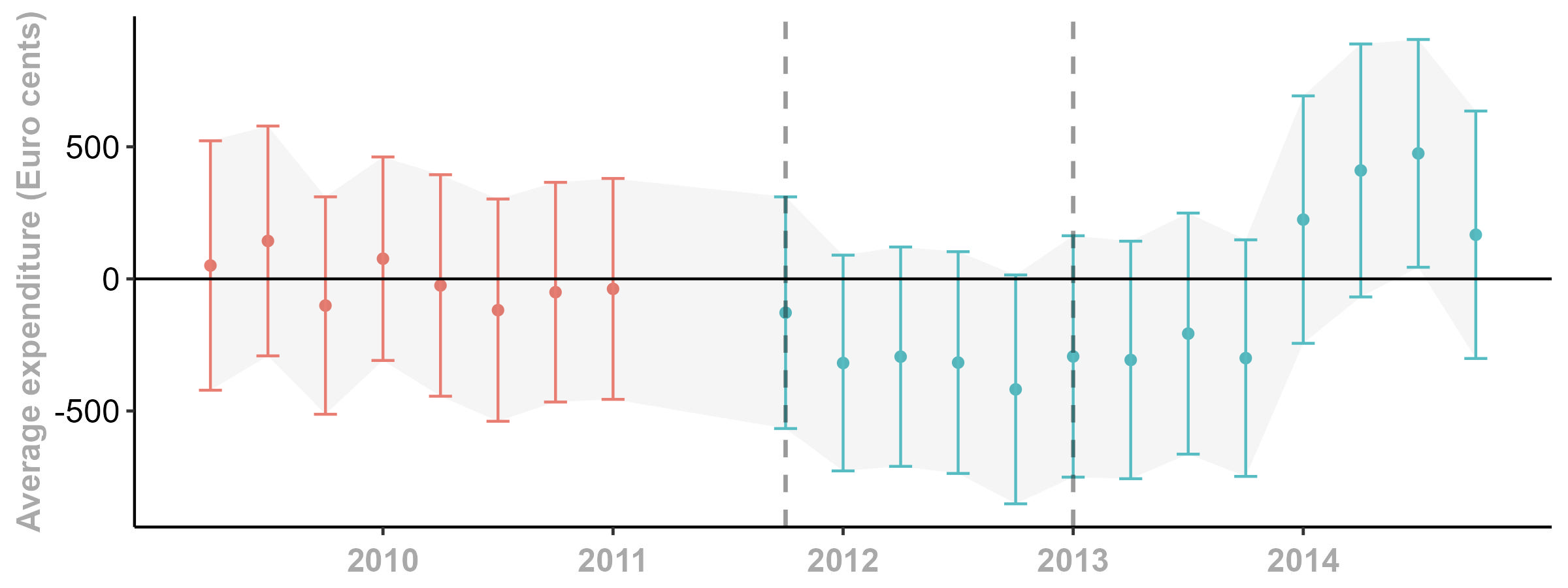}
        \caption{Cheese.}
    \end{minipage}
    \hfill
    \begin{minipage}{0.45\textwidth}
        \centering
        \includegraphics[width=\textwidth]{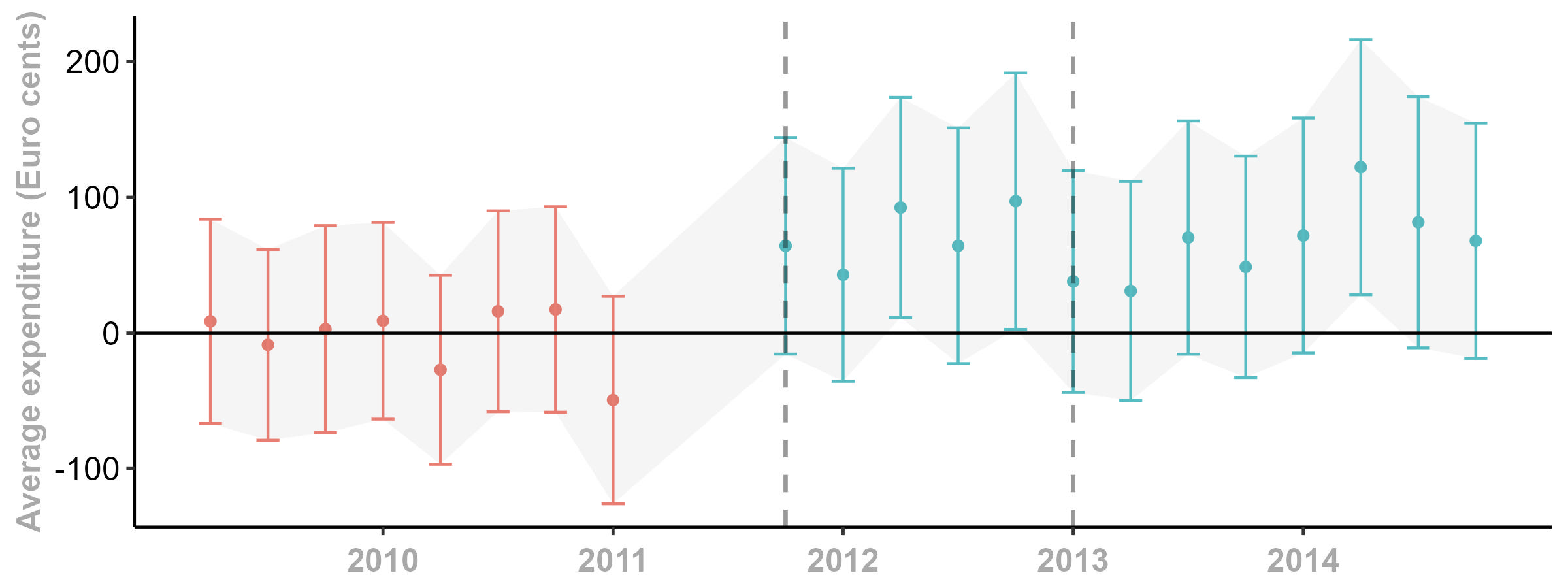}
        \caption{Cream.}
    \end{minipage}

    \hfill
     \begin{minipage}{0.45\textwidth}
        \centering
        \includegraphics[width=\textwidth]{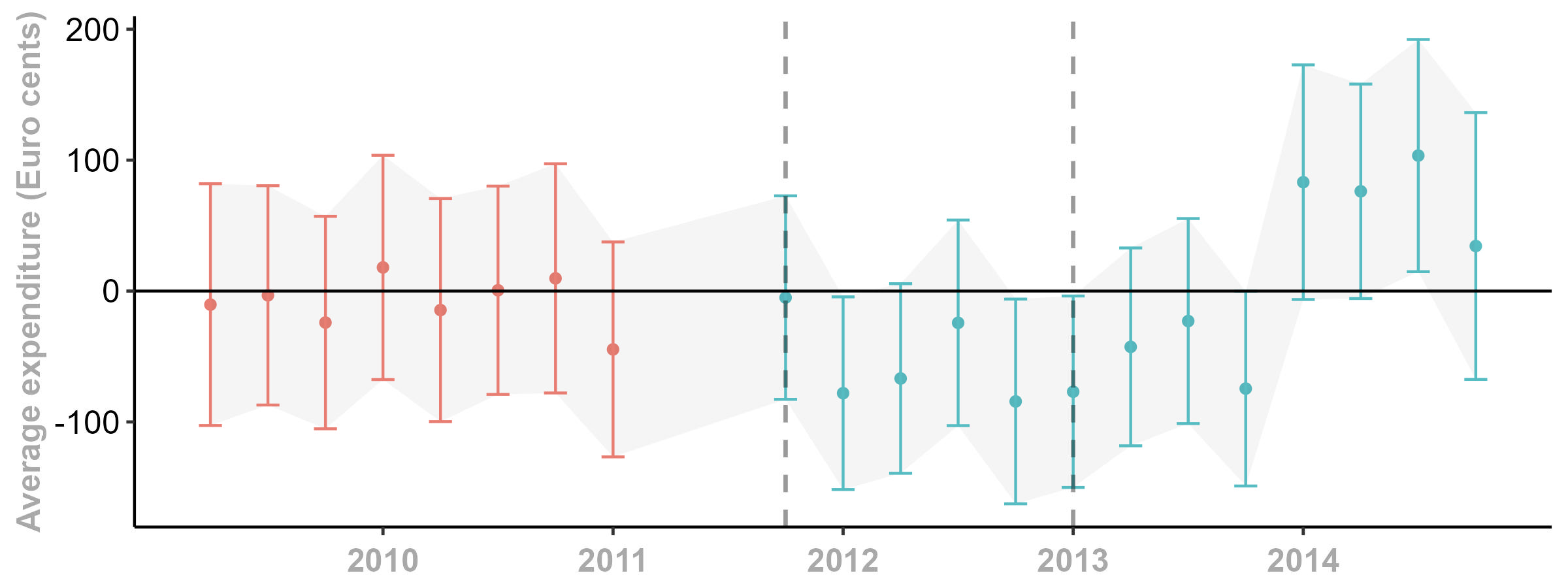}
        \caption{Liver sausage.}
    \end{minipage}
    \hfill
    \begin{minipage}{0.45\textwidth}
        \centering
        \includegraphics[width=\textwidth]{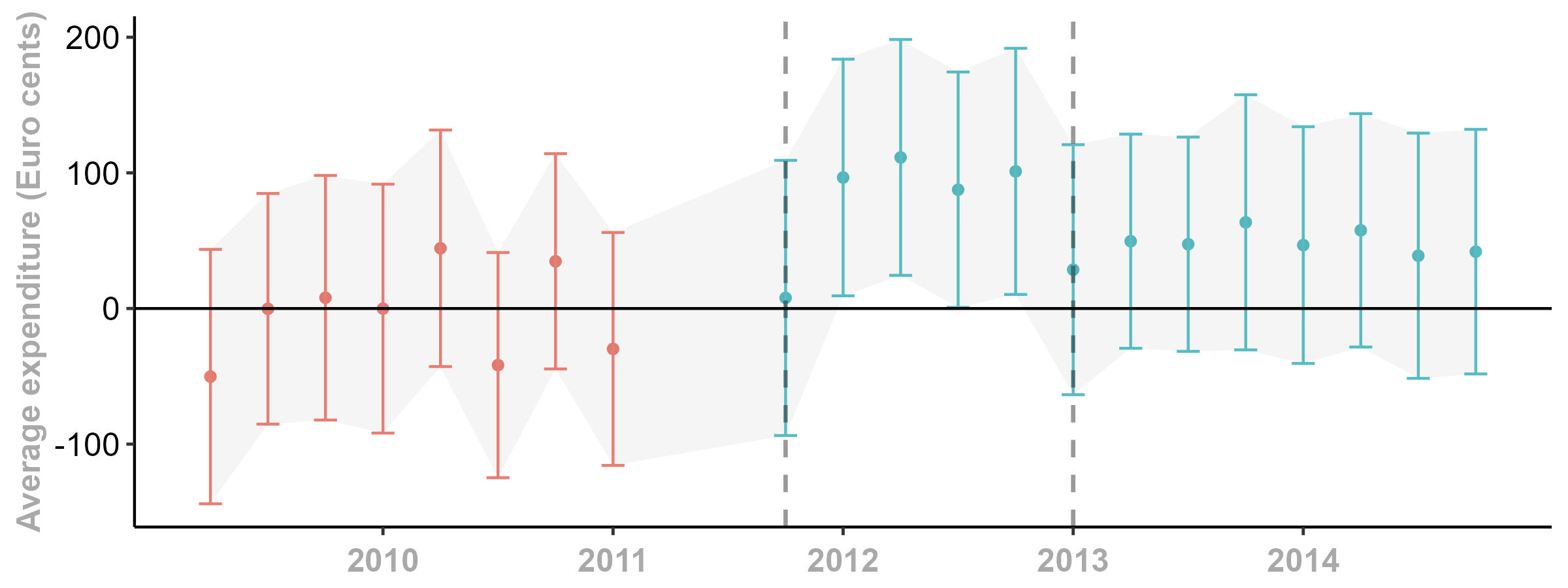}
        \caption{Margarine.}
    \end{minipage}

    \hfill
    \begin{minipage}{0.45\textwidth}
        \centering
        \includegraphics[width=\textwidth]{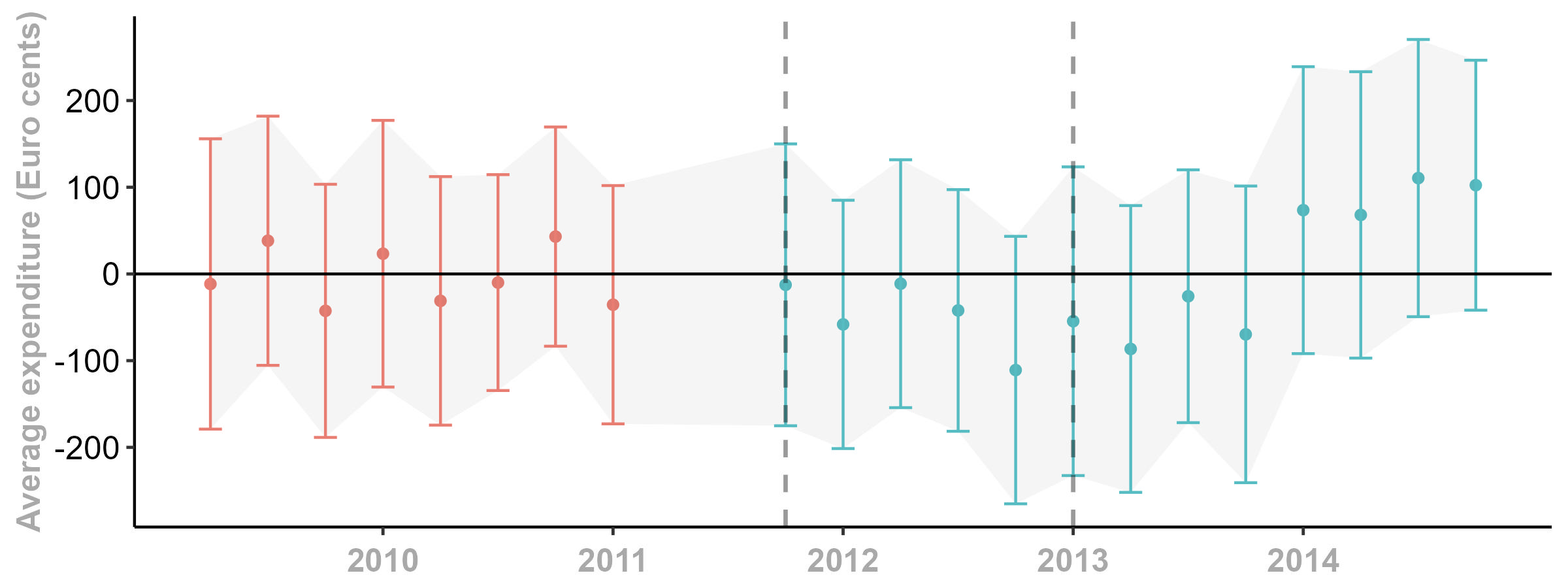}
        \caption{Salami.}
    \end{minipage}
    \hfill
    \begin{minipage}{0.45\textwidth}
        \centering
        \includegraphics[width=\textwidth]{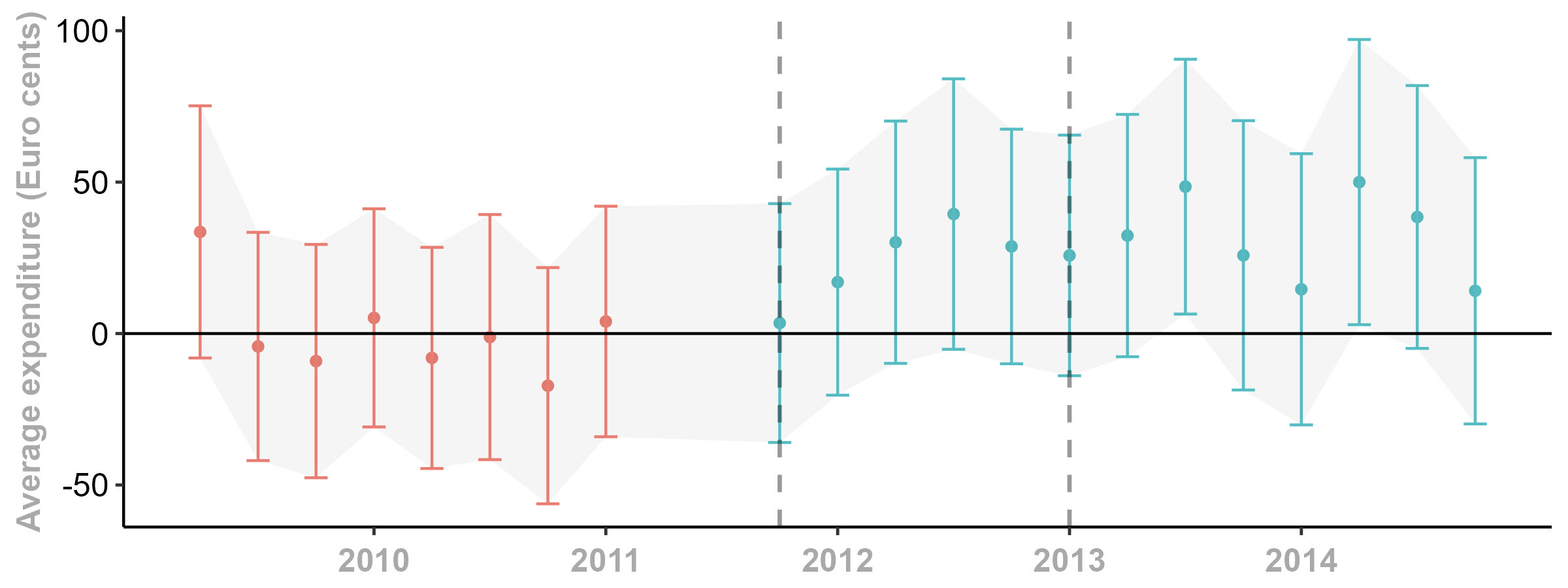}
        \caption{Sour cream.}
    \end{minipage}
    
    \caption{Average expenditure (Euro cents) for the different product categories. The figures depict the trends when unconditional parallel trends is assumed and provide insights into the strength of the influence of the household characteristics on the differences in trends between Northern-German and Danish households. The trends are similar to those observed under conditional parallel trends (see Appendix \ref{app:dynamic_treatments}), indicating that the household characteristics are not strongly influencing differences in consumption trends between Danish and Northern-German households.}
    \label{fig:unconditional_pta_expenditure}
\end{figure}

\end{appendices}

\end{document}